\documentclass[twocolumn,showpacs,preprintnumbers,amsmath,amssymb,pre]{revtex4}

\usepackage{graphicx}
\usepackage{dcolumn}
\usepackage{bm}

\newcommand{\bc}{\begin{center}}
\newcommand{\ec}{\end{center}}
\newcommand{\beq}{\begin{equation}}
\newcommand{\eeq}{\end{equation}}
\newcommand{\bea}{\begin{eqnarray}}
\newcommand{\eea}{\end{eqnarray}}
\newcommand{\ba}{\begin{array}}
\newcommand{\ea}{\end{array}}

\def\ie{\emph{i.e., }}
\def\etal{\emph{et al. }}

\def\sech{\ensuremath{\mathrm{sech}}}
\def\lapl{\ensuremath{\nabla^2}}
\def\Real{\ensuremath{\mathrm{Re}}}
\newcommand{\mat}[1]{\ensuremath{\mathbf #1}}
\newcommand{\sgn}[1]{\ensuremath{\mathrm{sgn}(#1)}}
\newcommand{\abs}[1]{\ensuremath{\left|#1\right|}}

\def\diag{\ensuremath{\mathrm{diag}}}

\def\eps{\epsilon}     
\def\tRC{\tilde{t}}    
\def\zdiff{z'}         

\def\ct{\tilde{c}}
\def\rhot{\tilde{\rho}}
\def\phit{\tilde{\phi}}
\def\psit{\tilde{\psi}}

\def\Psit{\tilde{\Psi}}
\def\qt{\tilde{q}}
\def\wt{\tilde{w}}

\def\phih{\hat{\phi}}

\def\cb{\bar{c}}
\def\rhob{\bar{\rho}}
\def\phib{\bar{\phi}}

\def\phih{\hat{\phi}}

\def\J{{\bf J}}
\def\F{{\bf F}}

\def\rholap{\check{\rho}}
\def\philap{\check{\phi}}
\def\qlap{\check{q}}

\begin{document}


\title{Nonlinear electrochemical relaxation around conductors}

\author{Kevin T. Chu$^{1,2}$ and Martin Z. Bazant$^1$}

\affiliation{
$^1$ Department of Mathematics, Massachusetts Institute of Technology,
Cambridge, MA 02139 \\
$^2$ Department of Mechanical and Aerospace Engineering, 
Princeton University, Princeton, NJ 08544 
}

\date{\today}

\begin{abstract}
  We analyze the simplest problem of electrochemical relaxation in
  more than one dimension -- the response of an uncharged, ideally
  polarizable metallic sphere (or cylinder) in a symmetric, binary
  electrolyte to a uniform electric field. In order to go beyond the
  circuit approximation for thin double layers, our analysis is based
  on the Poisson-Nernst-Planck (PNP) equations of dilute solution
  theory. Unlike most previous studies, however, we focus on the
  nonlinear regime, where the applied voltage across the conductor is
  larger than the thermal voltage. In such strong electric fields, the
  classical model predicts that the double layer adsorbs enough ions
  to produce bulk concentration gradients and surface conduction. Our
  analysis begins with a general derivation of surface conservation
  laws in the thin double-layer limit, which provide effective
  boundary conditions on the quasi-neutral bulk. We solve the
  resulting nonlinear partial differential equations numerically for
  strong fields and also perform a time-dependent asymptotic analysis
  for weaker fields, where bulk diffusion and surface conduction arise
  as first-order corrections. We also derive various dimensionless
  parameters comparing surface to bulk transport processes, which
  generalize the Bikerman-Dukhin number. Our results have basic
  relevance for double-layer charging dynamics and nonlinear
  electrokinetics in the ubiquitous PNP approximation.
\end{abstract}


\maketitle

\section{Introduction \label{sec:introduction}}
Diffuse-charge dynamics plays an important role in the response of
electrochemical and biological systems subject to time-dependent
voltages or electric fields~\cite{bazant2004}. The classical example
is impedance spectroscopy in
electrochemistry~\cite{sluyters1970,macdonald1990,parsons1990,geddes1997},
but electrochemical relaxation is also being increasingly exploited in
colloids and microfluidics~\cite{squires2005}.  For example, alternating
electric fields have been used to pump or mix liquid
electrolytes~\cite{ramos1999,ajdari2000,gonzalez2000,green2002,green2000b,
  ramos2003, studer2002,
  studer2004,nadal2002b,iceo2004b,iceo2004a,levitan2005}, to separate or
self-assemble colloids near electrodes~\cite{trau1997, yeh1997,
  faure1998, green2000a, marquet2002, nadal2002a, ristenpart2003}, and
to manipulate polarizable
particles~\cite{murtsovkin1996,iceo2004b,yariv2005,
  squires2006,rose2006b,rose2006,saintillon2006} or biological cells and
vesicles~\cite{helfrich1974, mitov1993,pethig1996}.

\begin{figure}
\bc
\includegraphics[width=3in]{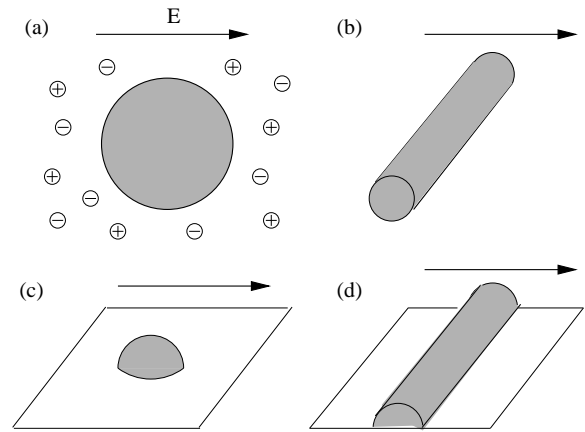}
\begin{minipage}[h]{2.9in}
\caption[Schematic diagram of model problem]{
\label{figure:metal_colloid_sphere_schematic}
(a) Schematic diagram of metallic colloidal sphere in a binary
electrolyte subjected to an applied electric field, which has the same
relaxation as a metallic hemisphere on a flat insulting
surface, shown in (c). We also consider the analogous
two-dimensional problems of a metallic cylinder (b), or a half
cylinder on an insulating plane (d).
}
\end{minipage}
\ec
\end{figure}

In this paper, we analyze some simple problems exemplifying the
nonlinear response of an electrolyte around an ideally polarizable
object due to diffusion and electromigration. As shown in
Figure~\ref{figure:metal_colloid_sphere_schematic}, we consider the
ionic relaxation around a metallic sphere (a) or cylinder (b) subject
to a suddenly applied uniform background electric field, as in
metallic colloids. Equivalently, we consider a metallic hemi-sphere
(c) or half-cylinder (d) on an insulating plane, to understand
relaxation around metallic structures on channel walls in
micro-electrochemical devices.  Although we do not consider fluid
flow, our analysis of nonlinear electrochemical relaxation is an
necessary first step toward understanding associated problems of
induced-charge electro-osmosis in the same
geometries~\cite{iceo2004b,iceo2004a,levitan2005,rose2006b,rose2006,saintillon2006},
and thus it also has relevance for the case of AC electro-osmosis at
planar electrode
arrays~\cite{ramos1999,ajdari2000,gonzalez2000,green2002,green2000b,
  ramos2003, studer2002, studer2004}.

In
electrochemistry~\cite{delahay_book,bockris_book,bard_book,newman_book}
and colloid science~\cite{hunter_book,lyklema_book_vol2}, it is common
to assume that the charged double-layer at a metal surface is very
thin and thus remains in quasi-equilibrium, even during charging
dynamics. As a result, for over a century~\cite{bazant2004}, the
standard model of electrochemical relaxation has been an equivalent
circuit, where the neutral bulk is represented by an Ohmic resistor
and the double layer by a surface
impedance~\cite{sluyters1970,macdonald1990,parsons1990,geddes1997},
which reduces to a linear capacitor at an ideally polarizable
surface. For our model problems, this ``RC circuit'' model was first
applied to electrochemical relaxation around a sphere by Simonov and
Shilov~\cite{simonov1977} and around a cylinder by Bazant and
Squires~\cite{iceo2004b,iceo2004a}. Similar RC-circuit analysis has
been applied extensively to planar electrode arrays in microfluidic
devices, following with Ramos et al.~\cite{ramos1999} and
Ajdari~\cite{ajdari2000}.

While convenient for mathematical analysis and often sufficiently
accurate, circuit models neglect the possibility of bulk concentration
gradients, which can arise at large applied voltages~\cite{bazant2004}
and/or when the surface is highly
charged~\cite{dukhin1969,hinch1983,hinch1984,shilov1970}, as well as
nonuniform surface transport of ions through the double
layer~\cite{bikerman1940,deryagin1969}. Dukhin and
Shilov~\cite{dukhin1969,shilov1970} and later Hinch, Sherwood, Chew,
and Sen~\cite{hinch1983,hinch1984} made significant progress beyond
the simple circuit model by including bulk diffusion in their studies
of double layer polarization around highly charged spherical particles
(of fixed charged density) in weak applied fields.  One of the main
results of their analysis is that for weak applied electric fields,
bulk concentration gradients appear as a small correction (on the
order of the applied electric field) to a uniform background
concentration.  In this work, we will lift the weak-field restriction
and consider the nonlinear response of the system to strong applied
fields, using the same mathematical model as in all prior work -- the
Poisson-Nernst-Planck (PNP) equations of dilute solution
theory~\cite{newman_book}.  As we shall see, nonlinear response
generally involves non-negligible bulk diffusion.

There are also difficulties with the traditional macroscopic picture
of the double layer as an infinitely thin surface impedance, or
possibly some more general nonlinear circuit element.  At the
microscopic level, the double layer has a more complicated structure
with at least two different regimes: a diffuse layer where the ions
move freely in solution and a compact surface layer, where ions may be
condensed in a Stern mono-layer with its own physical features (such
as surface capacitance, surface diffusivity, and surface
roughness)~\cite{delahay_book,bockris_book,bard_book}. The surface
capacitance may also include the effect of a dielectric coating, where
ions do not penetrate~\cite{ajdari2000}. Mathematically, in one
dimension the capacitor model can be derived as an effective boundary
condition for the neutral bulk (Nernst-Planck) equations by asymptotic
analysis of the PNP equations in the thin double-layer
limit~\cite{bazant2004}. Extending this analysis to higher dimensions,
however, requires allowing for tangential ``surface conduction''
through the double layer on the conductor for large applied electric
fields. Here, we derive effective boundary conditions for the neutral
bulk in the PNP model by following a general mathematical method for
surface conservation laws at microscopically diffuse
interfaces~\cite{chu_surf_cons_laws}.

The nonlinear response of electrochemical systems to strong applied
fields seems to be relatively unexplored. To our knowledge, the only
prior mathematical study of nonlinear relaxation with the PNP model is
the recent work of Bazant, Thornton, and Ajdari on the one-dimensional
problem of parallel-plate blocking electrodes applying a sudden pulsed
voltage~\cite{bazant2004}. (The same analysis has been recently
extended to ``modified-PNP'' models accounting for steric effects in
concentrated solutions~\cite{kilic2006}, which would also be an
important extension of our work.) For applied voltages in the
\emph{weakly nonlinear} regime, which is analogous to weak applied
fields in our problems, they find that the relaxation of the cell to
the steady state requires bulk diffusion processes that appear as a
small correction at $O(\eps)$, where the small parameter $\eps$ is the
ratio of the Debye length to a typical scale of the geometry. (In
contrast, in Refs.  ~\cite{dukhin1969,hinch1983,hinch1984,shilov1970},
it appears that primarily the the strength of the applied electric
field is assumed to be small, although the mathematical limit is not
explicitly defined.)  For applied voltages in the \emph{strongly
  nonlinear} regime, they show that bulk concentration gradients can
no longer be considered small, since $O(1)$ concentration variations
appear.  In both regimes, the absorption of neutral salt by the double
layer (and therefore build up of surface ion density) is the key
driving force for bulk diffusion. This positive salt adsorption in
response to an applied voltage, first noted in Ref.~\cite{bazant2004},
is opposite to the classical ``Donnan effect'' of salt
expulsion~\cite{lyklema_book_vol2}, which occurs if the surface
chemically injects or removes ions during the initial creation of the
double layer~\cite{lyklema2005}.

Bazant, \etal also emphasize the importance of both the charging \emph{and}
the diffusion time scales in the evolution of the electrochemical 
systems~\cite{bazant2004}.
Because circuit models inherently neglect diffusion processes, the only
characteristic dynamic time scale that appears is the so-called
RC charging time, $\tau_c = \lambda_D L/D$, where $\lambda_D$ is the 
Debye length, $L$ is the system size, and $D$ is the characteristic 
diffusivity of the ions~\footnote{Note that when charging is driven by 
an externally applied voltage or field, the relevant relaxation time for 
double layer charging is \emph{not} the often quoted Debye time, 
$\tau_D = \lambda_D^2/D$~\cite{hunter_book,lyklema_book_vol1}.  The Debye time 
is the correct characteristic response time for double layer charging only 
in the unphysical scenario where charge is instantaneously placed on the 
particle (as opposed to transported through the electrolyte)~\cite{ferry1948}.
This result has been discovered many times by different scientific 
communities~\cite{bazant2004,dukhin1993,dukhin1980,kornyshev1981,
macdonald1970} but only recently seems to be gaining widespread 
understanding.}.
However, when concentration gradients are present, diffusion processes 
and dynamics at the diffusion time scale, $\tau_L = L^2/D$ may be important.
Most theoretical analyses of electrochemical systems only consider
the dynamics at one of these two dominant time scales -- effectively 
decoupling the dynamics at the two time scales.
This decoupling of the dynamics is natural when one considers
the wide separation in the time scales that govern the evolution
of the system: $\tau_L \gg \tau_c$.  
An interesting discussion of how the two time scales are coupled using 
ideas related to time-dependent asymptotic matching is presented 
in ~Ref.\cite{bazant2004}.

The paper is organized as follows.  We begin in
section~\ref{sec:mathematical_model} by carefully considering the thin
double-layer limit of our model problems, which leads to effective
boundary conditions for the neutral-bulk equations in
section~\ref{sec:effective_bcs}. The most interesting new boundary conditions
are surface conservation laws, whose physical content we discuss in
detail in section~\ref{sec:surfproc}, where we also define
dimensionless parameters governing the importance of various surface
transport processes.  In section~\ref{sec:steady_response}, we explore
the steady response to large applied electric fields in our model
problems; a notable prediction is the formation of recirculating bulk
diffusion currents coupled to surface transport processes.  We then
turn to relaxation dynamics in the three regimes identified by Bazant
\etal~\cite{bazant2004}, using similar methods, albeit with nontrivial
modifications for two or three dimensions. We begin with the linear
response to a weak field in section ~\ref{sec:transient_response},
where we obtain exact solutions using transform methods for arbitrary
double-layer thickness and also consider AC electric fields. Next, in
section~\ref{sec:weakly_nonlinear_dynamics} we use boundary-layer
methods in space {\it and} time to analyze ``weakly nonlinear''
relaxation in somewhat larger fields, in the asymptotic limit of thin
double layers.  Finally, in
section~\ref{sec:strongly_nonlinear_dynamics} we comment on the
challenges of ``strongly nonlinear'' relaxation, where bulk diffusion
and surface conduction dominate the dynamics. We conclude in
section~\ref{sec:conclusions} by discussing limitations of the PNP
model and directions for future research.

\section{Mathematical Model \label{sec:mathematical_model}}

\subsection{ PNP Initial-Boundary-Value Problem }

As a model problem, we consider the response of an isolated, ideally
polarizable sphere (or cylinder) subjected to a uniform, applied
electric field, as shown in
Figure~\ref{figure:metal_colloid_sphere_schematic}.  For simplicity,
we focus only on a symmetric, binary electrolyte where both ionic
species have the same diffusivity and charge number.  In order to
study nonlinear effects and avoid imposing a time scale, we assume
that the uniform electric field is suddenly applied at $t = 0$.

As in most (if not all) prior work on electrochemical dynamics, we
assume the Poisson-Nernst-Planck equations of dilute solution
theory~\cite{newman_book}, 
\bea
 \frac{\partial C_+}{\partial t} &=&
   \nabla \cdot \left ( D \nabla C_+ + \frac{z_+eD}{kT} C_+ \nabla \Phi \right ) 
 \label{eq:dimensional_C+_eqn_bulk_diffusion_time} \\
 \frac{\partial C_-}{\partial t} &=&
   \nabla \cdot \left ( D \nabla C_- - \frac{z_+eD}{kT} C_- \nabla \Phi \right ) 
 \label{eq:dimensional_C-_eqn_bulk_diffusion_time} \\
 -\varepsilon_s \lapl \Phi &=& z_+e \left( C_+ - C_- \right)
 \label{eq:dimensional_poisson_eqn_bulk_diffusion_time}
 \eea where $D$ is the diffusivity, $z_+$ is the charge number of the
 positively charged species, $e$ is the charge of a proton, $k$ is
 Boltzmann's constant, $T$ is the absolute temperature, and
 $\varepsilon_s$ is the electric permittivity of the solution.  As
 usual, we have used the Nernst-Einstein relation to write the
 mobility in terms of the the diffusivity, $b = D/kT$. It is also
 useful to define the chemical potentials of the ions, 
\begin{equation}
\mu_\pm = kT \log C_\pm \pm z_+ e \Phi  
\label{eq:mu_dilute}
\end{equation}
from which their fluxes are defined as $\F_\pm = -b C_\pm \nabla
\mu_\pm$. 

At the conductor's surface,  we assume the same boundary conditions as
in Ref.~\cite{bazant2004}. We adopt a linear surface-capacitance condition
on the electrostatic potential~\cite{bazant2005},
\begin{equation}
 \Phi + \lambda_S
\frac{\partial \Phi}{\partial n} = V
    \label{eq:dimensional_stern_bc}
\end{equation}
where $\lambda_S$ is a length characterizing the compact-layer
surface capacitance (e.g. due to a Stern monolayer or a thin
dielectric coating), where $V$ is the potential of the conductor,
which is set either externally or by the condition of fixed total
charge~\cite{iceo2004b,iceo2004a,squires2006}.  (We will focus on the
case of zero total charge, where symmetry implies $V=0$ in our simple
geometries.) To focus on charging dynamics, we also assume an ideally
polarizable surface with no-flux boundary conditions: \bea D
\frac{\partial C_+}{\partial n} + \frac{z_+eD}{kT} C_+ \frac{\partial
  \Phi}{\partial n}
&=& 0 \label{eq:dimensional_C+_no_flux_bc} \\
D \frac{\partial C_-}{\partial n} - \frac{z_+eD}{kT} C_- \frac{\partial
  \Phi}{\partial n} &=& 0 \label{eq:dimensional_C-_rho_no_flux_bc}
\eea where the direction of the unit normal is taken to point inwards
towards the center of the sphere (\ie \emph{outwards} from the region
occupied by the electrolyte solution).  In the far field, we assume
that both the concentration and potential profiles tend toward their
initial conditions, given everywhere by 
\bea
C_{\pm}(R,\theta,\phi,t=0) &=& C_o \label{eq:dimensional_C_ic} \\
\Phi(R,\theta,\phi,t=0) &=& -E_o R \left( 1 - \frac{a^3}{R^3} \right)
\cos \theta
    \label{eq:dimensional_phi_ic}.
\eea
where $a$ is the radius of the sphere, $C_o$ is the bulk concentration
far away from the conductor, and $E_o$ is the applied electric field.
For the case of the cylinder, the initial conditions for the concentration 
profile remains the same and the initial electric potential takes the form
\bea
\Phi(R,\theta,t=0) &=& -E_o R \left( 1 - \frac{a^2}{R^2} \right)
\cos \theta
\eea

\subsection{Different Contributions to Ion Transport}

The transport equations
(\ref{eq:dimensional_C+_eqn_bulk_diffusion_time}) and
(\ref{eq:dimensional_C-_eqn_bulk_diffusion_time}) represent
conservation laws for the ionic species where the flux of each species
is a combination of diffusion and electromigration.  Because this
paper examines ionic fluxes in detail, let us take a moment to fix the
notation that we shall use to discuss different contributions to
charge and mass transport.  All fluxes will be denoted with the by the
variable $\F$ (or $F$ for scalar components of flux).  Superscripts
will be used to distinguish between the diffusion, $(d)$, and
electromigration, $(e)$, contributions to the flux.  Subscripts will
be used to denote the species or quantity with which the flux is
associated.  Finally, normal and tangential components of a flux will
denoted through the use of an extra subscript: $n$ for the normal
component and $t$ for the tangential component.  As examples,
$\F^{(e)}_+ = -\frac{z_+eD}{kT} C_+ \nabla \Phi$ represents the flux of
the positively charged species due to electromigration and
$F^{(d)}_{n,c} = -D \frac{\partial C}{\partial n}$ represents the flux
of the neutral salt concentration, $C = (C_+ + C_-)/2$, normal to a
surface arising from diffusion.  Table
\ref{tab:summary_of_bulk_fluxes} provides a summary of the various
bulk fluxes that shall arise in our discussion. (We also abuse
notation and use the same symbol $\mu$ for chemical potential with
dimensions and scaled to $kT$.)

\begin{table}
\caption{\label{tab:summary_of_bulk_fluxes}  Summary of notations for 
for the various ion fluxes and chemical potentials, before and after scaling.  }
\begin{ruledtabular}
\begin{tabular}{ccc}
  & Dimensional Formula~\footnote{
In these formulae, $\rho$ is half the charge density (\emph{not} the total 
charge density).  Also, we have abused notation and used the same variable 
$\rho$ for both the dimensional and dimensionless formulae.  $\rho$ in the 
dimensional formulae is equal to $C_o \rho$ in the dimensionless formulae.
}
& Dimensionless Formula \\
\hline \\
$\mu_\pm$ & $kT \log C_\pm \pm z_+e\Phi$ & $\log c_\pm \pm \phi$ \\[3pt]
\hline \\
 $\F_\pm$  & $-b C_\pm \nabla \mu_\pm$ & $-c_\pm \nabla \mu_\pm$ \\[3pt] 
\ & 
 $ -\left( D \nabla C_\pm \pm \frac{z_+eD}{kT} C_\pm \nabla \Phi \right )$ & 
 $ -\left( \nabla c_\pm \pm c_\pm \nabla \phi \right )$ \\[3pt]
 $\F^{(d)}_\pm$  & 
 $ -D \nabla C_\pm $ &
 $ -\nabla c_\pm $ \\[3pt]
 $\F^{(e)}_\pm$  & 
 $ \mp \frac{z_+eD}{kT} C_\pm \nabla \Phi $ & 
 $ \mp c_\pm \nabla \phi $ \\[3pt]
\hline \\
 $\F_c$  & 
 $ -\left ( D \nabla C + \frac{z_+eD}{kT} \rho \nabla \Phi \right )$ & 
 $ -\left( \nabla c + \rho \nabla \phi \right )$ \\[3pt]
 $\F^{(d)}_c$  & 
 $ -D \nabla C $ &
 $ -\nabla c $ \\[3pt]
 $\F^{(e)}_c$  & 
 $ -\frac{z_+eD}{kT} \rho \nabla \Phi$ & 
 $ -\rho \nabla \phi$ \\[3pt]
\hline \\
 $\F_\rho$  & 
 $ -\left( D \nabla \rho + \frac{z_+eD}{kT} C \nabla \Phi \right )$ & 
 $ -\left( \nabla \rho + c \nabla \phi \right )$ \\[3pt]
 $\F^{(d)}_\rho$  & 
 $ -D \nabla \rho $ & 
 $ -\nabla \rho$ \\[3pt]
 $\F^{(e)}_\rho$  & 
 $ -\frac{z_+eD}{kT} C \nabla \Phi$ & 
 $ -c \nabla \phi$ \\[3pt]
\end{tabular}
\end{ruledtabular}
\end{table}

\subsection{Dimensionless Formulation}
To facilitate the analysis of the model problem, it is convenient to 
nondimensionalize the governing equations and boundary conditions.
Scaling length by the radius of the sphere, $a$, the time to the diffusion 
time $\tau_D = a^2 / D$, and the electric potential by the thermal voltage 
divided by the cation charge number, $kT/z_+e$, the governing equations
(\ref{eq:dimensional_C+_eqn_bulk_diffusion_time}) --
(\ref{eq:dimensional_poisson_eqn_bulk_diffusion_time})
become
\bea
 \frac{\partial c_+}{\partial t} &=&
   \nabla \cdot \left ( \nabla c_+ + c_+ \nabla \phi \right ) 
 \label{eq:c+_eqn_bulk_diffusion_time} \\
 \frac{\partial c_-}{\partial t} &=&
   \nabla \cdot \left ( \nabla c_- - c_- \nabla \phi \right ) 
 \label{eq:c-_eqn_bulk_diffusion_time} \\
 -\eps^2 \lapl \phi &=& \left( c_+ - c_- \right) / 2
 \label{eq:poisson_eqn_bulk_diffusion_time}
\eea
where $c_\pm$, $\phi$, and $t$ are the dimensionless concentrations, 
electric potential and time, respectively, the spatial derivatives
are with respect to the dimensionless position, 
and $\eps$ is the ratio of the Debye screening length, 
$\lambda_D = \sqrt{\frac{\varepsilon_s k T}{2 z_+^2 e^2 C_o}}$, to the 
radius of the sphere.  The boundary conditions at the surface of the 
sphere and the initial conditions become
\bea
  \phi + \delta \eps \frac{\partial \phi}{\partial n} &=& v
    \label{eq:stern_bc} \\
  \frac{\partial c_+}{\partial n} + c_+ \frac{\partial \phi}{\partial n}
    &=& 0 \label{eq:c+_no_flux_bc} \\
  \frac{\partial c_-}{\partial n} - c_- \frac{\partial \phi}{\partial n}
    &=& 0 \label{eq:c-_no_flux_bc} \\
  c_\pm(r,\theta,\phi,t=0) &=& 1 \label{eq:c_pm_ic} \\
  \phi(r,\theta,\phi,t=0) &=& -E_o r \left( 1 - \frac{1}{r^3} \right) 
    \cos \theta \label{eq:phi_ic}.
\eea
In the far field, the dimensionless concentrations approach $1$ and the 
electric potential approaches $-E_o r \cos \theta$.
Note that in nondimensionalizing the surface capacitance boundary condition 
(\ref{eq:dimensional_stern_bc}), we have chosen to introduce a new
dimensionless parameter $\delta \equiv \lambda_S/\lambda_D$ which makes
it possible to study the effects of the Stern layer capacitance 
independently from the double layer thickness~\cite{bazant2005}.

Because the charge density and neutral salt concentration are important for
understanding the behavior of electrochemical transport at high applied 
fields, it is often useful to formulate the governing equations in terms
of the average concentration, $c = \left( c_+ + c_-  \right)/2$, and half the 
charge density, $\rho = \left( c_+ - c_-  \right)/2$ 
\cite{bazant2004,bazant2005, bonnefont2001}.  Using these definitions
(\ref{eq:c+_eqn_bulk_diffusion_time}) -- 
(\ref{eq:poisson_eqn_bulk_diffusion_time}) can be rewritten as
\bea
 \frac{\partial c}{\partial t} &=&
   \nabla \cdot \left ( \nabla c + \rho \nabla \phi \right ) 
 \label{eq:c_eqn_bulk_diffusion_time} \\
 \frac{\partial \rho}{\partial t} &=&
   \nabla \cdot \left ( \nabla \rho + c \nabla \phi \right ) 
 \label{eq:rho_eqn_bulk_diffusion_time} \\
 -\eps^2 \lapl \phi &=& \rho
 \label{eq:c_rho_poisson_eqn_bulk_diffusion_time}
\eea
Throughout our discussion, we shall alternate between this formulation and 
equations (\ref{eq:c+_eqn_bulk_diffusion_time}) -- 
(\ref{eq:poisson_eqn_bulk_diffusion_time}) depending on the context.
The initial and boundary conditions for this set of equations are easily 
derived from (\ref{eq:stern_bc}) -- (\ref{eq:phi_ic}).  Here we summarize
those boundary conditions that change:
\bea
  \frac{\partial c}{\partial n} + \rho \frac{\partial \phi}{\partial n}
    &=& 0 \label{eq:c_no_flux_bc} \\
  \frac{\partial \rho}{\partial n} + c \frac{\partial \phi}{\partial n}
    &=& 0 \label{eq:rho_no_flux_bc} \\
  c(r,\theta,\phi,t=0) &=& 1 \label{eq:c_ic} \\
  \rho(r,\theta,\phi,t=0) &=& 0 \label{eq:rho_ic}
\eea
For the remainder of this paper, we shall work primarily with 
dimensionless equations.  On occasion, we will mention the dimensional 
form of various expressions and equations to help make their physical
interpretation more apparent.

\subsection{Electroneutral Bulk Equations}
In the context of electrokinetics, it is desirable to reduce the complexity
of the electrochemical transport problem by replacing the PNP equations 
with a simpler set of equations that treats the bulk electrolyte and 
the double layer as separate entities.  
Circuit models \cite{bazant2004,iceo2004b,ramos2003,iceo2004a} have been 
used extensively to achieve this goal by reducing the transport problem to 
an electrostatics problem.
However, circuit models make the rather stringent assumption that bulk 
concentrations remain uniform at all times.  
Unfortunately, at high applied electric fields, this assumption is no longer 
valid because concentration gradients become important \cite{bazant2004}.  

In the present analysis, we consider an alternative simplification of the 
PNP equations that allows for bulk concentration variations.
Since we are interested in colloidal systems where particle diameters are 
on order of microns, $\eps$ is very small which suggests that 
we consider the thin double layer limit ($\eps \rightarrow 0$).  
In this limit, the bulk remains locally electroneutral, so it is 
acceptable to replace Poisson's equation with the local electroneutrality 
condition~\cite{newman_book,chu_thesis_2005}:
\bea
  \sum_i z_i c_i = 0 \label{eq:local_electroneutrality}.
\eea
For the case of symmetric, binary electrolytes, 
(\ref{eq:local_electroneutrality}) leads to the electroneutral Nernst-Planck 
equations:
\bea
 \frac{\partial c}{\partial t} &=& \nabla^2 c 
   \label{eq:dlc_c_eqn_bulk} \\
 0 &=& \nabla \cdot \left ( c \nabla \phi \right ) 
 \label{eq:dlc_rho_eqn_bulk} \\
 \rho &=& 0 \label{eq:dlc_LEN}.
\eea
Notice that the common practice of modeling the bulk electrolyte as a 
linear resistor obeying Ohm's law, $\nabla^2\phi=0$, arises from these
equations if the concentration profile is uniform.

We emphasize that these equations only describe the electrochemical 
system at the macroscopic level (\ie in the bulk region of the solution).  
The microscopic structure within the double layer, where local
electroneutrality breaks down, is completely neglected.  
Therefore, any physical effects of double layer structure can only be 
incorporated into the mathematical model via effective boundary conditions.

\section{Effective Boundary Conditions Outside the Double
  Layer \label{sec:effective_bcs}}
When local electroneutrality is used in place of Poisson's equation,
the physical boundary conditions imposed at at electrode surfaces must
be modified to account for the microscopic structure within the double
layer.  Fortunately, in the $\eps \rightarrow 0$ limit, the double
layer remains in quasi-equilibrium, so the Gouy-Chapman-Stern
model~\cite{bard_book} can be used to derive effective boundary
conditions for the system.  We emphasize that the GCS model is
\emph{not} an assumption; rather, it emerges as the leading-order
approximation in an asymptotic analysis of the thin double-layer
limit.  (See Ref.~\cite{bazant2004} for the history of this well known
result.)  Because the general form for the effective boundary
conditions of locally electroneutral electrochemical systems has not
been extensively discussed in the literature, we provide a detailed
derivation of these boundary conditions and discuss some associated
dimensionless parameters.

\subsection{Compact-layer Surface Capacitance}
We begin our discussion of effective boundary conditions by
considering the ``Stern boundary condition'' (\ref{eq:stern_bc}),
which describes the (linear) capacitance of a possible compact layer
on the surface.  Because Eq.~(\ref{eq:stern_bc}) involves the electric
potential and electric field at the inner edge of the diffuse charge
layer, it cannot be directly used as a boundary condition for the
locally electroneutral equations (\ref{eq:dlc_c_eqn_bulk}) --
(\ref{eq:dlc_LEN}).  However, by rearranging (\ref{eq:stern_bc}) and
using the GCS model, it is is possible to rewrite the Stern boundary
condition so that it only explicitly involves the electric potential
and average concentration at the macroscopic surface of the electrode
(\ie the outer edge of the diffuse charge
layer)~\cite{bazant2004,bazant2005}: 
\beq 
\zeta + 2 \delta \sqrt{c}
\sinh \left( \zeta/2 \right) = v - \phi.
  \label{eq:stern_bc_GCS}
\eeq
Here $\zeta$ is the potential drop across the diffuse part of the 
double layer (\ie zeta-potential), $v$ is the potential of the metal 
sphere in the thermal voltage scale, and $\phi$ and $c$ are the values 
of the electric potential and average concentration at the 
outer edge of the diffuse charge layer.
With dimensions, (\ref{eq:stern_bc_GCS}) is given by
\beq
  \tilde{\zeta} 
  + 2 \lambda_s \sqrt{\frac{2 k T C}{\varepsilon_s}} 
      \sinh \left( \frac{z_+ e \bar{\zeta}}{2 kT} \right)
  = V - \Phi
\eeq
where $\bar{\zeta}$ is the dimensional zeta-potential.  
Note that the GCS model for the double layer leads to the conclusion that 
the only dependence of the normal derivative of the electric potential 
on the double layer structure is through the zeta-potential.
A more detailed derivation of this form of the Stern boundary condition
can be found in~\cite{bazant2005}.

\subsection{Diffuse-layer Surface Conservation Laws \label{sec:ionic_fluxes}}
The effective boundary conditions for ionic fluxes are more 
complicated.  Because the physical domain for the macroscopic equations
(\ref{eq:dlc_c_eqn_bulk}) -- (\ref{eq:dlc_LEN}) excludes the diffuse
part of the double layer, the no-flux boundary conditions do not apply
-- it is possible for there to exist ion flux between the bulk region and 
the double layer.  Moreover, there is also the possibility of ion transport 
within the double layer itself (often neglected) which must be accounted for.  

\begin{figure}
\bc
\includegraphics[width=2.0in]{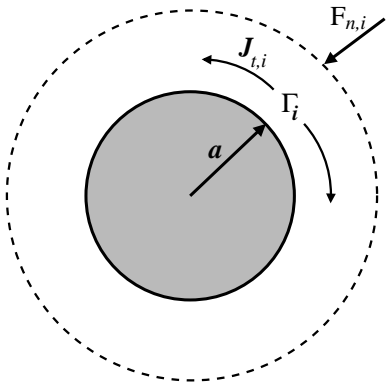}
\begin{minipage}[h]{3in}
\caption[Schematic diagram of normal and tangential fluxes.]{
\label{figure:surface_conservation_law}
Schematic diagram of normal and tangential fluxes involved in the
surface conservation law (\ref{eq:general_form_surf_conserv_law}).
The shaded circle represents the metal particle; the dotted circle
represents the outer ``edge'' of the double layer.
}
\end{minipage}
\ec
\end{figure}

The derivation of the effective flux boundary conditions
(\ref{eq:effective_flux_bc}) is based on a general theory of surface
conservation laws at microscopically diffuse interfaces, which we develop
in Ref.~\cite{chu_surf_cons_laws}.  The basic physical idea is to integrate
out the spatial variation within the double layer in the direction
normal to the electrode-electrolyte interface.  While intuitively
obvious, carrying out the integration involves careful use of
asymptotic analysis to address the mathematical subtleties of
integration over boundary layers.  The
theory tells us that effective flux
boundary conditions have the physically apparent form 
\beq \frac{\partial
  \Gamma_i}{\partial t} = -\nabla_s \cdot \J_{t,i} + F_{n,i}
  \label{eq:general_form_surf_conserv_law}
\eeq
where $(\nabla_s \cdot )$ denotes surface divergence, $\J_{t,i}$ is the 
tangential flux within and $F_{n,i}$ is the normal flux into the boundary 
layer (see Figure~\ref{figure:surface_conservation_law}).
The effective fluxes $\J_{t,i}$ and $F_{n,i}$ are directly related to the 
flux $\F_i$ for the transport process via
\bea 
  \J_{t,i} &=& \eps \int_0^\infty 
    \left( \tilde{\F}_{t,i} - \hat{\F}_{t,i} \right) dz
  \label{eq:general_boundary_layer_fluxes_s} \\
  F_{n,i} &=& \F_i \cdot \hat{n}
  \label{eq:general_boundary_layer_fluxes_n} 
\eea
where $\tilde{\F}$ and $\hat{\F}$ denote the flux within the boundary 
layer and the flux in the bulk just outside of the boundary layer and
the integration is over the entire boundary layer (\ie $z$ is
the inner variable of a boundary layer analysis).
For electrochemical transport, the flux is given by the Nernst-Planck 
equation 
\beq
  \F_i = - \left( \nabla c_i + z_i c_i \nabla \phi \right).
  \label{eq:nernst_planck_flux}
\eeq
Substituting this expression into 
(\ref{eq:general_boundary_layer_fluxes_s}) --
(\ref{eq:general_boundary_layer_fluxes_n}), rearranging a bit, 
and using the definition of a surface excess 
concentration~\cite{chu_surf_cons_laws,hunter_book,lyklema_book_vol2} 
\beq
  \Gamma_i \equiv \eps \int_0^\infty \gamma_i dz 
    = \eps \int_0^\infty \left( \ct_i - \hat{c}_i \right) dz,
  \label{eq:dlc_gamma_def}
\eeq
we obtain
\bea
  \J_{t,i} &=& 
     -\left( \nabla_s \Gamma_i + z_i \Gamma_i \nabla_s \hat{\phi}
                + \eps z_i \int_0^\infty \ct_i \nabla_s \psit dz \right)
  \label{eq:ion_transport_fluxes_s} \\
  F_{n,i} &=& -\frac{\partial c_i}{\partial n} 
          -z_i c_i \frac{\partial \phi}{\partial n}.
  \label{eq:ion_transport_fluxes_n}
\eea 
where $\psit \equiv \phit - \hat{\phi}$ is the excess electric potential
within the boundary layer and $\nabla_s$ denotes a surface 
gradient. 
As before, the tilde ($\tilde{\ }$) and hat ($\hat{\ }$) accents denote the 
quantities within boundary layer and the in the bulk immediately outside of 
the boundary layer.
Finally, the effective flux boundary conditions for electrochemical 
transport follow by using these results in 
(\ref{eq:general_form_surf_conserv_law}):
\bea
  \frac{\partial \Gamma_i}{\partial t} &=& 
     \nabla_s \cdot 
     \left[ 
        \nabla_s \Gamma_i + z_i \Gamma_i \nabla_s \phi 
      + \eps z_i \int_0^\infty \ct_i \nabla_s \psit ~dz
     \right] \nonumber \\
    &-& \left( \frac{\partial c_i}{\partial n} 
          +z_i c_i \frac{\partial \phi}{\partial n} \right)
  \label{eq:effective_flux_bc}
\eea
Note that even though our choice of electric potential scale eliminates the 
need to explicitly refer to the ionic charge numbers, $z_i$, we opt to 
continue using them in the present discussion so that it is clear where the 
charge number should appear for alternative choices of the electric potential 
scale; in the following discussion, the $z_i$ are essentially the sign of 
the ``dimensional'' ionic charge numbers.

There are a few important features of (\ref{eq:effective_flux_bc}) worth
mentioning.  First, the surface transport term (first term on the right
hand side) does not always contribute to the leading order effective flux 
boundary condition.  
Whether the surface conduction term must be retained at leading 
order depends on the magnitudes of $\Gamma_i$ (which in turn depends on
the zeta-potential) and the tangential component of the bulk electric 
field.  Interestingly, when the the surface transport term is significant,
the flux boundary condition depends explicitly on the small 
parameter~$\eps$.  Also, (\ref{eq:effective_flux_bc}) allow for two important 
physical effects that only arise for 2- and 3-D systems: 
(i) non-uniform double layer charging and 
(ii) surface transport within the double layer itself.
The presence of these effects lead to richer behavior for 2- and 3-D systems
compared to the 1D system studied in~\cite{bazant2004}.

To put (\ref{eq:effective_flux_bc}) into a more useful form, we use
the GCS model of the double layer to express the surface flux densities
in terms of the zeta-potential and the bulk concentration. 
From the GCS model~\cite{bard_book,newman_book}, we 
know that the excess concentration of each ionic species is given by 
\beq
  \gamma_i = \ct_i - \hat{c}_i = \hat{c} \left ( e^{- z_i \psit} - 1 \right )
  \label{eq:dlc_gamma}
\eeq
and that 
\beq
  \frac{\partial \psit}{\partial z} = 
    -2 \sqrt{\hat{c}} \sinh \left( \frac{z_+ \psit}{2} \right).
  \label{eq:dlc_dpsi_dz}
\eeq
Using these expressions, it is straightforward to show that the surface 
excess concentration is
\beq
  \Gamma_i = \frac{2 \eps \sqrt{\hat{c}}}{\abs{z_i}} 
    \left( e^{-z_i \zeta/2} - 1 \right).
  \label{eq:dlc_Gamma}
\eeq
Therefore, the first two surface flux density terms in 
(\ref{eq:effective_flux_bc}) can be written as
\bea
  \nabla_s \Gamma_i + z_i \Gamma_i \nabla_s \hat{\phi} &=& 
  \frac{\Gamma_i}{2} \nabla_s \log \hat{c}
  + z_i \Gamma_i \nabla_s \hat{\phi} \nonumber \\
  &-& \ \eps~\sgn{z_i} \sqrt{\hat{c}}~e^{- z_i \zeta/2} \nabla_s \zeta.
  \label{eq:dlc_surface_flux_terms_1_and_2}
\eea
To evaluate the last term in the surface flux density, we observe that
\beq
  \nabla_s \psit = \frac{\partial \psit}{\partial z} 
                   \left( -\frac{\nabla_s \zeta}
                   {2 \sqrt{\hat{c}}~\sinh \left( \zeta/2 \right)}
                 + \frac{z}{2} \nabla_s \log \hat{c} \right),
\eeq
which follows directly by comparing the normal and surface derivatives
of the leading order expression for the electric potential within 
the double layer~\cite{bazant2005,bonnefont2001,bazant2004}
\beq
\psit(z) = 4 \tanh^{-1}\left(\tanh(\zeta/4)
e^{-\sqrt{\hat{c}}z}\right). 
\label{eq:psi_gc_solution}
\eeq
Using this result, the integral in (\ref{eq:effective_flux_bc}) greatly 
simplifies and yields
\bea
  \eps z_i \int_0^\infty \ct_i \nabla_s \psit ~dz &=&
  \eps~\sgn{z_i} \sqrt{\hat{c}}~e^{-z_i \zeta/2} \nabla_s \zeta \nonumber \\
  &+& \frac{\Gamma_i}{2} \nabla_s \log \hat{c}
  \label{eq:dlc_surface_flux_term_3}.
\eea
Finally, combining (\ref{eq:dlc_surface_flux_terms_1_and_2}) and 
(\ref{eq:dlc_surface_flux_term_3}), the effective flux boundary condition
(\ref{eq:effective_flux_bc}) becomes 
\bea
  \frac{\partial \Gamma_i}{\partial t} &=& 
     \nabla_s \cdot 
     \left[ \Gamma_i \nabla_s \log \hat{c}
            + z_i \Gamma_i \nabla_s \hat{\phi} \right] \nonumber \\
    & & - \left( \frac{\partial c_i}{\partial n} 
          +z_i c_i \frac{\partial \phi}{\partial n} \right)
  \label{eq:effective_flux_bc_GCS}
\eea
where (dropping hats) $c$ and $\phi$ are understood to be in the bulk,
just outside the double layer.  Notice that the tangential gradients
in the zeta-potential have vanished in this equation so that the
surface flux density of the individual species is independent of
$\nabla_s \zeta$.

In terms of the (dimensionless) chemical potentials of the ions,
\beq
\mu_i = \log c_i + z_i \hat{\phi},
\eeq
the surface conservation law (\ref{eq:effective_flux_bc_GCS}) reduces
to a very simple form,
\begin{equation}
\frac{\partial \Gamma_i}{\partial t} =
\nabla_s\cdot(\Gamma_i\nabla\mu_i) - \hat{n}\cdot c_i\nabla\mu_i   \label{eq:mubc}
\end{equation}
where it is clear that the tangential gradient in the {\it bulk}
chemical potential just outside the double layer drives the transport
of the {\it surface} excess concentration of each ion, as well as the
bulk transport. This form of the surface conservation law should hold
more generally, even when the chemical potential does not come from
dilute solution theory (PNP equations).

Effective boundary conditions similar to
(\ref{eq:effective_flux_bc_GCS}) describing the dynamics of the double
layer have been known for some time.  In the late 1960s, Dukhin,
Deryagin, and Shilov essentially used (\ref{eq:effective_flux_bc_GCS})
in their studies of surface conductance and the polarization of the
diffuse charge layer around spherical particles with thin double
layers at weak applied fields~\cite{deryagin1969,dukhin1969,
  shilov1970}.  Later, Hinch, Sherwood, Chew, and Sen used similar
boundary conditions in their extension of the work of Dukhin \etal to
explicitly calculate the tangential flux terms for a range of large
surface potentials and asymmetric
electrolytes~\cite{hinch1983,hinch1984}.  A key feature of both of
these studies is the focus on small deviations from bulk equilibrium.
As a result, the effective boundary conditions used are basically
applications of (\ref{eq:effective_flux_bc_GCS}) for weak
perturbations to the background concentration and electric potential.
Our work differs from these previous analyses because we do not
require that the bulk concentration only {\em weakly} deviates from a
uniform profile, and we more rigorously justify the approximation
through the use of matched asymptotics.  In addition, our analysis is
not restricted to the use of the GCS model for the double layer;
equations (\ref{eq:ion_transport_fluxes_s}) --
(\ref{eq:effective_flux_bc}) are valid even for more general models of
the boundary layer.

Before moving on, we write the effective boundary conditions in 
dimensional form to emphasize their physical interpretation: 
\bea
  \frac{\partial \bar{\Gamma}_i}{\partial t} &=& 
  \nabla_s \cdot 
    \left[ D \bar{\Gamma}_i
               \nabla_s \log \left( \frac{C}{C_o} \right)
         + \frac{z_i e D}{kT} \bar{\Gamma}_i \nabla_s \Phi \right] \nonumber \\
   &-& \ \left( D_i \frac{\partial C_i}{\partial n} 
          + \frac{z_i e D}{kT} C_i \frac{\partial \Phi}{\partial n} \right),
\eea
where $\bar{\Gamma}_i$ is the dimensional surface excess concentration
of species $i$ and is defined as
\beq
  \bar{\Gamma}_i \equiv \int_{dl} \left( \tilde{C}_i - C_i \right) dZ
\eeq
where the integration is only over the double layer.  With the units 
replaced, it becomes clear that the effective boundary conditions is 
a two-dimensional conservation law for the excess surface concentration 
$\Gamma_i$ with a driving force for the flux and a source term that 
depend only on the dynamics away from the surface.  Thus, the effective
boundary conditions naturally generalize the simple capacitor picture of 
the double layer to allow for flow of ionic species tangentially along
the electrode surface.

\subsection{Surface Charge and Excess Salt Concentration}
Since the governing equations (\ref{eq:dlc_c_eqn_bulk}) -- (\ref{eq:dlc_LEN})
are formulated in terms of the salt concentration and charge density, 
it is convenient to derive boundary conditions that are directly related to 
these quantities.  Toward this end, we define $\eps q$ and $\eps w$ to be 
the surface charge density and surface excess salt concentration, 
respectively:
\bea
  q &=& \int_0^\infty \rhot dz 
     = \frac{1}{2} \int_0^\infty \left( \gamma_+ - \gamma_- \right) dz 
  \label{eq:dlc_q_def} \\
  w &=& \int_0^\infty \left( \ct - \hat{c} \right) dz 
     = \frac{1}{2} \int_0^\infty \left( \gamma_+ + \gamma_- \right) dz 
  \label{eq:dlc_w_def}.
\eea
By integrating the expressions for the diffuse layer salt 
concentration and charge density~\cite{hunter_book,bazant2004,newman_book,
bard_book}, 
\bea
  \ct &=& \hat{c} \cosh{\psit}
  \label{eq:c_diffuse_layer} \\
  \rhot &=& - \hat{c} \sinh{\psit}
  \label{eq:rho_diffuse_layer}
\eea
and
using (\ref{eq:dlc_dpsi_dz}), both $q$ and $w$ can be expressed as simple 
functions of the zeta-potential and the bulk concentration just outside 
of the double layer:
\bea
  q &=& -2 \sqrt{\hat{c}} \sinh(\zeta/2) 
  \label{eq:dlc_q_def_GCS} \\
  w &=& 4 \sqrt{\hat{c}} \sinh^2(\zeta/4).
  \label{eq:dlc_w_def_GCS} 
\eea
Thus, we can combine the effective flux boundary conditions for individual 
ions (\ref{eq:effective_flux_bc}) to obtain
\bea
  \eps \frac{\partial q}{\partial t} &=& 
     \eps \nabla_s \cdot 
     \left[ 
        \nabla_s q + w \nabla_s \hat{\phi}
      + \int_0^\infty \ct \nabla_s \psit ~dz
     \right] \nonumber \\
    & & -\ c \frac{\partial \phi}{\partial n} 
  \label{eq:q_evolution_eqn} \\
  \eps \frac{\partial w}{\partial t} &=& 
     \eps \nabla_s \cdot 
     \left[ 
        \nabla_s w + q \nabla_s \hat{\phi}
      + \int_0^\infty \rho \nabla_s \psit ~dz
     \right] \nonumber \\
    & & -\ \frac{\partial c}{\partial n}.
  \label{eq:w_evolution_eqn} 
\eea
Notice that, as in the PNP equations written in terms of $c$ and $\rho$, 
there is a symmetry between $q$ and $w$ in these equations.
As in the previous section, we can use the GCS model to rewrite
(\ref{eq:q_evolution_eqn}) and (\ref{eq:w_evolution_eqn}) solely
in terms of bulk field variables and the zeta-potential:
\bea
  \eps \frac{\partial q}{\partial t} &=&
     \eps \nabla_s \cdot 
     \left(
         q \nabla_s \log \hat{c}
       + w \nabla_s \hat{\phi}
     \right)
     - c \frac{\partial \phi}{\partial n} 
  \label{eq:q_evolution_eqn_GCS} \\
  \eps \frac{\partial w}{\partial t} &=& 
     \eps \nabla_s \cdot 
     \left(
         w \nabla_s \log \hat{c}
       + q \nabla_s \hat{\phi}
     \right)
     - \frac{\partial c}{\partial n},
  \label{eq:w_evolution_eqn_GCS} 
\eea
which is the form of the effective flux boundary conditions 
we shall use in our analysis below.

\section{ Surface Transport
  Processes}
\label{sec:surfproc}

Before proceeding to the analysis, we discuss the relative importance
of the various surface transport processes, compared to their neutral
bulk counterparts. For clarity, we summarize our results for
tangential surface fluxes in
Table~\ref{tab:summary_of_surface_fluxes}, with and without
dimensions. As with the associated bulk fluxes summarized in
Table~\ref{tab:summary_of_bulk_fluxes}, we introduce different notations
for contributions by diffusion and electromigration (superscripts
$(d)$ and $(e)$, respectively) to the tangential (subscript $t$)
surface fluxes of cations, anions, charge, and neutral salt
(subscripts $+$, $-$, $q$, and $w$, respectively). This allows us to
define dimensionless parameters comparing the various contributions.

\begin{table*}
\caption{\label{tab:summary_of_surface_fluxes}  Summary of surface 
flux formulae for electrochemical transport. }
\begin{ruledtabular}
\begin{tabular}{ccc}
  & Dimensional Formula$^{a,b,c}$
\footnotetext[1]{
We have abused notation and used the same variable $\Gamma_i$ for both 
the dimensional and dimensionless formulae.  $\Gamma_i$ in the 
dimensional formulae is equal to $C_o a \Gamma_i$ in the dimensionless 
formulae.}
\footnotetext[2]{ The dimensional surface charge density, $Q$,
is defined by $Q \equiv C_o \lambda_D q$.}
\footnotetext[3]{ The dimensional excess 
neutral salt concentration, $W$, is defined by $W \equiv C_o \lambda_D w$.}
& Dimensionless Formula \\
\hline \\
 $\J_{t,\pm}$  & 
 $ - b  \Gamma_\pm \nabla_s \mu_\pm$ & $- \Gamma_\pm \nabla_s \mu_\pm$ \\[3pt]
\ &
 $ -\left [ 
   D \Gamma_\pm \nabla_s \log \left( C/C_o \right) 
   \pm \frac{z_+eD}{kT} \Gamma_\pm \nabla_s \Phi \right ]$ & 
 $ -\left( \Gamma_\pm \nabla_s \log c \pm \Gamma_\pm \nabla_s \phi
 \right )$ \\[3pt] 
 $\J^{(d)}_{t,\pm}$  & 
 $ -D \Gamma_\pm \nabla_s \log \left( C/C_o \right) $ &
 $ -\Gamma_\pm \nabla_s \log c $ \\[3pt]
 $\J^{(e)}_{t,\pm}$  & 
 $ \mp \frac{z_+eD}{kT} \Gamma_\pm \nabla_s \Phi $ & 
 $ \mp \Gamma_\pm \nabla_s \phi $ \\[3pt]
\hline \\
 $\J_{t,q}$  & 
 $ -\left [ D Q \nabla_s \log \left( C/C_o \right) 
   + \frac{z_+eD}{kT} W \nabla_s \Phi \right ]$ & 
 $ -\eps \left( q \nabla_s \log c + w \nabla_s \phi \right )$ \\[3pt]
 $\J^{(d)}_{t,q}$  & 
 $ -D Q \nabla_s \log \left (C/C_o \right)$ &
 $ -\eps q \nabla_s \log c $ \\[3pt]
 $\J^{(e)}_{t,q}$  & 
 $ -\frac{z_+eD}{kT} W \nabla_s \Phi$ & 
 $ -\eps w \nabla_s \phi$ \\[3pt]
\hline \\
 $\J_{t,w}$  & 
 $ -\left [ D W \nabla_s \log \left( C/C_o \right) 
   + \frac{z_+eD}{kT} Q \nabla_s \Phi \right ]$ & 
 $ -\eps \left( w \nabla_s \log c + q \nabla_s \phi \right )$ \\[3pt]
 $\J^{(d)}_{t,w}$  & 
 $ -D W \nabla_s \log \left (C/C_o \right)$ &
 $ -\eps w \nabla_s \log c$ \\[3pt]
 $\J^{(e)}_{t,w}$  & 
 $ -\frac{z_+eD}{kT} Q \nabla_s \Phi$ & 
 $ -\eps q \nabla_s \phi$ \\[3pt]
\end{tabular}
\end{ruledtabular}
\end{table*}

J. J. Bikerman pioneered the experimental and theoretical study of
double-layer surface transport~\cite{bikerman1933,bikerman1935} and
first defined the dimensionless ratio of surface current to bulk
current, across a geometrical length scale, $a$, for a uniformly
charged double layer and a uniform bulk
solution~\cite{bikerman1940}. Using our notation, the Bikerman number
is
\begin{equation}
Bi = \frac{J_{t,q}}{a J_0}  \label{eq:alphaB}
\end{equation}
where $J_0 = z_+e F_0$ is a reference bulk current in terms of the
typical diffusive flux, $F_0 = DC_0/a$.  B. V. Deryagin and
S. S. Dukhin later added contributions from electro-osmotic flow,
using the GCS model of the double layer (as did
Bikerman)~\cite{deryagin1969}, and Dukhin and collaborators then used
this model to study electrophoresis of highly charged particles (with
large, but nearly uniform surface
charge)~\cite{dukhin1969,shilov1970,dukhin1993}. As in other
situations, surface-conduction effects in electrophoresis are
controlled by $Bi$, which thus came to be known in the Russian
literature as the ``Dukhin number'', $Du$. (Dukhin himself denoted it
by $Rel$). 

Recently, Bazant, Thornton and Ajdari pointed out
that the (steady) Bikerman-Dukhin number is equal to the ratio of
the excess surface salt concentration to its bulk counterpart at the
geometrical scale $a$,
\begin{equation}
Bi = \frac{\Gamma_+ + \Gamma_i}{2 a C_0}
\end{equation}
and they showed its importance in a one-dimensional problem of
electrochemical relaxation between parallel-plate
electrodes~\cite{bazant2004}. This surprising equivalence (in light
of the definition of $Bi$) demonstrates that surface conduction
becomes important relative to bulk conduction simply because a
significant number of ions are adsorbed in the double layer compared to
the nearby bulk solution. This means that salt adsorption leading to
bulk diffusion should generally occur at the same time as surface
conduction, if the double layer becomes highly charged during the
response to an applied field or voltage, as in our model problems
below.

Unlike prior work, however, our multi-dimensional nonlinear analysis
allows for nonuniform, time-dependent charging of the double layer, so
the Bikerman-Dukhin number can only be defined locally and in
principle could vary wildly across the surface.  Moreover, since we
separate the contributions to surface transport from diffusion and
electromigration, we can define some new dimensionless numbers. From
Eqs.~(\ref{eq:q_evolution_eqn_GCS})-(\ref{eq:w_evolution_eqn_GCS}),
we find that the  following surface-to-bulk flux ratios are related to
the excess surface-to-bulk ratio of neutral salt concentration
\begin{eqnarray}
\alpha &=& \epsilon w =  4
\frac{\lambda_D}{a}\sqrt{\frac{C}{C_0}}\sinh^2
  \left(\frac{z_+e\bar{\zeta}}{4kT}\right) \label{eq:alphadef} \\
&\sim& \frac{|J_{t,q}^{(e)}|}{a J_0} \sim
  \frac{|J_{t,w}^{(d)}|}{a F_0}. 
\label{eq:alpha}
\end{eqnarray}
Note that when surface diffusion is neglected, as in most prior work,
then $\alpha = Bi$, since surface currents arise from electromigration
alone.  The other surface-to-bulk flux ratios are given by the
surface-to-bulk ratio of the charge density,
\begin{eqnarray}
\beta &=& \epsilon |q| =  2
\frac{\lambda_D}{a}\sqrt{\frac{C}{C_0}}\sinh\left(\frac{z_+e\bar{\zeta}}{2kT}\right) \label{eq:betadef}  \\
&\sim& \frac{|J_{t,q}^{(d)}|}{a J_0} \sim
\frac{|J_{t,w}^{(e)}|}{a F_0} 
\end{eqnarray}
For thin double layers ($\lambda_D \ll a$), we see that the
surface-to-bulk flux ratios, $\alpha$ and $\beta$, are only
significant when $\bar{\zeta}$ significantly exceeds the thermal voltage
$kT/e$. 

To better understand how the charge and neutral-salt fluxes are
carried, it is instructive to form the ratio of these numbers,
\begin{eqnarray}
\frac{\alpha}{\beta} &=&  \tanh\left(\frac{z_+e\bar{\zeta}}{4kT}\right) \\
&\sim& \frac{|J_{t,w}^{(d)}|}{|J_{t,w}^{(e)}|}
\sim \frac{|J_{t,q}^{(e)}|}{|J_{t,q}^{(d)}|} 
\end{eqnarray}
For weakly charged double layers, $z_+e\bar{\zeta} \ll 4kT$, where $\alpha \ll
\beta$, the (small) surface flux of salt is dominated by
electromigration, while the surface flux of charge (surface current)
is dominated by diffusion (if bulk concentration gradients exist). For
highly charged double layers $z_+e\bar{\zeta} > 4kT$ where $\alpha \sim
\beta$, the contributions to each flux by diffusion and
electromigration become comparable, as counterions are completely
expelled ($q \sim w$).

\section{Nonlinear Steady Response for Thin Double Layers
         \label{sec:steady_response}} 

\subsection{ Effective Equations }

Using the mathematical model developed in the previous section, we now
examine the steady response of a metal sphere or cylinder with thin
double layers subjected to a large, uniform applied electric field.
At steady-state, the unsteady term is eliminated from the governing
equations (\ref{eq:dlc_c_eqn_bulk}) -- (\ref{eq:dlc_LEN}), so we have
\bea 
0 &=& \nabla^2 c
    \label{eq:c_eqn_steady} \\
  0 &=& \nabla \cdot \left( c \nabla \phi \right)
    \label{eq:phi_eqn_steady}.
\eea
Similarly, the flux boundary conditions (\ref{eq:q_evolution_eqn_GCS}) -- 
(\ref{eq:w_evolution_eqn_GCS}) become
\bea
  0 &=&
     \eps \nabla_s \cdot 
     \left(
         q \nabla_s \log c
       + w \nabla_s \phi
     \right)
     - c \frac{\partial \phi}{\partial n} 
  \label{eq:q_bc_steady} \\
  0 &=&
     \eps \nabla_s \cdot 
     \left(
         w \nabla_s \log c 
       + q \nabla_s \phi
     \right)
     - \frac{\partial c}{\partial n}.
  \label{eq:w_bc_steady}
\eea
Notice that we have retained the surface transport terms in 
these boundary conditions even though they appear at $O(\eps)$.
At large applied fields, it is no longer valid to order 
the terms in an asymptotic expansions by $\eps$ alone.  
We also need to consider factors of the form $\eps e^{\zeta}$
or $\eps \sinh(\zeta)$ since $e^{\zeta}$ may be as large as $O(1/\eps E)$
for large applied fields.  Since both $q$ and $w$ contain factors
which grow exponentially with the zeta-potential, we cannot discard
the surface conduction terms in (\ref{eq:q_bc_steady}) and
(\ref{eq:w_bc_steady}).
Finally, the Stern boundary condition (\ref{eq:stern_bc_GCS}) 
remains unchanged because it does not involve any time derivatives.

As mentioned earlier, the steady problem exhibits interesting features
that have not been extensively explored.  Unfortunately, the 
nonlinearities present in the governing equations and boundary 
conditions make it difficult to proceed analytically, so we use
numerical  methods to gain insight into the behavior of the system.
In this section, we first briefly describe the numerical model we use to 
study the system.   We then use the numerical model to study the 
development of $O(1)$ bulk concentration variations and their impact 
on transport around metal colloid spheres.

\subsection{Numerical Model}
To solve (\ref{eq:c_eqn_steady}) -- (\ref{eq:phi_eqn_steady}) numerically
in a computationally efficient manner, we use a pseudospectral 
method~\cite{trefethen_spectral_book,boyd_spectral_book,
fornberg_spectral_book}.  For problems in electrochemical transport,
pseudospectral methods are particularly powerful because they naturally 
resolve boundary layers by placing more grid points near boundaries of
the physical domain~\cite{trefethen_spectral_book,fornberg_spectral_book,
boyd_spectral_book}.  
We further reduces the computational complexity and cost of the numerical
model by taking advantage of the axisymmetry of the problem to reduce 
the numerical model to two dimensions (as opposed to using a fully 3-D 
description).  

For the computational grid, we use a tensor product grid of a 
uniformly spaced grid for the polar angle direction and a
shifted semi-infinite rational Chebyshev grid for the radial 
direction~\cite{boyd_spectral_book}.
To obtain the discretized form of the differential operators on this grid, 
we use Kronecker products of the differentiation matrices for the 
individual one-dimensional grids \cite{trefethen_spectral_book}.
The numerical model is then easily constructed using collocation
by replacing field variables and continuous operators in the 
mathematical model by grid functions and discrete operators.  
The resulting nonlinear, algebraic system of equations for the values of 
the unknowns at the grid points is solved using a standard Newton iteration 
with continuation in the strength of the applied electric field.  
The Jacobian for the Newton iteration is computed exactly by using a 
set of simple \emph{matrix}-based differentiation rules derived 
in the appendix of~\cite{chu_thesis_2005}.  By using the exact Jacobian,
the convergence rate of the Newton iteration is kept low; typically 
less than five iterations are required for each value of the continuation
parameter before the residual of the solution to the discrete system of 
equations is reduced to an absolute tolerance of 
$10^{-8}$~\footnote{The residual is computed using the $L^\infty$ norm.}.
Directly computing the Jacobian in matrix form had the additional advantage
of making it easy to implement the numerical model in MATLAB, a high-level
programming language with a large library of built-in functions.
It is also worth mentioning that we avoid the 
problem of dealing with infinite values of the electric potential by 
formulating the numerical model in terms of $\phi + E r \cos \theta$, 
the deviation of the electric potential from that of the uniform applied 
electric field, rather than $\phi$ itself.  

\begin{figure}[htb]
\bc
\includegraphics[width=1.6in]{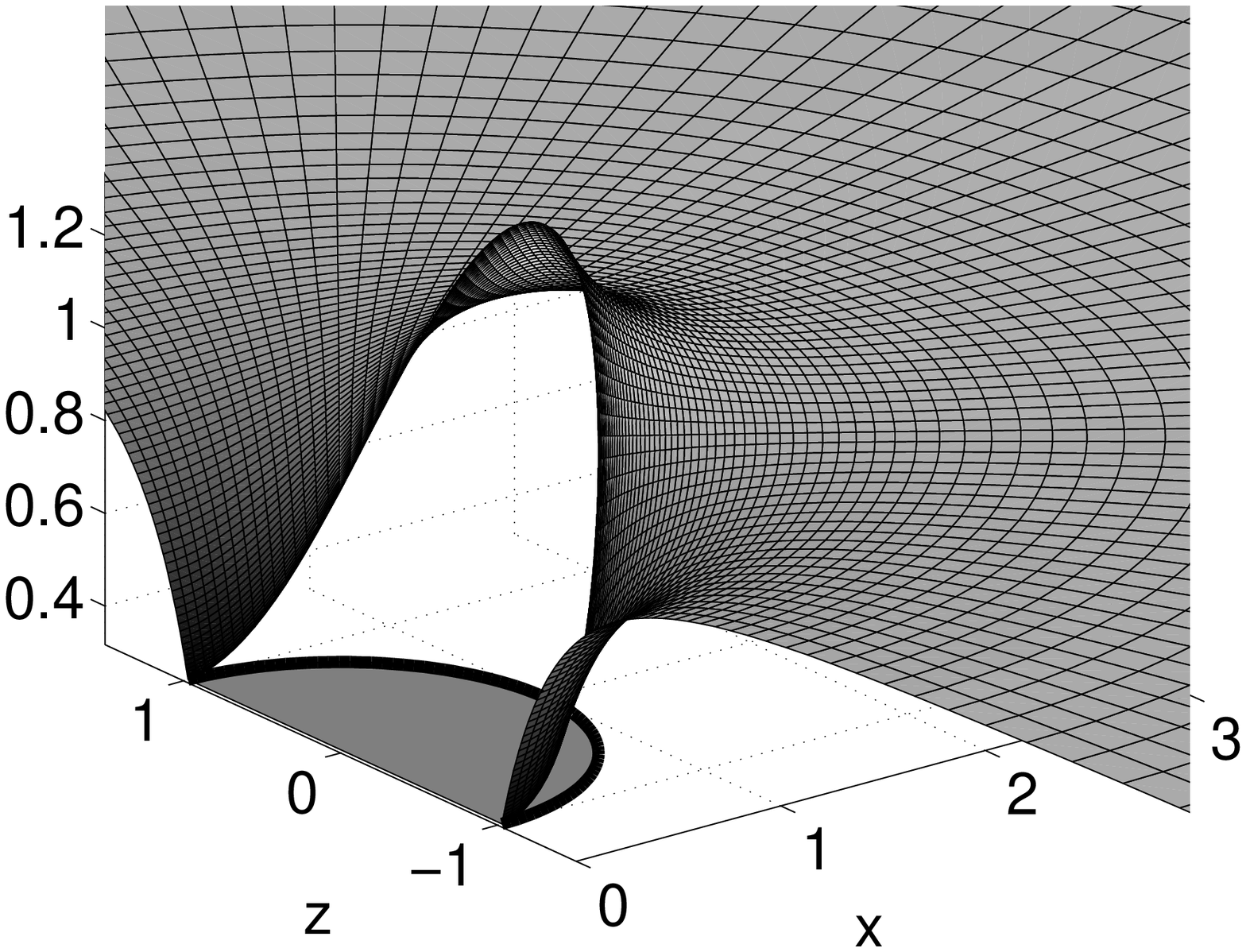}
\includegraphics[width=1.6in]{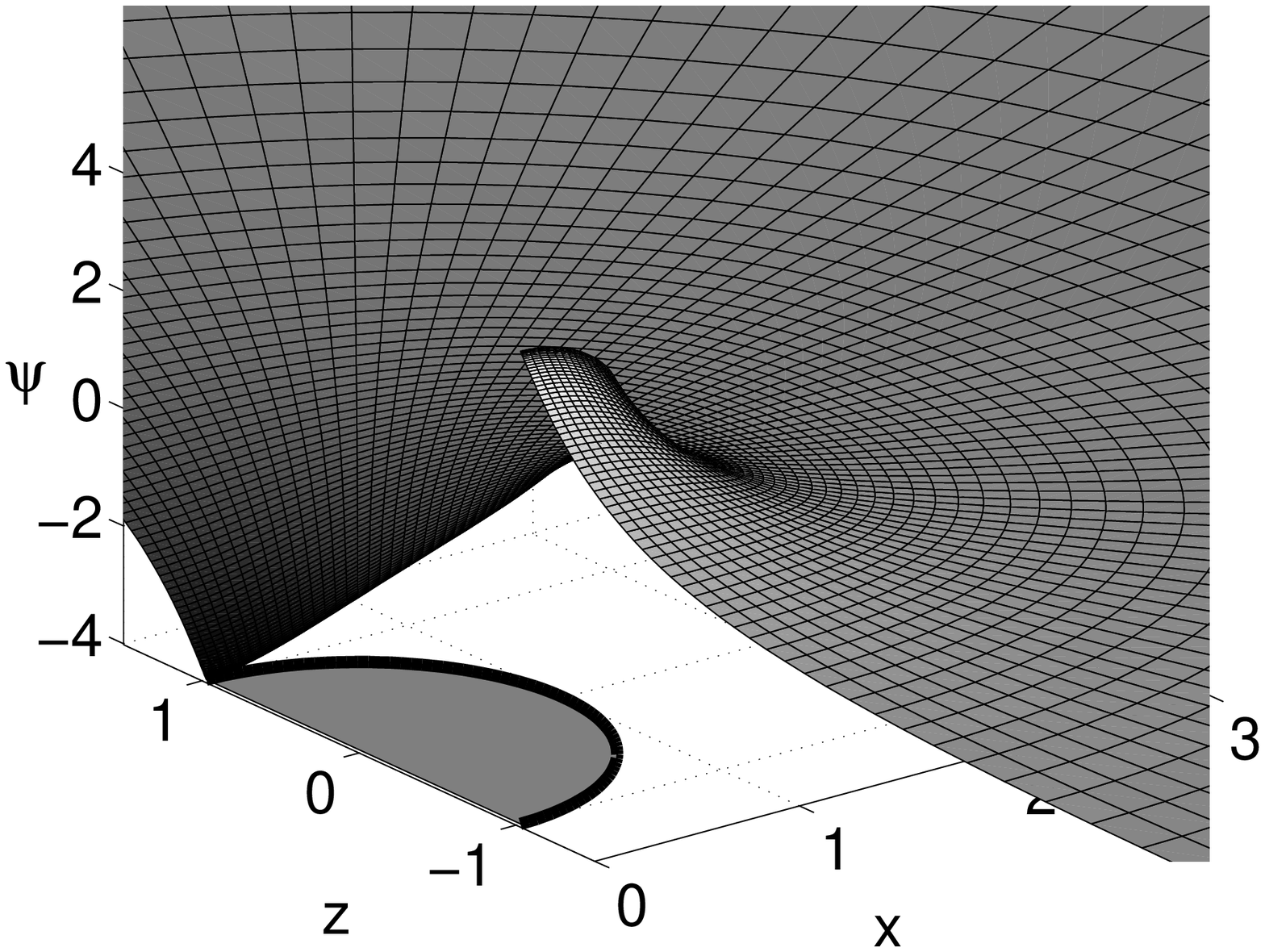}
\begin{minipage}[h]{3in}
\caption[Numerical solutions for the concentration and excess potential 
for $E = 10$]{
\label{figure:c_and_psi_full_profiles}
Numerical solutions for the concentration $c$ (left) and excess electric 
potential $\psi$ (right) for $E = 10$, $\eps = 0.01$, $\delta = 1$.  Notice 
the large gradients in the concentration profile near the surface of the 
sphere.
}
\end{minipage}
\ec
\end{figure}

\begin{figure}
\bc
\includegraphics[width=1.6in,height=1.4in]{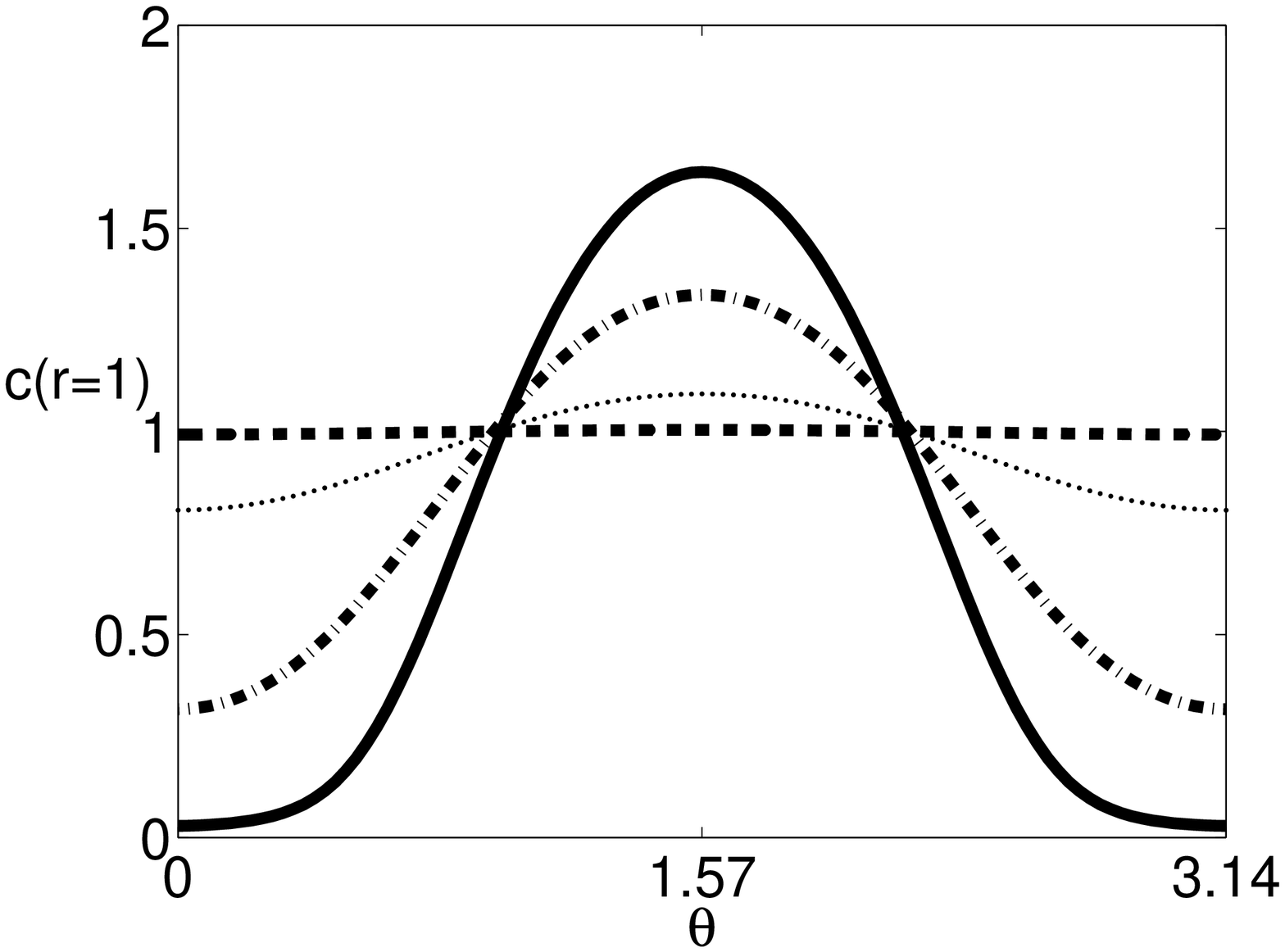}
\includegraphics[width=1.6in,height=1.4in]{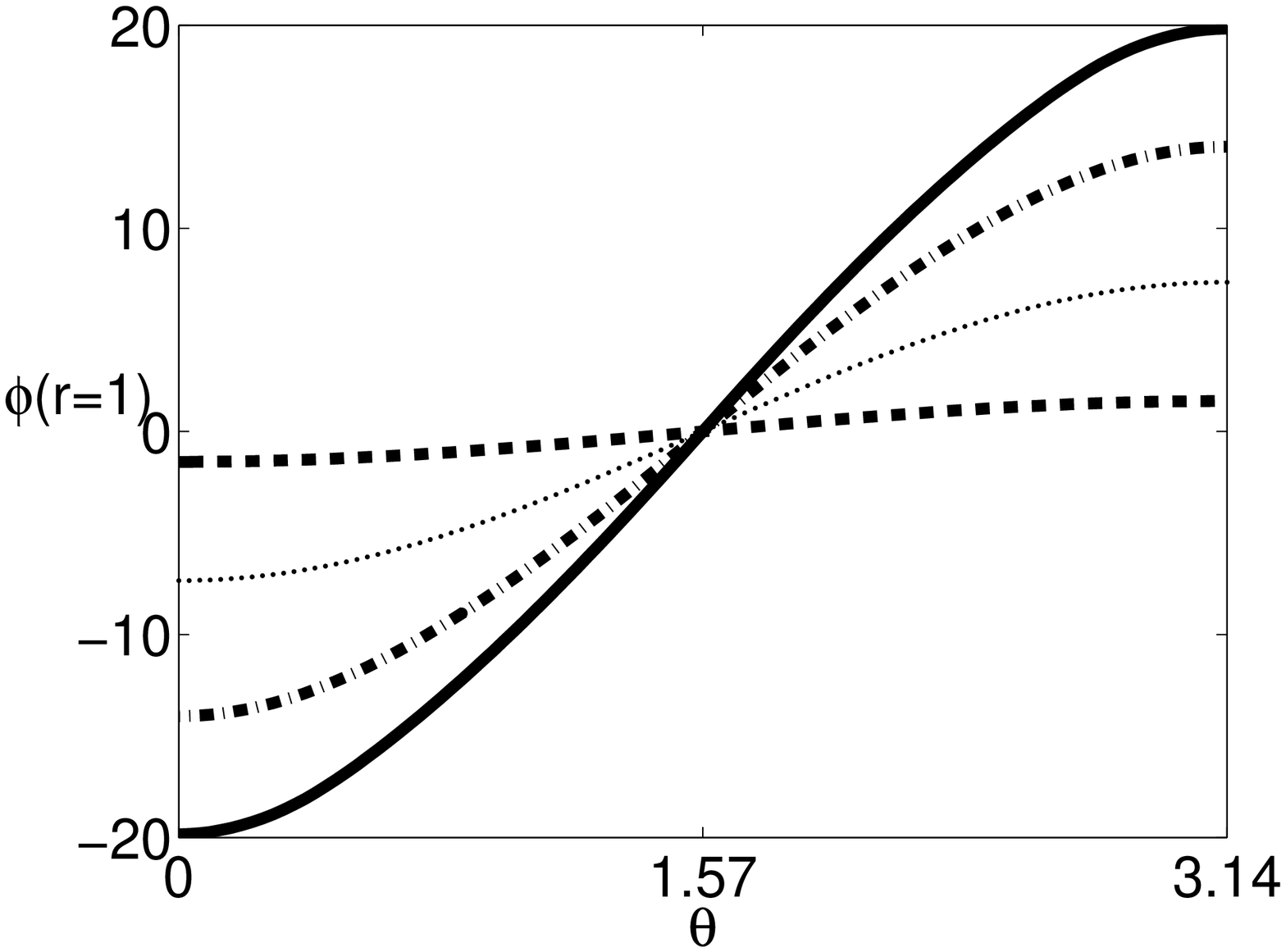}
\begin{minipage}[h]{3in}
\caption[Bulk concentration and electric potential profiles at surface of 
sphere for varying values of the applied electric field]{
\label{figure:c_s_and_phi_s}
Bulk concentration $\hat{c}$ and electric potential $\hat{\phi}$ at the 
surface of the sphere for varying values of the applied electric field.  
In these figures, $\eps = 0.01$ and $\delta = 1$.  Notice that for $E=15$, 
the poles ($\theta = 0$ and $\theta=\pi$) are about to be depleted of ions 
(\ie $\hat{c} \approx 0$).
}
\end{minipage}
\ec
\end{figure}

In our numerical investigations, we used the numerical method described
above with $90$ radial and $75$ angular grid points.  This grid resolution 
balanced the combined goals of high accuracy and good computational 
performance.  Figure~\ref{figure:c_and_psi_full_profiles} shows typical
solutions for the concentration and electric potential (relative to the
background applied potential) for large applied electric fields.
A comparison of the concentration and electric potential at the 
surface of the sphere for varying values of $E$ are shown in 
Figure~\ref{figure:c_s_and_phi_s}.

\subsection{Enhanced Surface Excess Concentration and Surface Conduction}
\begin{figure}
\bc
\includegraphics[width=1.6in,height=1.4in]{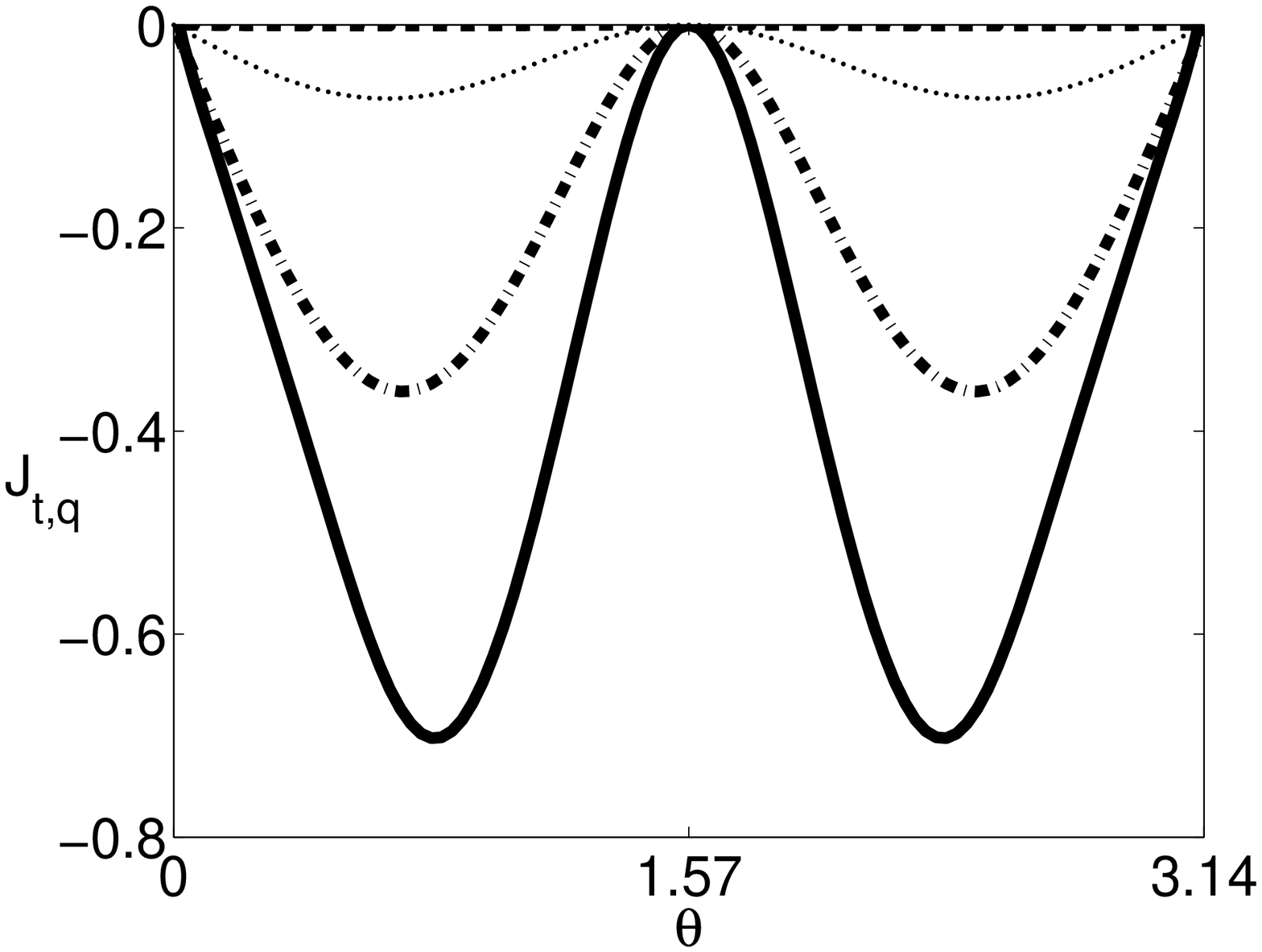}
\includegraphics[width=1.6in,height=1.4in]{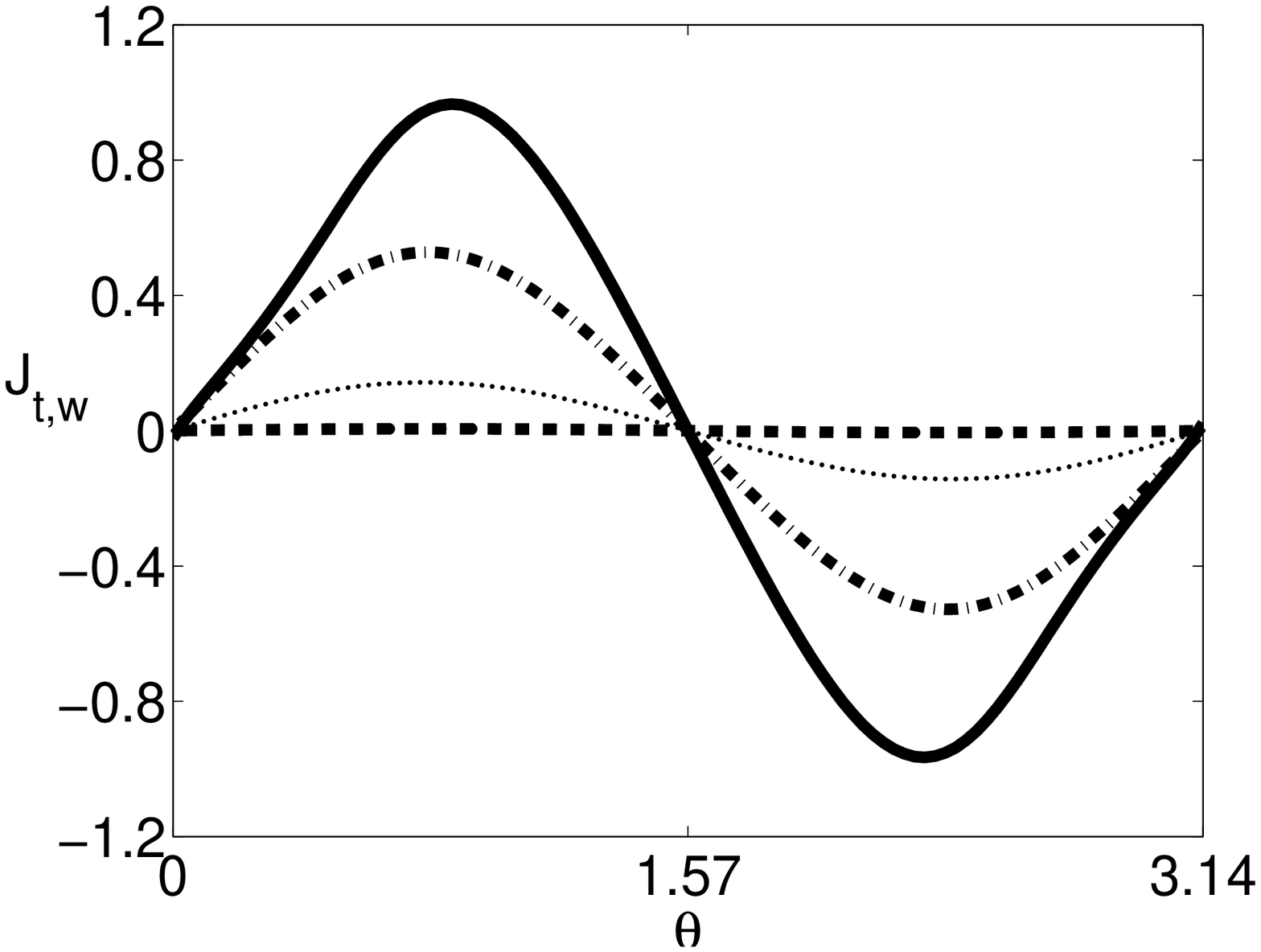}
\begin{minipage}[h]{3in}
\caption[Tangential surface fluxes for the surface charge density and 
excess surface concentration of neutral salt for varying values of the 
applied electric field]{
\label{figure:Fq_and_Fw}
Tangential surface fluxes for the surface charge density $|\J_{t,q}|$ (left) 
and excess surface concentration of neutral salt $|\J_{t,w}|$ (right) for 
varying values of the applied electric field. 
In these figures, $\eps = 0.01$ and $\delta = 1$.  Notice that for large
applied fields, the surface fluxes are $O(1)$ quantities and make a 
non-negligible leading-order contribution in (\ref{eq:q_evolution_eqn_GCS}) 
and (\ref{eq:w_evolution_eqn_GCS}). 
}
\end{minipage}
\ec
\end{figure}

\begin{figure}
\bc
\includegraphics[width=1.6in,height=1.4in]
  {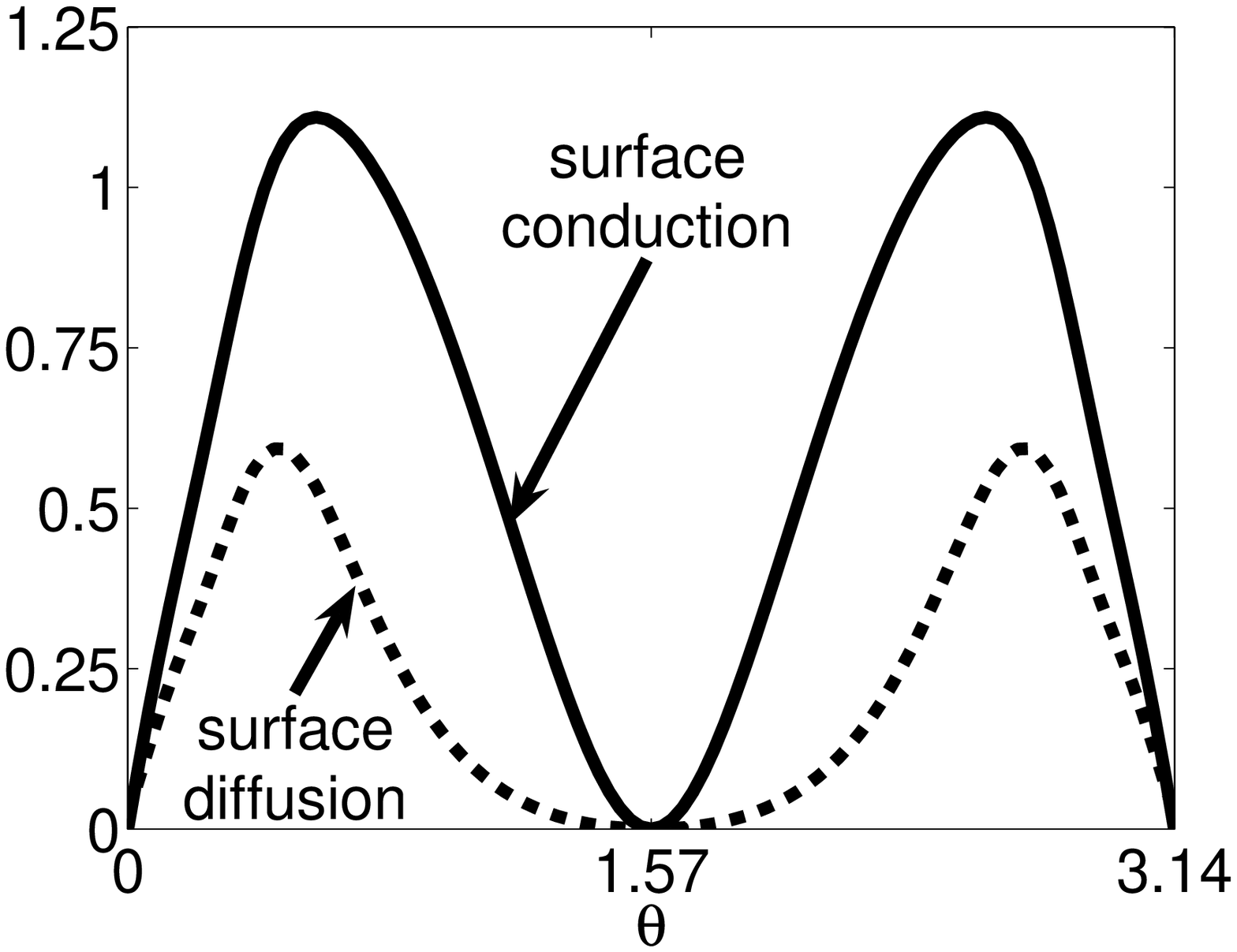}
\includegraphics[width=1.6in,height=1.4in]
  {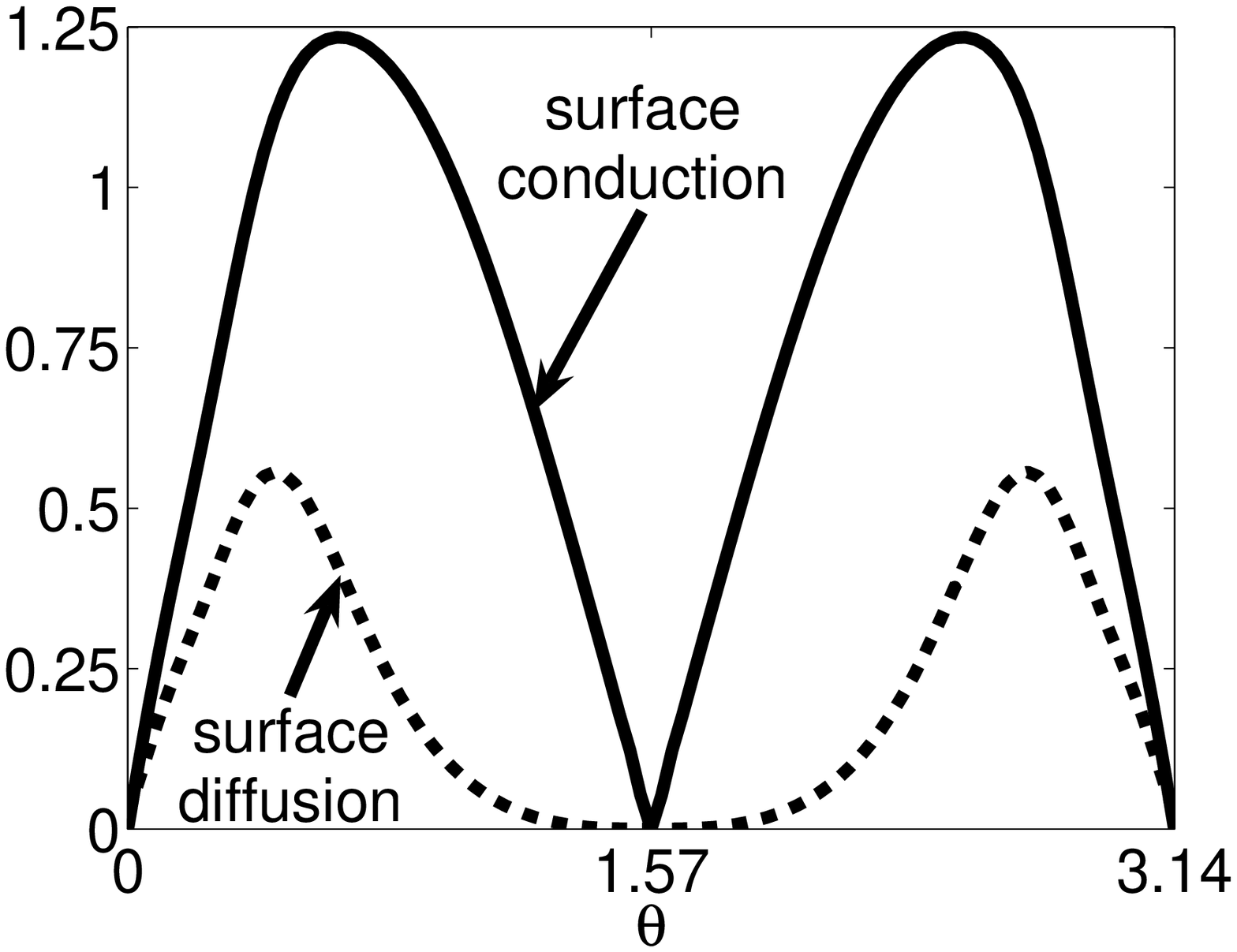}
\begin{minipage}[h]{3in}
\caption[Comparison of the magnitudes of surface conduction and
surface diffusion for the tangential fluxes of $q$ and $w$]{
\label{figure:compare_surf_conduction_diffusion}
Comparison of the magnitudes of $|\J^{(d)}|$ (solid 
lines) and $|\J^{(e)}|$ (dashed lines) for the excess surface fluxes of 
$\eps q$ (left) and $\eps w$ (right) for an applied electric field value of 
$E = 15$.  In these figures, $\eps = 0.01$ and $\delta = 1$.   
Notice that in both cases, the surface diffusion is on the order
of $1/\hat{c} E$ times the surface conduction.
}
\end{minipage}
\ec
\end{figure}

Perhaps the most important aspects of the steady response at high applied 
electric fields are the enhanced surface excess ion concentration and
surface transport within the double layer. 
\begin{figure}
\bc
\includegraphics[width=1.6in,height=1.4in]{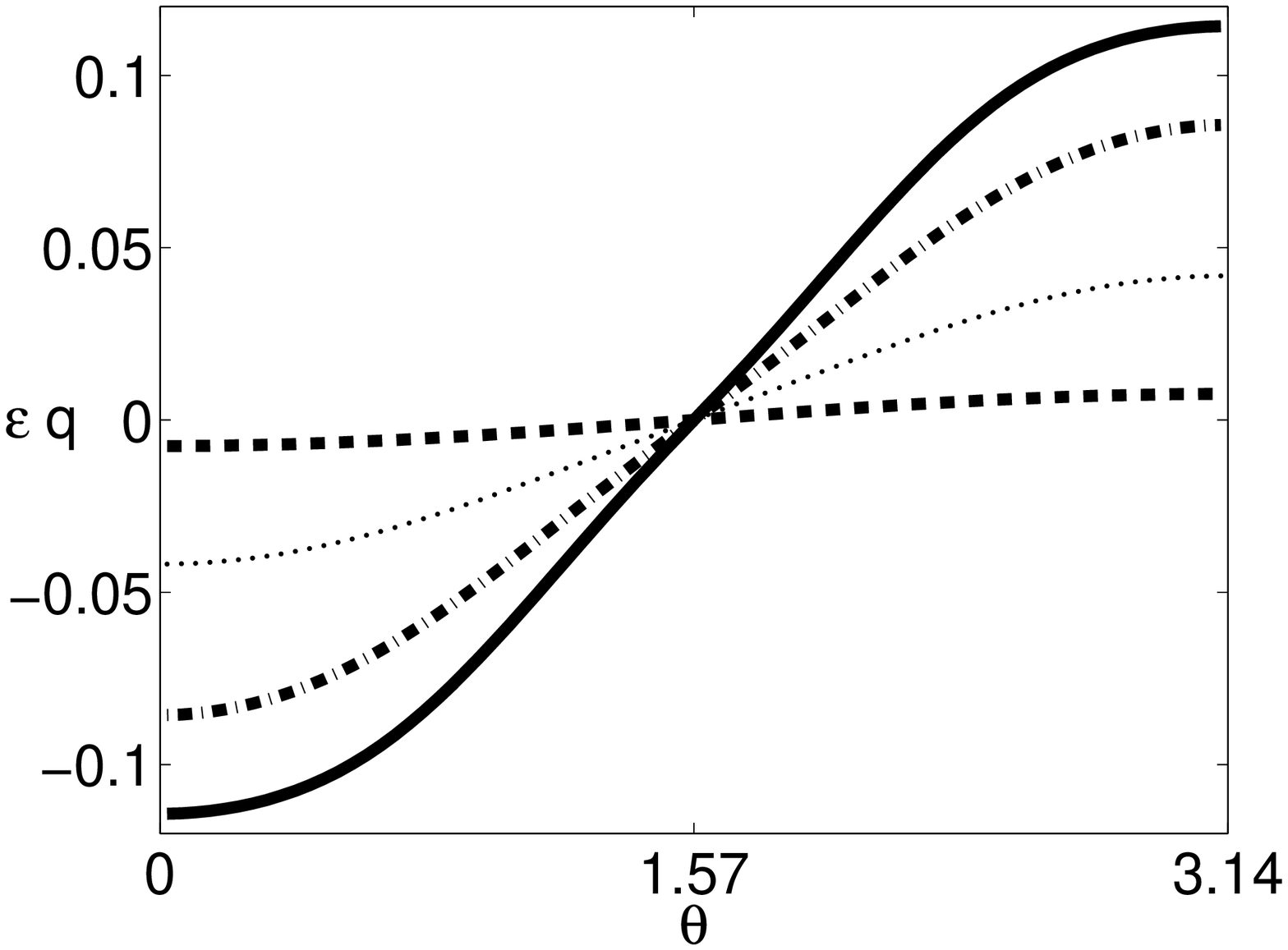}
\includegraphics[width=1.6in,height=1.4in]{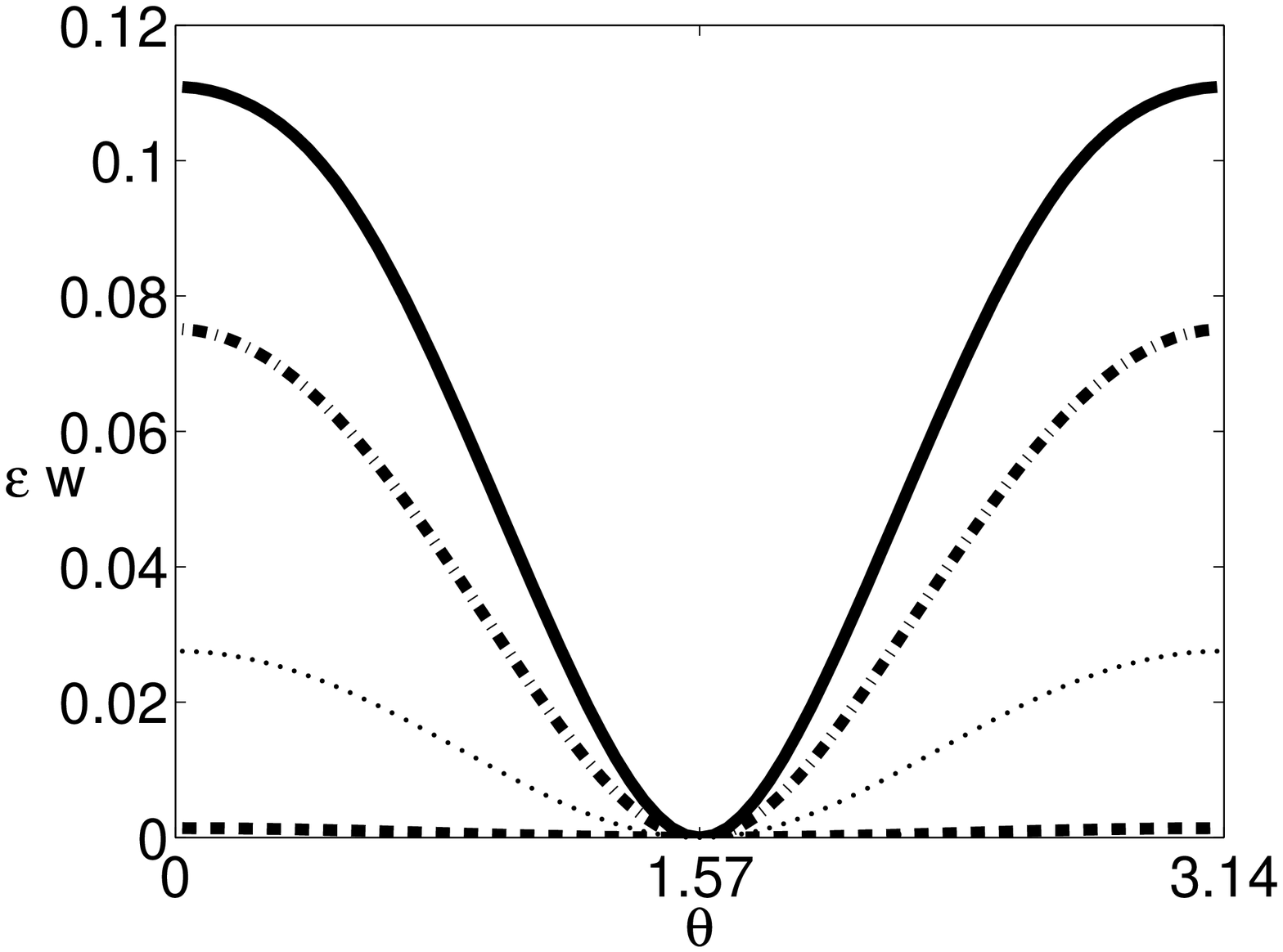}
\begin{minipage}[h]{3in}
\caption[Surface charge density and excess surface concentration of 
neutral salt and for varying values of the applied electric field]{
\label{figure:q_and_w}
Surface charge density $\eps q$ (left) and excess surface concentration of 
neutral salt $\eps w$ (right) and for varying values of the applied electric 
field.  In these figures, $\eps = 0.01$ and $\delta = 1$.  Notice that for 
large applied fields, $\eps w = O(1/E)$ and $\eps q = O(1/E)$ so that the 
surface conduction terms in (\ref{eq:q_evolution_eqn_GCS}) 
and (\ref{eq:w_evolution_eqn_GCS}) are $O(1)$.
}
\end{minipage}
\ec
\end{figure}
As shown in Figure~\ref{figure:q_and_w}, at high applied fields, 
the excess surface concentrations is $O \left( 1/E \right)$, so surface 
transport within the double layer (shown in Figure~\ref{figure:Fq_and_Fw}) 
becomes non-negligible in the leading order effective flux boundary 
conditions (\ref{eq:q_bc_steady}) -- (\ref{eq:w_bc_steady}).
Interestingly, surface conduction, $\J^{(e)}_{t,q}$ and $\J^{(e)}_{t,w}$, 
are the dominant contributions to surface transport
(see Figure~\ref{figure:compare_surf_conduction_diffusion}).
While there are clearly surface gradients in concentration, surface
diffusion is smaller than surface migration by a factor on the order of
$1/\hat{c} E$.  
Also, we reiterate that the driving force for surface transport is 
solely from surface gradients of the bulk concentration and bulk electric 
potential; gradients in the zeta-potential do not play a role because 
they are completely canceled out.

\begin{figure}
\bc
\includegraphics[width=2.5in,height=1.5in]{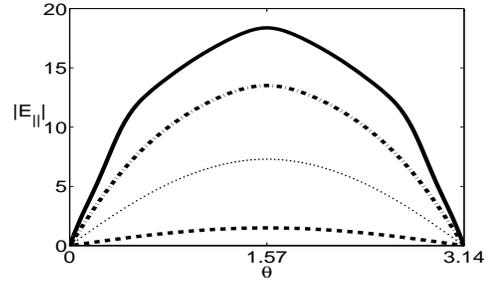}
\begin{minipage}[h]{3in}
\caption[Tangential component of bulk electric field at surface of sphere 
for varying values of the applied electric field]{
\label{figure:E_t}
Tangential component of bulk electric field, $|E_t|$ at surface of sphere 
for varying values of the applied electric field. 
In these figures, $\eps = 0.01$ and $\delta = 1$.  
}
\end{minipage}
\ec
\end{figure}

\begin{figure}
\bc
\includegraphics[width=1.6in,height=1.4in]{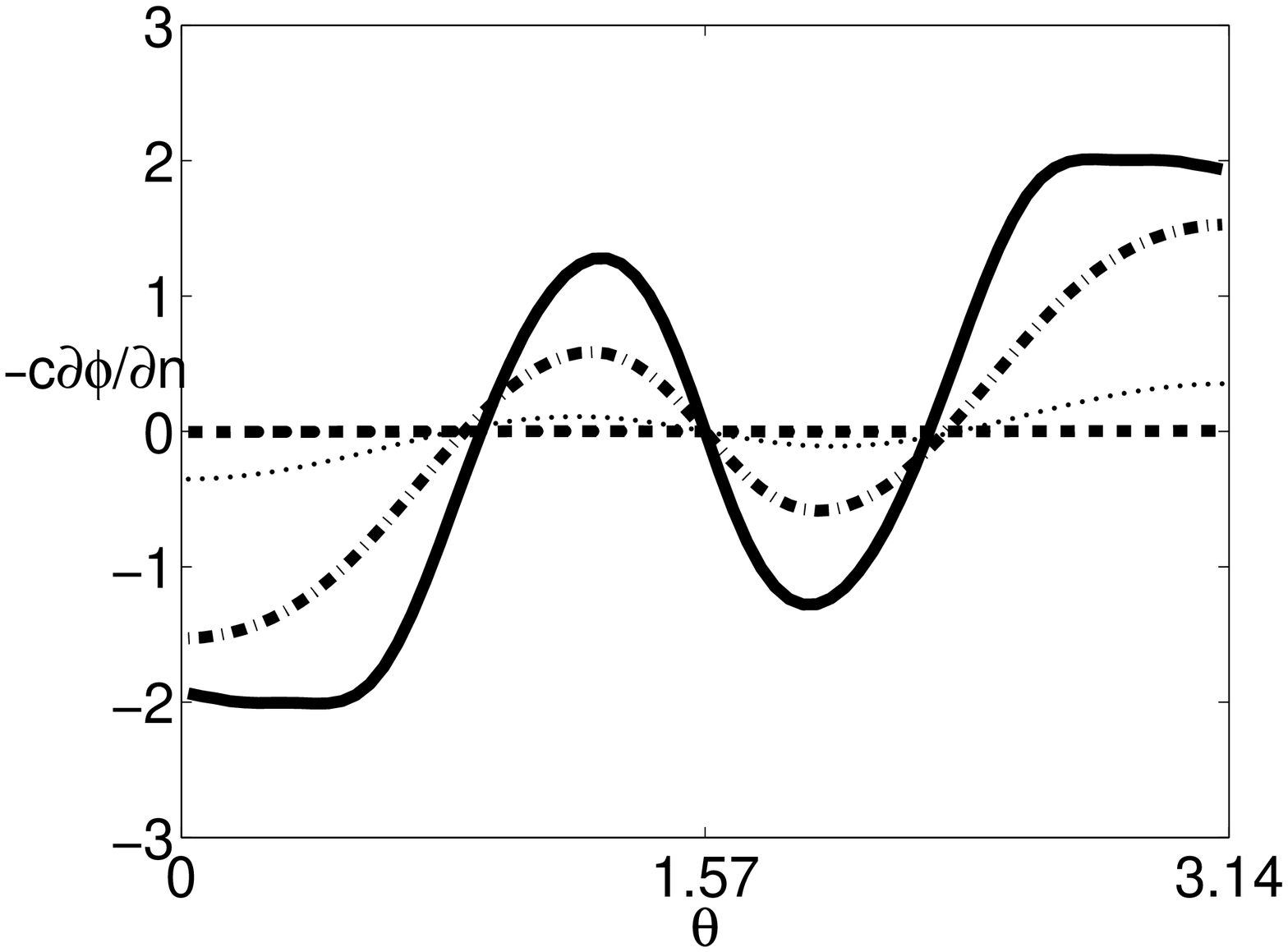}
\includegraphics[width=1.6in,height=1.4in]{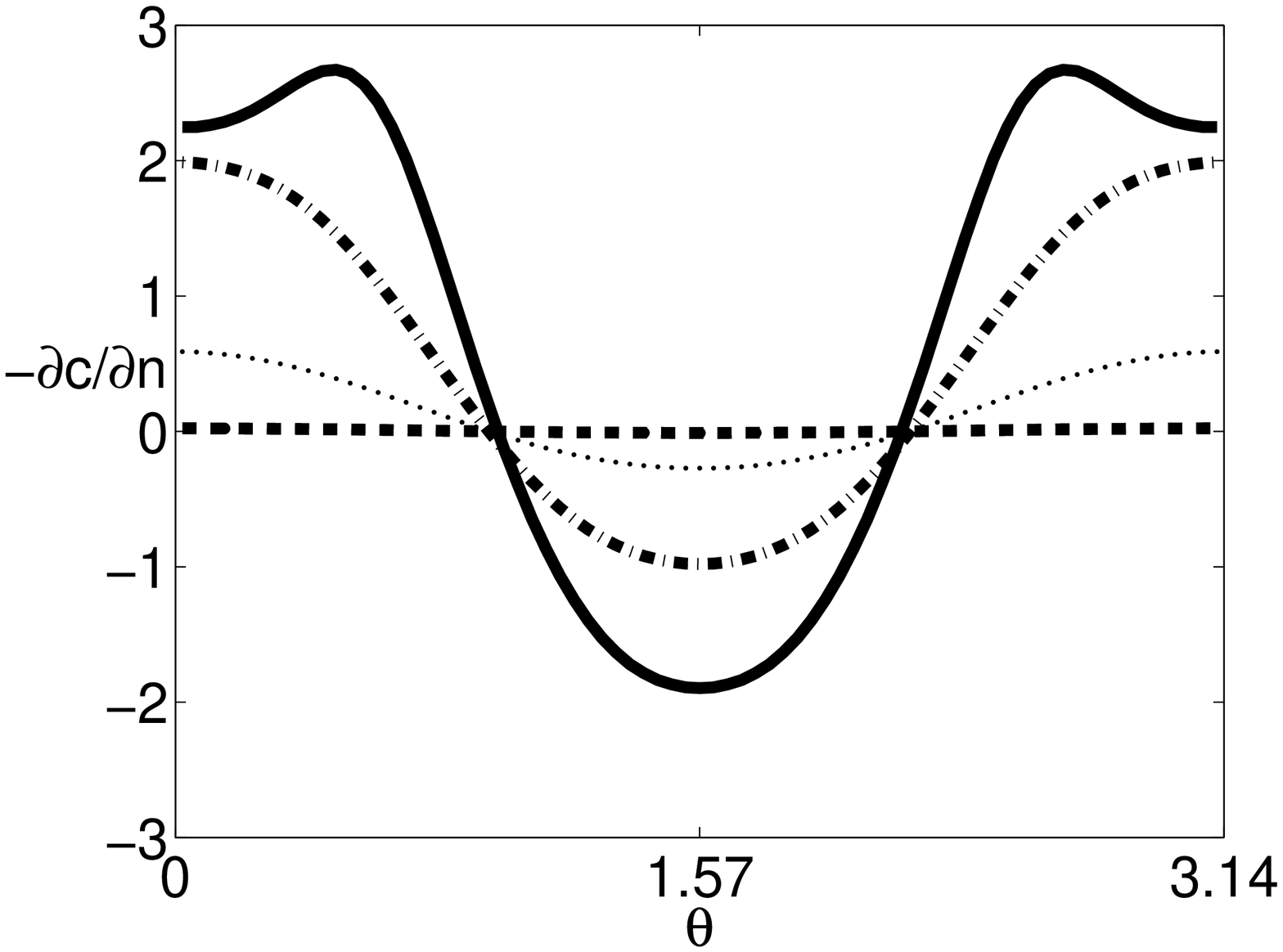}
\begin{minipage}[h]{3in}
\caption[Normal flux of current and neutral salt into the double
layer for varying values of the applied electric field]{
\label{figure:normal_fluxes_from_bulk}
Normal flux of current $c \frac{\partial \phi}{\partial n}$ (left) 
and neutral salt $\frac{\partial c}{\partial n}$ (right) into the double
layer for varying values of the applied electric field.
In these figures, $\eps = 0.01$ and $\delta = 1$.   
}
\end{minipage}
\ec
\end{figure}

An important feature of the surface fluxes, $\J_{t,q}$ and $\J_{t,w}$, shown 
in Figure~\ref{figure:Fq_and_Fw} is that they are non-uniform.  
This non-uniformity is strongly influenced by the non-uniformity in the  
tangential electric field (see Figure~\ref{figure:E_t}).
The non-uniformity of the surface excess salt concentration and 
charge surface density also play a role but to a lesser extent.
For the steady problem, the surface excess concentration of ions 
remains constant in time, so the non-uniformity in the surface fluxes 
leads to non-uniform normal fluxes of current and neutral salt from the 
bulk into the double layer (see Figure~\ref{figure:normal_fluxes_from_bulk}).
Notice that the normal flux of neutral salt into the double layer, which 
is given by $-\partial c/ \partial n$, shows an injection of neutral salt 
at the poles ($-\partial c/ \partial n > 0$), where the charging
is strongest, and an ejection of neutral salt at the equator 
($-\partial c/\partial n < 0$), where there is essentially no excess
neutral salt build up.   This configuration of fluxes leads to the neutral 
salt depletion at the poles and accumulation at the equator shown in 
Figures~\ref{figure:c_and_psi_full_profiles} and
\ref{figure:c_s_and_phi_s}. 
Similarly, the normal current density $-\hat{c} \partial \phi/ \partial n$, 
shows an influx of negative (positive) current density at the north (south) 
pole and a positive (negative) current density closer to the equator.
At the equator, there is no normal current density because the
normal diffusion currents of cations and anions exactly balance and 
there is no normal electric field to drive a migration current.

\subsection{Bulk Diffusion and Concentration Gradients}
One major consequence of surface conduction is transport of large amounts of
neutral salt into the double layer.  These cause strong concentration 
gradients to appear near the surface of the sphere 
(Figures~\ref{figure:c_and_psi_full_profiles} and 
\ref{figure:c_s_and_phi_s}), indicating that the usual assumption of 
a uniform concentration profile is invalid at high electric fields.  
The presence of these large concentration gradients at relatively low 
electric fields ($E \approx 5$) should not be surprising since it is 
well-known that large voltage effects often begin with voltages as low as
a few times the thermal voltage \cite{bazant2004}.  The dramatic influence
of the voltage arises from the exponential dependence of double layer
concentrations on the zeta-potential.  

\begin{figure}
\bc
\includegraphics[width=2in,height=2in]{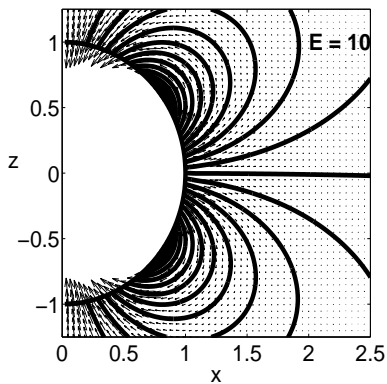}
\begin{minipage}[h]{3in}
\caption[Diffusion currents]{
\label{figure:diffusion_currents}
Diffusion currents drive transport of neutral salt near the surface 
of the sphere.  In this figure, $E = 10$, $\eps = 0.01$ and $\delta = 1$.  
Notice that streamlines of neutral salt are closed; current lines 
start on the surface of the sphere near the equator and end closer 
to the poles.
}
\end{minipage}
\ec
\end{figure}

Since the transport of neutral salt is driven by concentration gradients,
the presence of these strong variations leads to diffusion currents 
(see Figure \ref{figure:diffusion_currents}).  
An important feature of these diffusion currents is that 
they are closed; current lines start on the surface of the sphere near
the equator where neutral salt is ejected into the bulk (as a result of 
neutral salt transport within the double layer) and end close to the
poles where neutral salt is absorbed by the double layer.  
These recirculation currents are important because they allow the system
to conserve the total number of cation and anions \emph{locally}.  
Without them, the local depletion and accumulation of ions would require 
global changes to the bulk concentration (\ie the concentration at infinity 
would be affected).

While the presence of diffusion currents is interesting, we must be 
careful in how they are interpreted in terms of the motion of individual ion 
molecules.  In actuality, no ions are moving purely under the influence of
diffusion.  Rather, the cation and anion migration flux densities are 
slightly imbalanced due to the presence of a concentration gradient which 
results in a net transport of neutral salt concentration.

\subsection{Individual Ion Currents}
\begin{figure}
\bc
\includegraphics[width=1.6in,height=1.4in]
  {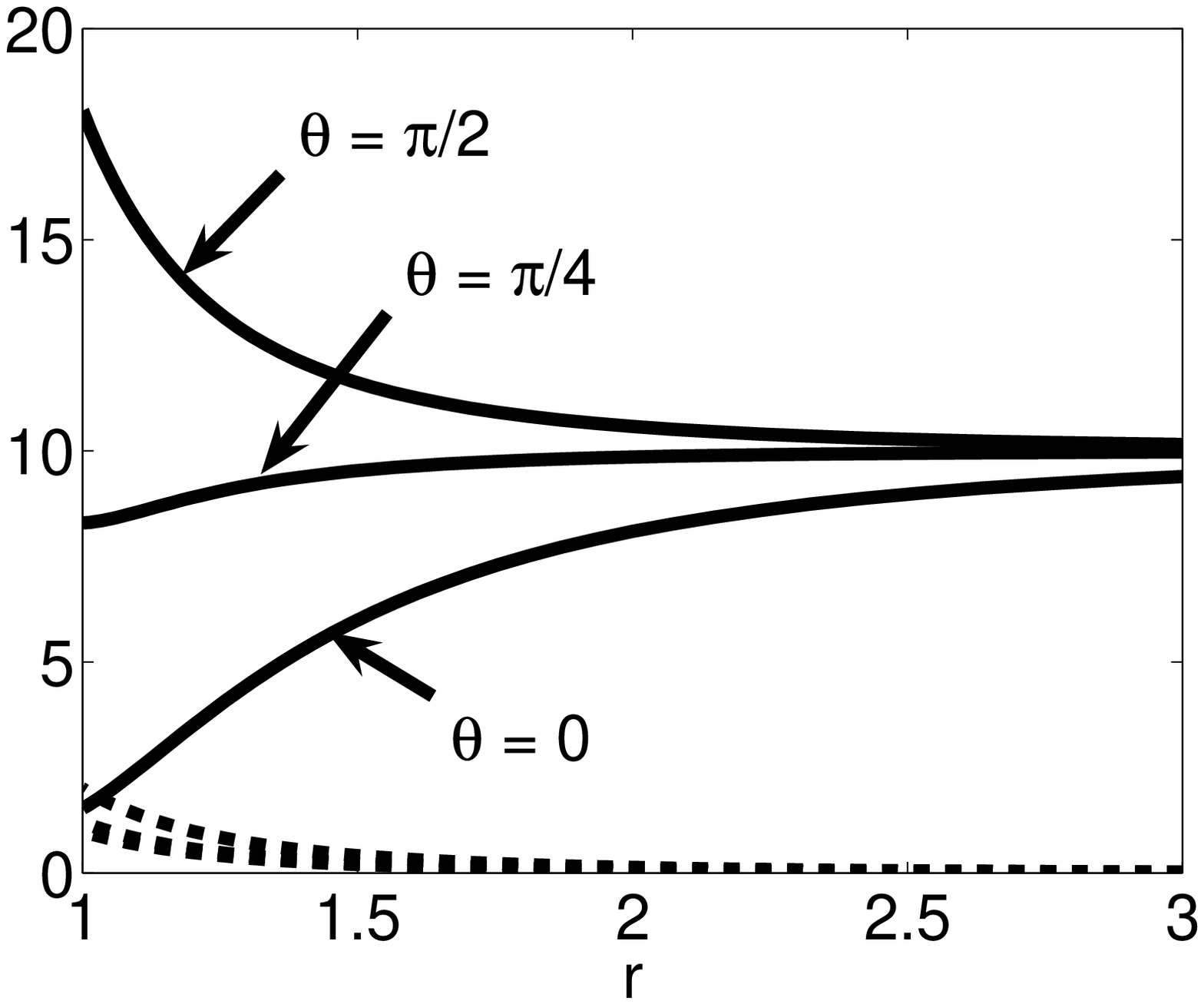}
\includegraphics[width=1.6in,height=1.4in]
  {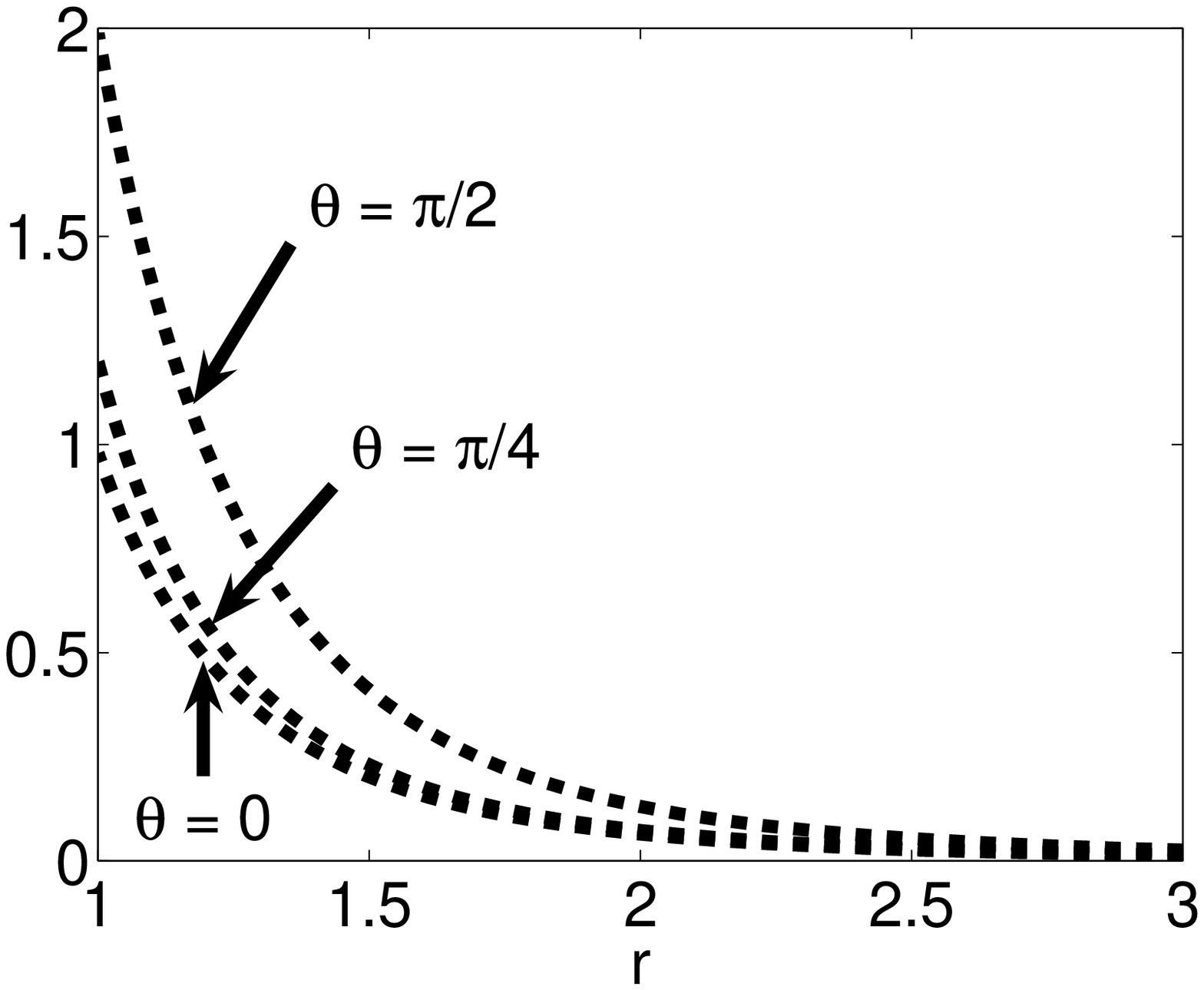}
\begin{minipage}[h]{3in}
\caption[Comparison of the magnitudes of bulk electromigration and 
diffusion at various positions on the surface of sphere]{
\label{figure:bulk_migration_vs_diffusion}
Comparison of the magnitudes of bulk electromigration 
$|\F_\rho^{(e)}|$ (solid 
lines) and diffusion $|\F_c^{(d)}|$ (dashed lines) fluxes as a function of 
distance from the surface of the sphere at $\theta = 0, \pi/4$, and  
$\pi/2$ (left).  
The figure on the right zooms in on the diffusion flux.  
In these figure, $E = 10$, $\eps = 0.01$ and $\delta = 1$.  
Notice that magnitude of the electromigration dominates diffusion and that 
the electromigration term itself becomes dominated by the contribution from 
the applied electric field a short distance from the surface of the sphere.
}
\end{minipage}
\ec
\end{figure}

Since cations and anions are the physical entities that are transported 
through the electrolyte, it is useful to consider the cation and anion 
flux densities individually.
As shown in Figure~\ref{figure:bulk_migration_vs_diffusion}, in the bulk,
the contribution of electromigration to transport dominates diffusion.
Moreover, within a short distance from the sphere, the electromigration 
term itself becomes dominated by the contribution from the
applied electric field. 
Thus, the concentration gradient only serves to slightly bias the flux 
densities so that cation (anion) motion is slightly retarded near the north 
(south) pole.  

\begin{figure}[htb]
\bc
\includegraphics[width=1.6in,height=1.4in]{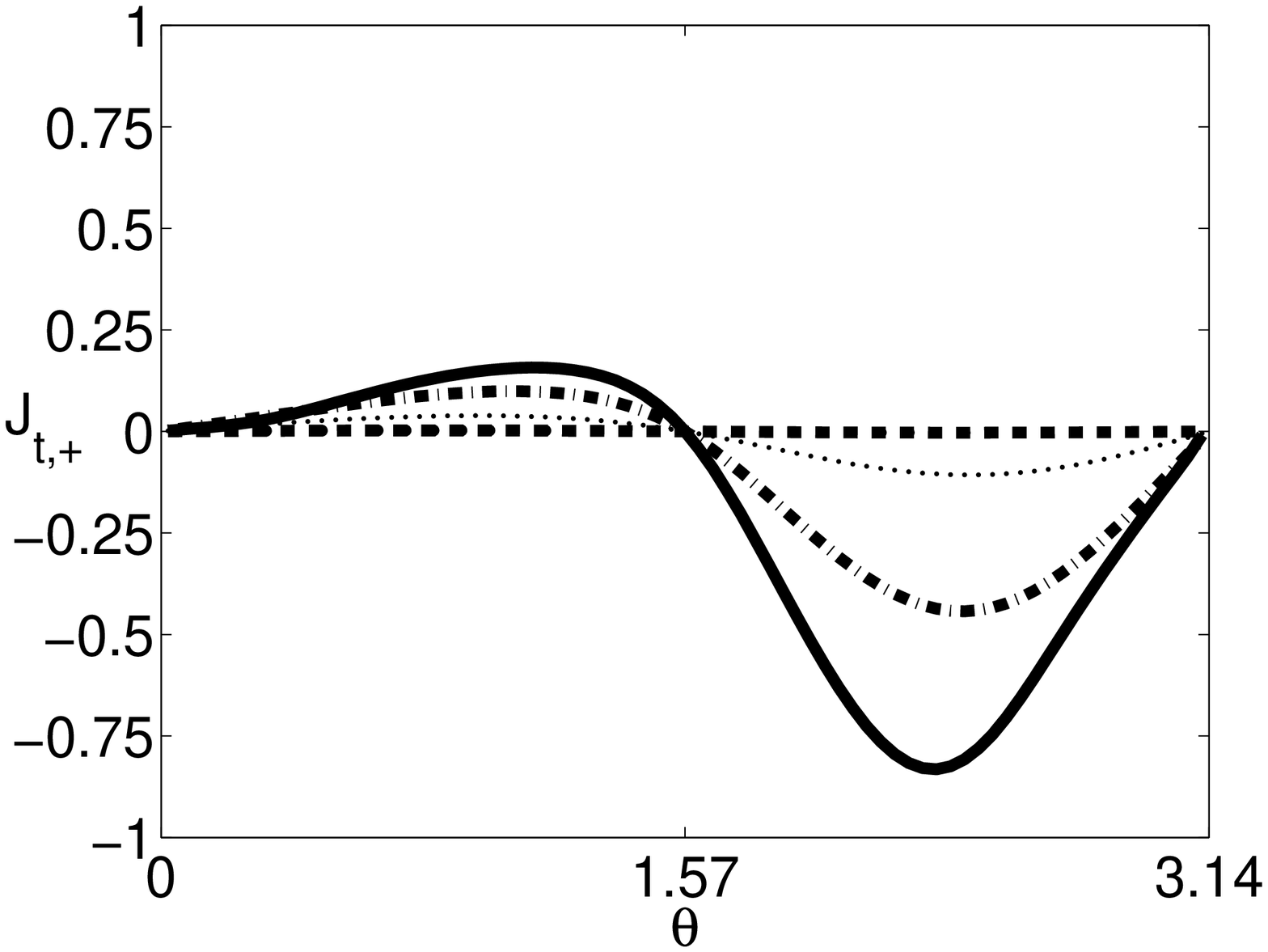}
\includegraphics[width=1.6in,height=1.4in]{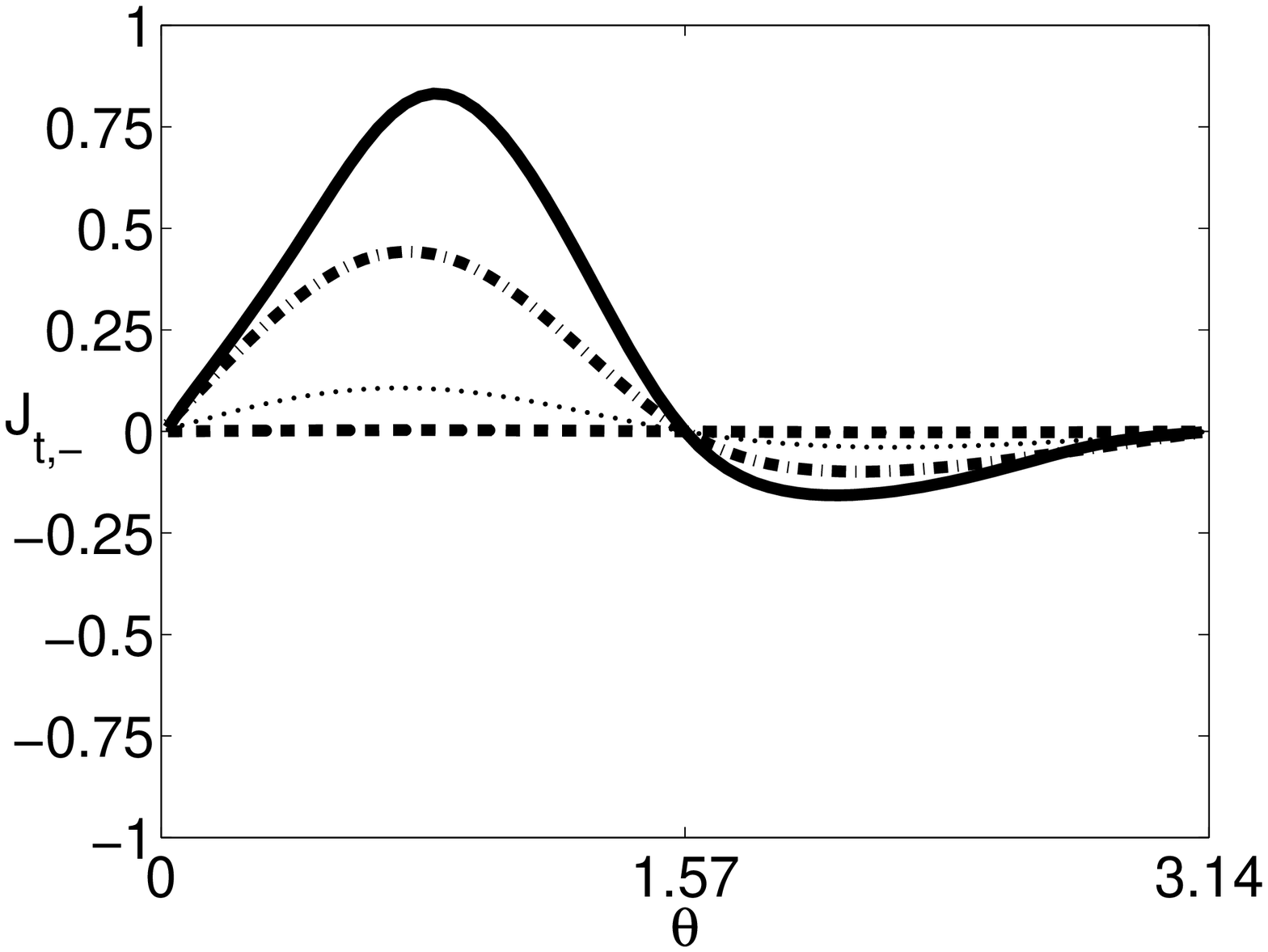}
\begin{minipage}[h]{3in}
\caption[Tangential surface fluxes of cations and anions
for varying values of the applied electric field]{
\label{figure:J_cplus_and_J_cminus}
Tangential surface fluxes for the cation $|\J_{t,+}|$ (left) and 
anion $|\J_{t,-}|$ (right) for varying values of the applied electric field. 
In these figures, $\eps = 0.01$ and $\delta = 1$.  Notice that for large
applied fields, the surface fluxes are $O(1)$ quantities (in the
appropriate hemisphere) which leads to a non-negligible leading-order 
contribution in (\ref{eq:effective_flux_bc_GCS}).
}
\end{minipage}
\ec
\end{figure}

It is more interesting to consider the surface transport of the individual 
ions within the double layer.  In the northern hemisphere, the double layer 
is dominated by anions; similarly, the southern hemisphere is dominated by 
cations.  
As a result, transport in each hemisphere is primarily due to
only one species (see Figure \ref{figure:J_cplus_and_J_cminus}).
This observation provides a direct explanation for the depletion and 
accumulation regions in the concentration profile in terms of the
the motion of individual ions. 
As mentioned earlier, the transport is from the poles to the 
equator because surface conduction is driven by the tangential component of 
the bulk electric field (see Figure~\ref{figure:E_t}).  
Since the double layer in the northern hemisphere is predominantly anions, 
the surface ion transport is from the poles towards the equator.  
A similar argument in the southern hemisphere shows that surface transport 
of the majority ion is again from the poles towards the equator.  
Thus, influx of ions into the double layer occurs in the regions near
the poles and outflux occurs by the equator.  

The dominance of a single species within the double layer for each hemisphere
leads to a small conundrum:  how does the bulk electrolyte near the 
surface of the sphere remain locally electroneutral?  In the northern 
hemisphere, it would seem that more anion should be absorbed at the pole
and ejected at the equator leading to bulk charge imbalance.  Analogous
reasoning involving cation leads to the same conclusion in the southern
hemisphere.  The resolution to the conundrum comes from remembering that 
both diffusion and electromigration contribute to transport.  
The imbalance in the normal flux required to sustain an imbalanced 
concentration profile in the double layer is achieved by carefully
balancing diffusion (which drives both species in the same direction)
and electromigration (which drives the two species in opposite directions)
so that the normal flux of the appropriate species dominates.  
For example, at the north pole, an electric field pointing away from
the surface of the sphere will suppress the cation normal flux towards 
the surface while enhancing the anion normal flux.  

In this situation, the electric field plays a similar role as the diffusion
potential for electrochemical transport in an electroneutral solution.  
Recall that when the cation and anion have different diffusivities, the 
electric field acts to slow down the species with the higher diffusivity 
and speed up the species with the lower diffusivity in such a way that both 
species have equal flux densities.  
In the current situation, the electric field serves to create the 
necessary imbalance in the cation and anion flux densities so
that the surface excess concentrations within the double layer can be 
maintained while preserving local electroneutrality in the bulk.

\section{Linear Relaxation for Arbitrary Double Layer Thickness 
\label{sec:transient_response}} 

\subsection{ Debye-Falkenhagen Equation }

Although the focus of this paper is on nonlinear relaxation, leading
to the steady state described in the previous section, it is
instructive to first consider linear response to a weak applied field,
where exact solutions are possible. Moreover, the linear analysis also
has relevance for the early stages of nonlinear relaxation before
significant double layer charge has accumulated, such that
$\max\{\zeta(\theta)\} \ll kT/e$, as long as the metal surface is
uncharged when the field is applied. The linear response also allows a
more general analysis, including AC periodic response, in addition to
our model problem with sudden DC forcing.

When the potential drop across the particle is much smaller than the 
thermal voltage ($E_o \ll 1$), it is possible to analyze the response of 
the system \emph{without} assuming that the double layers are thin; that is, 
we need not assume that $\eps$ is small and may describe the system
using the full unsteady PNP equations, 
(\ref{eq:c_eqn_bulk_diffusion_time}) -- (\ref{eq:rho_no_flux_bc}).
Instead, we assume that the response of the system is only a small
deviation from the equilibrium solution: 
\bea
c \equiv 1 \ , \ \ \rho \equiv 0 \ , \ \ \phi = E r \cos \theta.
\eea 
In this limit, we can linearize the ionic concentrations around 
a uniform concentration profile so that $c = 1 + \delta c$.  
Using this expression in (\ref{eq:rho_eqn_bulk_diffusion_time}) and
making use of Poisson's equation (\ref{eq:poisson_eqn_bulk_diffusion_time}) to 
eliminate the electric potential, we find that the charge density, 
$\rho = \left( c_+-c_- \right)/2 = \left( \delta c_+ - \delta c_- \right)/2$, 
obeys the (dimensionless) Debye-Falkenhagen equation \cite{debye1928}:
\beq
  \frac{\partial \rho}{\partial t} \approx \lapl\rho - \frac{1}{\eps^2} \rho
  \label{eq:debye_falkenhagen}.
\eeq
Similarly, the flux boundary condition corresponding to this equation
reduces to 
\beq
  \frac{\partial \rho}{\partial n} + \frac{\partial \phi}{\partial n} = 0
  \label{eq:linear_response_rho_bc}.
\eeq
Note that (\ref{eq:debye_falkenhagen}) and Poisson's equation 
(together with the no-flux and Stern boundary conditions) are a linear 
system of partial differential equations. 
Thus, we can take advantage of integral transform techniques.

\subsection{Transform Solutions for Arbitrary $\eps$ and $\delta$}
Since for weak applied fields, the model problem is a linear, initial value 
problem, it is natural to carry out the analysis using Laplace transforms 
in time.  Transforming the Debye-Falkenhagen and Poisson equations, we obtain
\bea
  \lapl \rholap &=& \beta^2 \rholap \\
  -\eps^2 \lapl \philap &=& \rholap
  \label{eq:poisson_eqn_LT}
\eea
where 
\beq
  \beta(s)^2 = s + \frac{1}{\eps^2}
\eeq
and the check accents $(\check{\ })$ are used to denote Laplace transformed 
variables.  Similarly, the boundary conditions become
\bea
  \frac{\partial \rholap}{\partial n} + 
    \frac{\partial \philap}{\partial n} &=& 0 
  \label{eq:lin_response_rho_bc} \\ 
  \philap + \delta \eps \frac{\partial \philap}{\partial n} &=& v s^{-1} 
  \label{eq:lin_response_phi_bc} \\
  -\nabla \philap \rightarrow E_o s^{-1} \ &\textrm{as}& \ 
    r \rightarrow \infty.
  \label{eq:lin_response_phi_bc_infinity}
\eea

To solve the resulting boundary value problem, we take advantage
of the spherical geometry to write the solution in terms of 
spherical harmonics.  Since $\rholap$ satisfies the modified Helmholtz
equation, we can expand it in a series with terms that are products of
spherical harmonics, $Y_l^m(\theta,\phi)$, and modified spherical Bessel 
functions, $k_l(\beta r)$.  Moreover, we can reduce the series to 
a single term 
\beq
  \rholap(r,\theta, \phi) = R~k_1(\beta r)~Y_1^0(\theta,\phi) 
    = R~k_1(\beta r)~\cos \theta 
  \label{eq:rholap_general_soln}
\eeq
by taking into account the symmetries of the charge density:
(i) axisymmetry, (ii) antisymmetry with respect to $\theta = \pi/2$, 
and (iii) the dipolar nature of the externally applied field.
Note that we have only retained the term involving the modified spherical 
Bessel functions that decays as $r \rightarrow 0$ because $\rholap$ vanishes 
at infinity.
Similarly, the general solution for $\philap$ may be expressed as 
\beq
  \philap(r,\theta, \phi) =
    - E_o s^{-1} r \cos \theta 
    + A + B \frac{\cos \theta}{r^2}
    + C~k_1(\beta r)~\cos \theta
  \label{eq:philap_general_soln}
\eeq
where the first term accounts for the boundary condition on the electric 
field at infinity, the next two terms are the general solution to Laplace's 
equation possessing the required symmetries, and the last term 
is the particular solution to Poisson's equation.  Note that
we have left out the monopolar term in the potential because it is
only necessary for charged spheres.  Our analysis is not made any less 
general by neglecting this term; the case of a charged sphere in an
weak applied field is handled by treating it as the superposition of 
a charged sphere in the absence of an applied field with an uncharged sphere 
subjected to an applied field. 

The coefficients in (\ref{eq:rholap_general_soln}) and 
(\ref{eq:philap_general_soln}) are determined by enforcing Poisson's 
equation and the boundary conditions (\ref{eq:lin_response_rho_bc}) --
(\ref{eq:lin_response_phi_bc}).  Plugging 
(\ref{eq:rholap_general_soln}) and (\ref{eq:philap_general_soln}) 
into Poisson's equation (\ref{eq:poisson_eqn_LT}), we obtain
\beq
   C = -\frac{R}{\left( \beta \eps \right)^2}.
\eeq
To apply the boundary conditions, note that on the sphere
$\frac{\partial}{\partial n}  = - \left . \frac{\partial}{\partial r} 
\right |_{r=1}$ so that
\beq
  \frac{\partial \philap}{\partial n} = 
    E_o s^{-1} \cos \theta
    + 2 B \cos \theta
    + \frac{1}{\left( \beta \eps \right)^2} 
      \left . \frac{\partial \rholap}{\partial r} 
      \right |_{r = 1}
\eeq
where the last term was obtained by using the relation between $C$ and $R$. 
Thus, the no-flux boundary condition (\ref{eq:lin_response_rho_bc})
becomes
\bea
  0 &=& 
    E_o s^{-1} \cos \theta + 2 B \cos \theta
    + \left (\frac{1}{\left( \beta \eps \right)^2} -1 \right ) 
    \left .\frac{\partial \rholap}{\partial r} \right |_{r=1}
    \nonumber \\
  &=& 
    \left \{ 
      \begin{array}{l}
        E_o s^{-1} + 2 B \\
        + \ R \left (\frac{1}{\left( \beta \eps \right)^2} -1 \right ) 
        \beta k_1'(\beta) 
       \end{array}
     \right \} \cos \theta
  \label{eq:lin_response_noflux_bc_series}
\eea
Similarly, the Stern boundary condition (\ref{eq:lin_response_phi_bc})
becomes
\bea
  v s^{-1} &=&  A \ + \nonumber \\
    & & \left \{ 
      \begin{array}{l}
       \left(\delta \eps - 1\right) E_o s^{-1} 
      + \left(1+2\delta \eps \right) B \\
      + \ \frac{R}{( \beta \eps )^2} 
        \left [ -k_1(\beta) + \delta \eps \beta k_1'(\beta) \right ]
      \end{array}
    \right \} \cos \theta
  \label{eq:lin_response_stern_bc_series}
\eea
By independently equating the coefficients of the different spherical 
harmonics in (\ref{eq:lin_response_noflux_bc_series}) and 
(\ref{eq:lin_response_stern_bc_series}), we obtain (after a little algebra)
\bea
  A &=& v s^{-1} 
    \label{eq:lin_response_A}  \\
  R &=& \frac{-3 E_o s^{-1} \left( \beta \eps \right)^2}
    {2 k_1(\beta) + \left [ 1 - \left( \beta \eps \right)^2 
                            \left(1 + 2 \delta \eps \right) \right ] 
      \beta k_1'(\beta)} 
    \label{eq:lin_response_R} \\
  B &=& -\frac{E_o s^{-1}}{2} 
        -\frac{R}{2} \left( \frac{1}{\left( \beta\eps \right)^2} - 1\right)
        \beta k_1'(\beta) 
    \label{eq:lin_response_B}.
\eea
Finally, by writing $k_1(x)$ in terms of elementary 
functions~\cite{weisstein_modified_spherical_bessel_function},
\beq
  k_1(x) = \frac{e^{-x} (x+1)}{x^2},
\eeq
we can express $R$ as
\beq
  R = \frac{-3 E_o s^{-1}}
  { e^{-\beta} 
    \left [ 
    \left ( 1 + \frac{2}{\beta} + \frac{2}{\beta^2} \right )
    \left ( 1 + 2 \delta \eps \right )
    - \frac{1}{\left( \beta \eps \right)^2}
    \right ]
  }. \label{eq:lin_response_R_expanded} 
\eeq

Following Bazant, Thornton, and Ajdari~\cite{bazant2004}, we focus on
times that are long relative to the Debye time ($t = O(\eps^2)$ in 
dimensionless units).  In this limit, $s \ll 1/\eps^2$ so that
the charge density on the surface of the sphere is given by 
\beq
  \rholap(r = 1,\theta,s) \sim 
    \left ( \frac{-K_\rho s^{-1}}{1+\tau_\rho s} \right) \cos \theta
\eeq
with 
\bea
  K_\rho &=& \frac{3 E_o (1+\eps)} {2 \gamma} \\
  \tau_\rho &=& \eps
     \left [ 
       \frac{(1 + 2 \delta \eps) (1 + \eps) - \delta \eps}
       {2 \gamma (1+\eps)}
    \right ] \\
  \gamma &=& \left ( 1 + 2 \delta \eps \right ) 
             \left ( 1 + \eps \right ) + \delta.
\eea

Inverting the Laplace transform, we see that at long times,
the charge at the surface of the sphere has an exponential
relaxation with a characteristic time on the order of $\eps$: 
\beq
  \rho(r = 1,\theta,t) \sim 
  -K_\rho \left ( 1 - e^{-t/\tau_\rho} \right ) \cos \theta.
  \label{eq:lin_response_rho_relaxation_long_times}
\eeq
Note that $\tau_\rho$ is on the order of the dimensionless RC time, $\eps$, 
which is much larger than $\eps^2$, the dimensionless Debye time.

To obtain the linear response of cylinders, we follow the same procedure
as above.  The main differences are that the series solution is written 
in terms of a cosine series with the radial dependence given by modified 
Bessel functions.  Without going through the algebra, the results for
cylinders are given in Table~\ref{tab:linear_response} along side the
analogous results for spheres.

\begin{table*}
\caption{\label{tab:linear_response}  Table of formulae for 
the linear response of metallic cylinders and spheres to weak applied 
electric fields.
}
\begin{ruledtabular}
\begin{tabular}{ccc}
  & Cylinder\footnote{$\theta$ measured from vertical axis.} & Sphere \\
\hline \\[1pt]
 $\rholap$ & $R K_1(\beta r) \cos \theta$ 
           & $R k_1(\beta r) \cos \theta$              \\[12pt]

 $\philap$ & $vs^{-1} -E_o s^{-1} r \cos \theta 
             + \left [\frac{B}{r} - \frac{R K_1(\beta r)}{(\beta \eps)^2} 
               \right ] \cos \theta$ 
             & $vs^{-1} -E_o s^{-1} r \cos \theta 
             + \left [\frac{B}{r^2} - \frac{R k_1(\beta r)}{(\beta \eps)^2} 
               \right ] \cos \theta$  \\[12pt]

 $R$ & $\frac{-2 E_o s^{-1} \left( \beta \eps \right) ^2}
    {K_1(\beta) 
    + \left [ 1 - (\beta \eps)^2 (1+\delta \eps) \right ] \beta K_1'(\beta)}$
     & $\frac{-3 E_o s^{-1}} { e^{-\beta} 
    \left [ 
    \left ( 1 + \frac{2}{\beta} + \frac{2}{\beta^2} \right )
    \left ( 1 + 2 \delta \eps \right )
    - \frac{1}{\left( \beta \eps \right)^2}
    \right ]}$ \\[16pt]

 $B$ & $-E_o s^{-1}
        -R \left( \frac{1}{\left( \beta\eps \right)^2} - 1\right)
        \beta K_1'(\beta)$
     & $-\frac{1}{2} E_o s^{-1}
        -\frac{1}{2} R \left( \frac{1}{\left( \beta\eps \right)^2} - 1\right)
        \beta k_1'(\beta)$ \\[16pt]

 $K_\rho$ & $\frac{2 E_o K_1(1/\eps)}{K_1(1/\eps) - \delta K_1'(1/\eps)}$
          & $\frac{3 E_o (1+\eps)} {2 
   \left [ \left( 1+2\delta\eps \right) \left( 1+\eps \right) + \delta
   \right]}$ \\[16pt]

 $\tau_\rho$ & $-\eps \left \{ 
       \frac{2 \eps K_1(1/\eps) 
     + \left[2+\delta \eps - \delta \frac{K_1'(1/\eps)}{K_1(1/\eps)} \right]
       K_1'(1/\eps)
     + \delta K_1''(1/\eps)}{2 \left[K_1(1/\eps) - K_1'(1/\eps) \right]} 
     \right \}$
             & $\eps \left \{ 
       \frac{(1 + 2 \delta \eps) (1 + \eps) - \delta \eps} {2 (1 + \eps)
     \left [ \left( 1+2\delta\eps \right) \left( 1+\eps \right) + \delta
     \right]} \right \}$ \\[16pt]

 $K_q$ & $\frac{2 E_o \eps K_0(1/\eps)}{K_1(1/\eps) - \delta K_1'(1/\eps)}$
       & $\frac{3 E_o \eps}
         {\left( 1+2\delta\eps \right) \left( 1+\eps \right) + \delta}$ 
       \\[16pt]

 $\tau_q$ & $-\eps \left \{ 
       \frac{\left(\eps + \frac{K_0'(1/\eps)}{K_0(1/\eps)} \right) K_1(1/\eps) 
     + \left[1+2\delta\eps-\delta \frac{K_0'(1/\eps)}{K_0(1/\eps)} \right]
       K_1'(1/\eps)
     + \delta K_1''(1/\eps)}{2 \left[K_1(1/\eps) - K_1'(1/\eps) \right]} 
     \right \}$
          & $\eps \left \{
    \frac{(1 + 2 \delta \eps) (1 + \eps) }{2 
    \left [ \left( 1+2\delta\eps \right) \left( 1+\eps \right) + \delta 
    \right]} \right \}$ \\[12pt]

\end{tabular}
\end{ruledtabular}
\end{table*}

\subsection{Response to a Weak, Oscillatory Field}
Due to the close relationship between Fourier and Laplace transforms,
the algebra involved in analyzing the response of the sphere to
a weak, oscillatory field is almost identical to the response to 
a suddenly applied field.  Thus, for sufficiently low frequencies 
($\omega \ll 1/\eps^2$), we can immediately write down the response to 
a weak, oscillatory field of the form 
$E = E_o \Real \left ( e^{i \omega t} \right )$:
\beq
  \rho(r = 1,\theta,t) = -K_\rho 
    \Real \left ( \frac{e^{i \omega t}}{1 + i \omega \tau_\rho} \right )
    \cos \theta
  \label{eq:lin_response_rho_oscillatory_field}.
\eeq

\subsection{Accumulated Surface Charge Density}
Because of its importance in many physical processes, the accumulated
surface charge density, $\qlap$, is an interesting quantity to consider.
It is easily computed from the (volume) charge density $\rho$ by 
integrating it in the radial direction from $r=1$ to $r = \infty$.  
While the identification of this integral with a surface charge density 
really only makes sense in the thin-double layer limit, the 
accumulated surface charge density is still worth examining.  
Fortunately, the integral is straightforward because 
$k_1(z) = -k_0'(z)$~\cite{weisstein_modified_spherical_bessel_function}
and the radial dependence of the charge density is independent of the angle. 
Using these observations, the surface charge density is given by 
\bea
  \qlap(\theta,s) &=& \frac{R~k_0(\beta)}{\beta} \cos \theta
  \nonumber \\ 
    &=& \frac{R~e^{-\beta}}{\beta^2}  \cos \theta.
\eea
In the long time limit $s \ll 1/\eps^2$, we find that the 
surface charge density an exponential relaxation 
$q(\theta,t) = -K_q \left (1 - e^{-t/\tau_q} \right ) \cos \theta$
with 
\bea
  K_q &=& \frac{3 E_o \eps}{\gamma} \\
  \tau_q &=& \eps \left [
    \frac{(1 + 2 \delta \eps) (1 + \eps) }{2 \gamma}
    \right ].
\eea
As with $\rho$, we see that the characteristic relaxation 
time for $q$ is on the order of the dimensionless RC time.

\subsection{Time Scales for Linear Response}
\begin{figure}[htb]
\bc
\includegraphics[width=1.6in,height=1.4in]{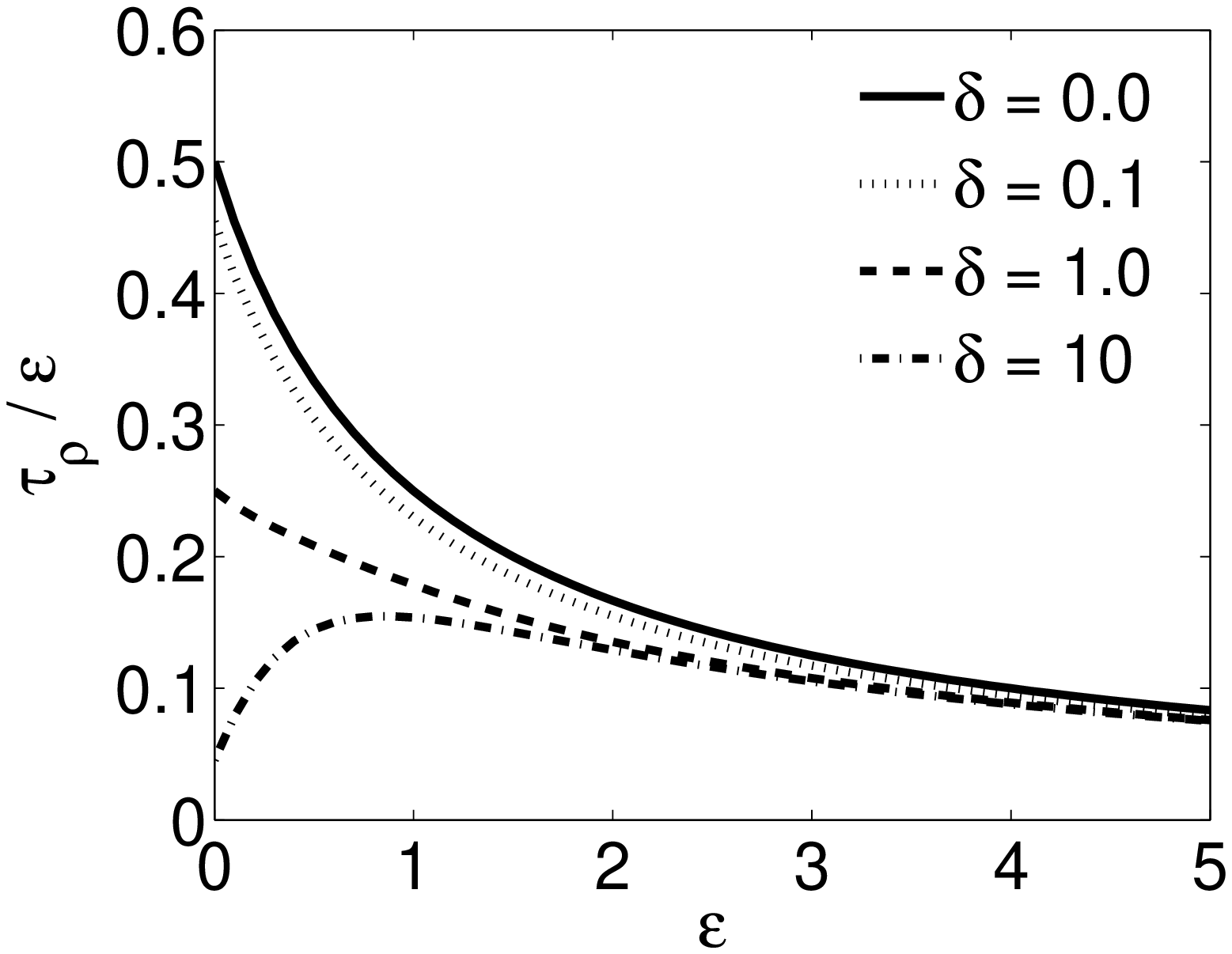}
\includegraphics[width=1.6in,height=1.4in]{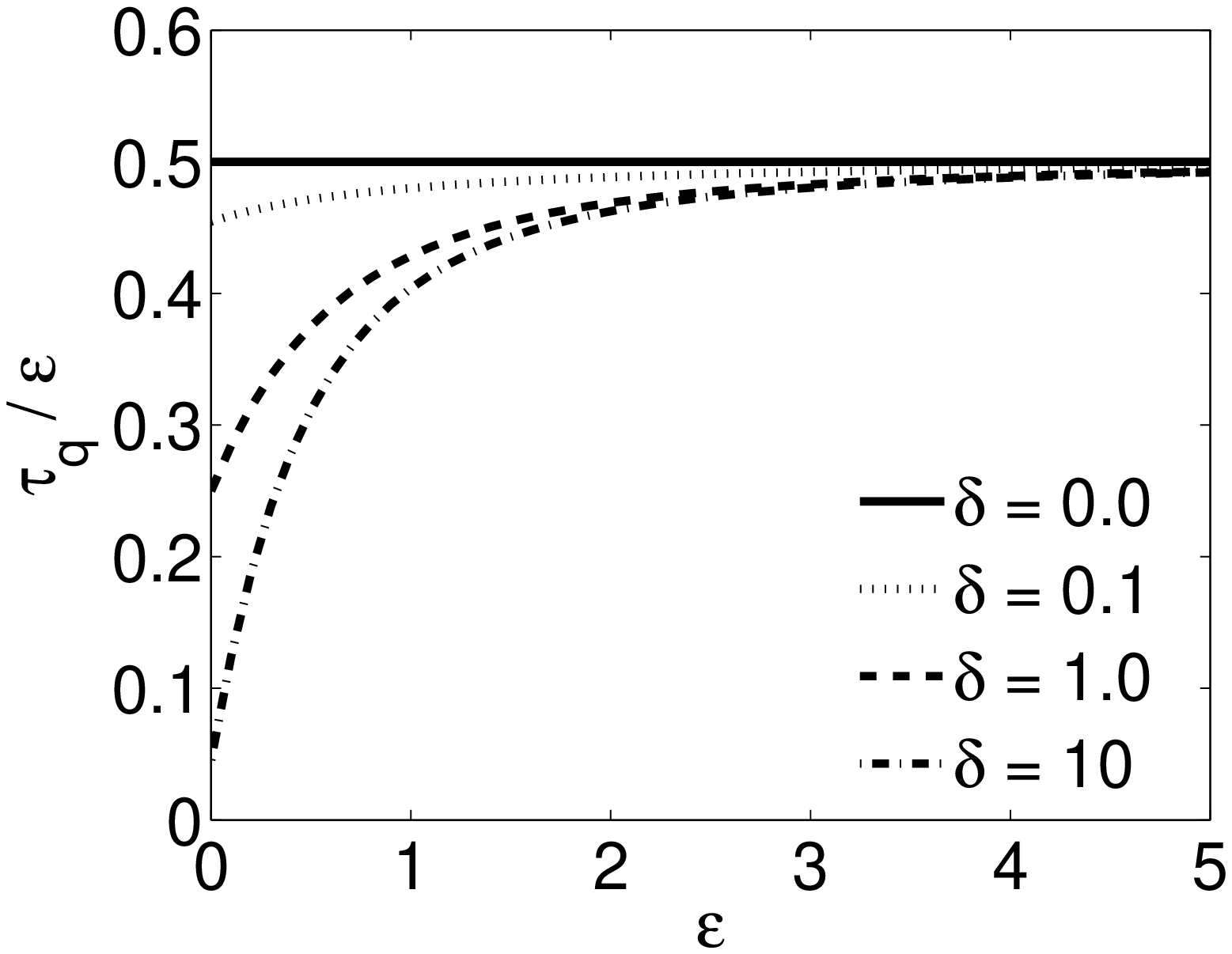}
\begin{minipage}[h]{3in}
\caption[Exponential relaxation time constants for the charge density
and accumulated surface charge density at weak applied fields as a function
of $\eps$ and $\delta$]{
\label{figure:3D_lin_relax_time}
Exponential relaxation time constants for the charge density $\rho$ and 
the accumulated surface charge density $q$ at weak applied fields
as a function of $\eps$ and $\delta$.
The left panel shows the relaxation time constant for the charge 
density at the surface of the sphere, $\tau_\rho(r=1)$.
The right panel shows the relaxation time constant for the accumulated
surface charge density, $\tau_q$.  
}
\end{minipage}
\ec
\end{figure}

We have seen that at long times both $\rho$ and $q$ relax exponentially 
with characteristic time scales on the order of the RC time, $\eps$.  
However, as Figure \ref{figure:3D_lin_relax_time} shows, the relaxation times 
for the two quantities are not exactly the same and have a nontrivial 
dependence on the diffuse layer thickness, $\epsilon$, and Stern capacitance, 
$\delta$. 
Notice that for infinitely thin double layers ($\eps = 0$) the relaxation 
times for the surface charge density and the accumulated charge density 
are identical.  This behavior is expected since for thin double 
layers, almost all of the charge density in the diffuse layer is located 
very close to the surface of the particle.  In this limit, the relaxation
time has a strong dependence on the Stern capacitance. 
For thick double layers ($\eps \gg 1$), this dependence on the Stern
capacitance disappears and the relaxation time curves for all $\delta$
values converge.  

Physically, the difference in the relaxation times for the surface charge 
and the accumulated charge densities for nonzero $\eps$ values is an 
indication of the complex spatio-temporal structure of the double layer 
charging.   For thin double layers, the difference $\tau_\rho$ and $\tau_q$
is relatively small because the charge in the double layer is restricted 
to a thin region.  
For thick double layers, however, the difference in relaxation times is 
accentuated because the charge in the double layer is spread out over a 
larger spatial region, which does not necessarily charge uniformly.  
In fact, that $\tau_\rho(r=1)$ is smaller than $\tau_q$ for larger values
of $\eps$ (Figure \ref{figure:3D_lin_relax_time}) suggests that regions 
closer to the surface of the sphere charge faster than regions that
are further away.

\section{Weakly Nonlinear Relaxation for Thin Double Layers}
\label{sec:weakly_nonlinear_dynamics}

\subsection{ Dynamical Regimes in Space-Time }

For weak applied fields in linear response, the complicated dependence
of the Laplace transform solution for large $s$ (short times) above
hints at the presence of multiple time scales in the charging
dynamics.  In this section, using boundary-layer methods, we derive
asymptotic solutions in the thin double-layer limit ($\eps \ll 1$) at
somewhat larger electric fields (defined below) for the concentrations
and electric potential by solving the leading order equations at the
two dominant time scales: (1) the RC time $\lambda_D a/D$ and (2) the
bulk diffusion time $a^2/D$.  We proceed by seeking the leading order
term (and in some cases, the first-order correction) to the governing
equations (\ref{eq:c_eqn_bulk_diffusion_time}) --
(\ref{eq:c_rho_poisson_eqn_bulk_diffusion_time}) with both the spatial
\emph{and} the time coordinate scaled to focus on the space-time
region of interest.

\begin{figure}
\bc
\includegraphics[width=3.1in]{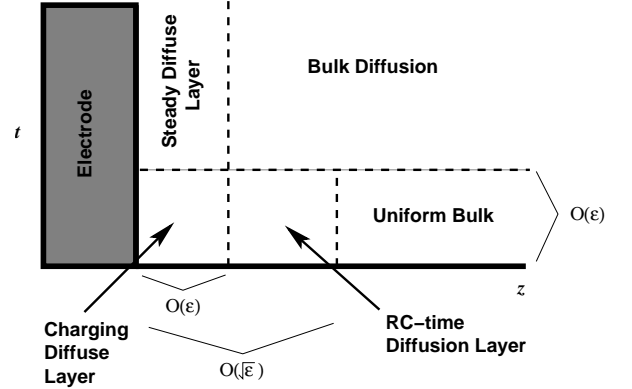}
\begin{minipage}[h]{3in}
\caption[Five dominant regions of space-time that define the dynamic
response of a metal colloid sphere to an applied electric field]{
\label{figure:five_space_time_regions}
Five asymptotically distinct regions of space-time that govern the 
dynamic response of a metal colloid sphere to an applied electric field.  
Note the nested spatial boundary layers at the RC time ($t = O(\eps)$).
}
\end{minipage}
\ec
\end{figure}

For the analysis, it is important to realize that the space-time domain is 
divided into five asymptotically distinct regions 
(see Figure \ref{figure:five_space_time_regions}).
At the RC time, there exist three spatially significant regions:
(i) an $O(\eps)$ quasi-steady double layer, (ii) an $O(\sqrt{\eps})$ 
dynamically active diffusion layer, and (iii) a quasi-equilibrium, uniform 
bulk electrolyte layer with time-varying harmonic electric potential.  
At this time scale, the charging dynamics are completely driven by the 
$O(\sqrt{\eps})$ diffusion layer.  
At the diffusion time, there are only two important spatial regimes: 
(i) a quasi-steady double layer and 
(ii) a dynamic bulk that evolves through locally electroneutral,
diffusion processes.

\subsection{Dynamics at the RC Time}

\subsubsection{Uniform Bulk and Equilibrium Double Layers}
To examine the dynamics at the RC time, we rewrite 
(\ref{eq:c_eqn_bulk_diffusion_time}) -- 
(\ref{eq:c_rho_poisson_eqn_bulk_diffusion_time})
by rescaling time to the RC time using $\tRC = t/\eps$:
\bea
 \frac{\partial c}{\partial \tRC} &=&
   \eps \nabla \cdot \left ( \nabla c + \rho \nabla \phi \right ) 
 \label{eq:c_eqn_bulk_RC_time} \\
 \frac{\partial \rho}{\partial \tRC} &=&
   \eps \nabla \cdot \left ( \nabla \rho + c \nabla \phi \right ) 
 \label{eq:rho_eqn_bulk_RC_time} \\
 -\eps^2 \lapl \phi &=& \rho
 \label{eq:poisson_eqn_bulk_RC_time}.
\eea
Since the spatial coordinate is scaled to the bulk length, the leading order 
solution of these equations describes the dynamics of the bulk during the 
double layer charging phase.  Substituting a regular asymptotic expansions 
of the form 
\beq
  c(x,t) \sim c_0 + \eps c_1 + \eps^2 c_2 + \ldots
\eeq 
into equations 
(\ref{eq:c_eqn_bulk_RC_time})-(\ref{eq:poisson_eqn_bulk_RC_time})
yields a hierarchy of partial differential equations.
By sequentially solving the equations using the initial conditions, 
it is easy to show that the ``outer'' solutions at the RC charging time 
scale are
\bea
  c_0~(x,t) &\equiv& 1 \\
  c_j(x,t) &\equiv& 0 \ ,\ \ j = 1,2,3, \ldots  \\
  \rho_j(x,t) &\equiv& 0 \ ,\ \ j = 0,1,2, \ldots 
  \label{eq:c_and_rho_bulk_RC_time}
\eea 
with $\phi_j$ is harmonic at all orders.  In other words, the bulk 
solution can be completely expressed (\emph{without} a series expansion) 
as a uniform concentration profile, $c~(x,t) \equiv 1$, and 
a time-varying harmonic electric potential, $\phi$.  
Note that by taking advantage of spherical geometry and axisymmetry in 
our problem, we can write the potential as a series in spherical harmonics 
with zero zonal wavenumber (\ie Legendre polynomials in $\cos \theta$):
\bea
  \phi(r,\theta,t) &=& -E_o r \cos \theta 
     + \sum_{l=0}^{\infty} P_l(\cos \theta) \frac{A_{l}(t)}{r^{l+1}} 
  \label{eq:phi_bulk_RC_time}
\eea
where the radial dependence of each term has been selected so that 
$\phi$ automatically satisfies Laplace's equation at all times.

At the RC charging time, the $O(\eps)$ double layer is in quasi-equilibrium,
which is easily verified by rescaling the spatial coordinate normal 
to the particle surface by the dimensionless Debye length $\eps$.  
As mentioned earlier, a quasi-equilibrium double layer possesses a structure
described by the Gouy-Chapman-Stern model.  For convenience, we repeat
the GCS solution here:
\bea
  \ct_\pm = \hat{c} e^{\mp\psit} \ &,& 
  \ \ \ct = \hat{c} \cosh{\psit} \ ,
  \ \ \rhot = - \hat{c} \sinh{\psit} \nonumber \\
  \psit(z) &=& 4 \tanh^{-1}\left(\tanh(\zeta/4) e^{-\sqrt{\hat{c}}z}\right).
  \label{eq:double_layer_structure}
\eea 
Recall that $\psit$ is the excess voltage relative to the bulk, 
$\psit = \phit - \hat{\phi}$ and that the tilde and hat accents are
used to indicate quantities within and just outside of the double layer,
respectively.  Also, we reiterate that unlike the 1D case, $\zeta$ is 
\emph{not} a constant but is a function of spatial along the electrode 
surface.

Since the bulk concentration at the RC time is uniform, the double layer
structure is given by (\ref{eq:double_layer_structure}) with $\hat{c}$ set 
equal to $1$.  
Notice that at the RC time, bulk concentration gradients have not yet had 
time to form, so the variation of the diffuse layer concentration and 
charge density along the surface of the sphere is solely due to a 
non-uniform zeta-potential.

\subsubsection{$O \left( \sqrt{\eps} \right)$ Diffusion Layer}
The analysis in the previous section leads us to an apparent paradox.
The dynamics of the system seem to have been lost since both the bulk and 
the boundary layers are in quasi-equilibrium at leading order; neither
layer drives its own dynamics.
As discussed in \cite{bazant2004}, the resolution to this paradox lies in 
the time-dependent flux matching between the bulk and the boundary layer.  
Unfortunately, it is inconsistent to directly match the bulk to the
boundary layer; there \emph{must} exist a nested
$O \left( \sqrt{\eps} \right)$ diffusion layer in order to account for the 
build up of both surface charge \emph{and} surface excess neutral salt 
concentration.  

Mathematically, the presence of the diffusion layer at the RC time scale
appears as a dominant balance in the transport equations by rescaling
the spatial coordinate in the normal direction to the surface by 
$\sqrt{\eps}$ to obtain:
\bea
 \frac{\partial \cb}{\partial \tRC} &=&
   \eps^2 \nabla_s \cdot \left ( \nabla_s \cb + \rhob \nabla_s \phib \right ) 
   \nonumber \\
   & & + \ \frac{\partial}{\partial \zdiff} 
   \left ( \frac{\partial \cb}{\partial \zdiff}
     +\rhob \frac{\partial \phib}{\partial \zdiff} 
   \right )
 \label{eq:c_eqn_diffusion_RC_time} \\
 \frac{\partial \rhob}{\partial \tRC} &=&
   \eps^2 \nabla_s \cdot \left ( \nabla_s \rhob + \cb \nabla_s \phib \right ) 
   \nonumber \\
   & & + \ \frac{\partial}{\partial \zdiff} 
   \left ( \frac{\partial \rhob}{\partial \zdiff}
     +\cb \frac{\partial \phib}{\partial \zdiff} 
   \right ) 
 \label{eq:rho_eqn_diffusion_RC_time} \\
 - \eps^2 \lapl_s \phib - \eps \frac{\partial^2 \phib}{\partial \zdiff^2} 
    &=& \rhob
 \label{eq:poisson_eqn_diffusion_RC_time}.
\eea
Here the bar accent denotes the ``diffusion layer'' solution at the RC time
and $z' = Z/\sqrt{\eps}$ is the spatial coordinate in the direction
normal to the surface.  Notice that at this length scale, the system
is \emph{not} in quasi-equilibrium as it is at the bulk and Debye length
scales.  It is, however, locally electroneutral at leading order
as a result of (\ref{eq:poisson_eqn_diffusion_RC_time}).

As the double layer charges, it absorbs an $O(\eps)$ amount of charge and 
neutral salt from the $O(\sqrt{\eps})$ diffusion layer.  Therefore, we 
expect that concentration changes within the diffusion to be on the order 
of $\sqrt{\eps}$, which motivates the use of an asymptotic expansion of 
the form
\beq
  \cb(x,t) \sim \cb_0 + \eps^{1/2}~\cb_{1/2} + \eps~\cb_1 + \ldots.
\eeq 
Using this expansion in (\ref{eq:c_eqn_diffusion_RC_time}) -- 
(\ref{eq:poisson_eqn_diffusion_RC_time}), we find that the leading
order equations are:
\bea
 \frac{\partial \cb_0}{\partial \tRC} &=&
   \frac{\partial^2 \cb_0}{\partial \zdiff^2}
   \label{eq:O1_c_eqn_diffusion_RC_time} \\
 \frac{\partial}{\partial \zdiff} 
   \left( \cb_0 \frac{\partial \phib_0}{\partial \zdiff} \right)
   &=& 0
   \label{eq:O1_rho_eqn_diffusion_RC_time} \\
 \rhob_0 &=& 0
   \label{eq:O1_poisson_eqn_diffusion_RC_time}.
\eea
The initial conditions and boundary condition as 
$\zdiff \rightarrow \infty$ for these equations are 
$\cb_0 (t = 0) \equiv 1$
and $\cb_0 (\zdiff \rightarrow \infty)= 1$, respectively.  
The boundary condition at $\zdiff = 0$ is given by flux matching 
with the double layer.  
Rescaling space and time in (\ref{eq:q_evolution_eqn_GCS}) and 
(\ref{eq:w_evolution_eqn_GCS}), we find that the appropriate flux 
boundary conditions for the diffusion layer are
\bea
  \frac{\partial q}{\partial \tRC} &=&
     \eps \nabla_s \cdot 
     \left(
         q \nabla_s \log \cb
       + w \nabla_s \phib
     \right)
     + \frac{1}{\sqrt{\eps}~}~\cb \frac{\partial \phib}{\partial \zdiff} 
  \label{eq:q_flux_bc_diffusion_layer} \\
  \frac{\partial w}{\partial \tRC} &=& 
     \eps \nabla_s \cdot 
     \left(
         w \nabla_s \log \cb
       + q \nabla_s \phib
     \right)
     + \frac{1}{\sqrt{\eps}~}~\frac{\partial \cb}{\partial \zdiff}
  \label{eq:w_flux_bc_diffusion_layer}. 
\eea
Thus, the leading order flux boundary conditions, 
which appear at $O(1/\sqrt{\eps})$, are
\bea
  \cb_0 \frac{\partial \phib_0}{\partial \zdiff} &=& 0 \\
  \frac{\partial \cb_0}{\partial \zdiff} &=& 0. 
\eea
The leading order solutions in the diffusion layer, 
$\cb_0 \equiv 1$ and $\phib_0 = \phi~(Z \rightarrow 0)$,
are easily determined by applying the initial and boundary conditions 
to (\ref{eq:O1_c_eqn_diffusion_RC_time}) --
(\ref{eq:O1_poisson_eqn_diffusion_RC_time}).

To obtain dynamics, we must examine the first-order correction to
the solution.  At the next higher order, the governing equations are
\bea
 \frac{\partial \cb_{1/2}}{\partial \tRC} &=&
   \frac{\partial^2 \cb_{1/2}}{\partial \zdiff^2}
   \label{eq:Osqrteps_c_eqn_diffusion_RC_time} \\
   \frac{\partial^2 \phib_{1/2}}{\partial \zdiff^2}
   &=& 0
   \label{eq:Osqrteps_rho_eqn_diffusion_RC_time} \\
 \rhob_{1/2} &=& 0
   \label{eq:Osqrteps_poisson_eqn_diffusion_RC_time}
\eea
where we have made use of the leading order solution to simplify
(\ref{eq:Osqrteps_rho_eqn_diffusion_RC_time}).  Again, the initial 
conditions and boundary condition at infinity are simple: 
$\cb_{1/2} (t = 0) \equiv 0$ and 
$\cb_{1/2} (\zdiff \rightarrow \infty)= 0$.
The flux boundary conditions at $\zdiff = 0$, however, are more interesting
because they involve the charging of the double layer:
\bea
  \frac{\partial q}{\partial \tRC} &=&
     \left . \frac{\partial \phib_{1/2}}{\partial \zdiff}  \right |_{z=0}
  \label{eq:O1_q_flux_bc_diffusion_layer} \\
  \frac{\partial w}{\partial \tRC} &=& 
     \left . \frac{\partial \cb_{1/2}}{\partial \zdiff} \right |_{z=0}
  \label{eq:O1_w_flux_bc_diffusion_layer}. 
\eea
Thus, simple diffusion of neutral salt at $x = O(\sqrt{\eps})$ driven by 
absorption into the $O(\eps)$ double layer dominates the dynamics of the 
diffusion layer.  From (\ref{eq:Osqrteps_rho_eqn_diffusion_RC_time})
and (\ref{eq:O1_q_flux_bc_diffusion_layer}), we see that the electric 
potential possesses a linear profile with slope given by rate of 
surface charge build up in the double layer:
\beq
  \phib \sim \phi~(Z \rightarrow 0) 
    + \eps^{1/2}~\left( \frac{\partial q}{\partial \tRC} \right) \zdiff
  \label{eq:phi_diffusion_layer}.
\eeq
Note that constant term at $O(\sqrt{\eps})$ is forced to be zero by 
matching with the bulk electric potential since all higher order corrections 
to the bulk potential are identically zero.

\subsubsection{Effective Boundary Conditions Across Entire Diffusion Layer} 
It is precisely the fact that the electric potential has the form
(\ref{eq:phi_diffusion_layer}) that justifies the common approach
of asymptotic matching directly between the bulk and the double layer
\cite{iceo2004b, bazant2004, iceo2004a}.  For instance, 
the time-dependent matching used by Bazant, Thornton, and 
Ajdari \cite{bazant2004} rests on the implicit assumption that
the leading order electric field in the diffusion layer is a constant
and appears at $O(\sqrt{\eps})$ so that
\beq
  \left. \frac{\partial \phi}{\partial Z} \right|_{Z=0} = 
  \left. \frac{1}{\sqrt{\eps}} 
     \left( \frac{\partial \phib}{\partial \zdiff} \right) 
  \right|_{\zdiff \rightarrow \infty}
  \sim \left. \frac{\partial \phib_{1/2}}{\partial \zdiff}
    \right|_{\zdiff \rightarrow \infty}
  = \frac{\partial q}{\partial \tRC}
  \label{eq:double_layer_charging_eqn_RC_time}.
\eeq
Similarly, the definition of the zeta-potential and the Stern boundary 
condition (\ref{eq:stern_bc_GCS}) across the entire diffusion layer 
require that $\phib \sim \phi~(Z \rightarrow 0)$ at leading order.

\subsubsection{Leading-order Dynamics}
Using the results of our discussion, we now derive the leading order 
equations that govern the charging dynamics on the surface of the sphere.  
Since charging is non-uniform over the electrode surface, the equations take 
the form of partial differential equations on the surface of the sphere.
Defining $\Psi(\theta,\phi) \equiv v - \phi(r=1,\theta,\phi)$,
we can write the Stern boundary condition (\ref{eq:stern_bc_GCS})
and the double layer charging equation 
(\ref{eq:double_layer_charging_eqn_RC_time}) as
\bea
  \zeta + 2 \delta \sinh \left( \zeta/2 \right) &=& \Psi 
    \label{eq:effective_stern_bc_RC_time} \\
  C (\Psi) \frac{\partial \Psi}{\partial \tRC}
    &=& \frac{\partial \phi}{\partial n}
   \label{eq:effective_charging_eqn_RC_time}
\eea
where we have introduced the leading order differential double layer 
capacitance $C (\Psi) = -\partial q / \partial \Psi$ and used the
fact that $\hat{c} = 1$ at the RC time.
Together with (\ref{eq:dlc_q_def_GCS}), 
(\ref{eq:effective_stern_bc_RC_time}) and
(\ref{eq:effective_charging_eqn_RC_time}) form a complete set of 
boundary conditions for the leading order electric potential in the bulk 
region.  
To compute the double layer capacitance, we can combine 
(\ref{eq:dlc_q_def_GCS}) with (\ref{eq:effective_stern_bc_RC_time}) to 
obtain 
\beq
  C = \frac{1}{\sech \left(\zeta/2\right) + \delta}
  \label{eq:double_layer_capacitance}.
\eeq
Since $C$ depends on $\Psit$ (via $\zeta$), the charging equation
(\ref{eq:effective_charging_eqn_RC_time}) is nonlinear which makes 
the problem analytically intractable.

For small applied fields (where $\zeta \approx E_o$ is a 
reasonable approximation), analytical progress can be made by linearizing 
the  the double-layer capacitance around $\zeta = 0$ to obtain
$C \sim 1/(1+\delta)$.   This approximation leads to a linear charging 
equation making it possible to solve the equations as an expansion 
in spherical harmonics.  Substituting the expansion~(\ref{eq:phi_bulk_RC_time})
into (\ref{eq:effective_charging_eqn_RC_time}) and the definition of
$\Psi$, we obtain a decoupled system of ordinary differential equations for 
the expansion coefficients:
\bea 
  \frac{d A_0}{d\tRC} + (1+\delta) A_0 &=& \frac{d v}{d\tRC} \nonumber \\
  \frac{d A_1}{d\tRC} + 2 (1+\delta) A_1 &=& 
    -(1+\delta) E_o + \frac{d E_o}{d\tRC} \nonumber \\
  \frac{d A_l}{d\tRC} + (1+l) (1+\delta) A_l &=& 0 \ , \ l > 1. 
  \label{eq:harmonic_coef_eqns}
\eea
To retain generality, we have allowed for the possibility that the applied 
electric field and surface potential are time-varying.  
For the case of a steady surface potential and uniform applied field, 
we find that the bulk electric potential is
\beq
  \phi = \frac{v}{r} e^{-(1+\delta) \tRC} 
    - E_o r \cos \theta \left [
    1 + \frac{1}{2 r^3} \left ( 1 - 3 e^{-2 (1+\delta) \tRC} \right )
    \right ]
  \label{eq:phi_bulk_low_field_RC_time}
\eeq
where we have assumed that both the surface potential and the electric field 
are suddenly switched on at $\tRC = 0$.  The initial condition in this 
situation is that of a conducting sphere at potential $v$ in a uniform 
applied electric field $E_o$: 
$\phi = v - E_o r \cos \theta \left( 1-1/r^3\right)$.  

Using (\ref{eq:phi_bulk_low_field_RC_time}), the double layer potential and 
total diffuse charge are easily determined to be
\bea
  \Psi &=& v \left ( 1 - e^{-(1+\delta) \tRC} \right ) \nonumber \\
   & & + \ \frac{3}{2}E_o \cos \theta 
         \left ( 1 - e^{-2 (1+\delta) \tRC} \right )
  \label{eq:Psi_low_field_RC_time} \\
  q &\sim& -\frac{v}{1+\delta} \left ( 1 - e^{-(1+\delta) \tRC} \right ) 
    \nonumber \\
    & & - \ \frac{3}{2 \left(1+\delta\right) }
      E_o \cos \theta \left ( 1 - e^{-2 (1+\delta) \tRC} \right )
  \label{eq:q_low_field_RC_time}.
\eea
These results are consistent with the calculations by Bazant and 
Squires~\cite{iceo2004b,iceo2004a}, which is expected since the low applied 
field limit corresponds to the regime where the total diffuse charge is 
linearly related to the zeta-potential (and therefore the total double layer 
potential).  It is worth mentioning that in the common situation where 
the surface potential is set well before $t = 0$, then the first term 
in each of the above three expressions is not present. 
Analogous double layer charging formulae for the cylinders are shown in 
Table~\ref{tab:linear_charging_RC_time}.
\begin{table*}
\caption{\label{tab:linear_charging_RC_time}  Table of formulae for 
double layer charging of metallic cylinders and spheres at weak applied 
electric fields at the RC-time.  In these formulae, the potential, $v$, 
of the particle is set to zero.
}
\begin{ruledtabular}
\begin{tabular}{ccc}
  & Cylinder\footnote{$\theta$ measured from vertical axis.} & Sphere \\ 
\hline \\[1pt]
 $\phi$ & $-E_o r \cos \theta 
           \left [ 1 + \frac{1}{r^2} \left( 1-2e^{-(1+\delta)t} \right) 
           \right ]$
        & $-E_o r \cos \theta 
           \left [ 1 + \frac{1}{2r^3} \left( 1-3 e^{-2(1+\delta)t} \right) 
           \right ]$ \\[8pt]

 $\Psi$ & $ 2 E_o \cos \theta \left( 1 - e^{-(1+\delta) t} \right) $
        & $ \frac{3}{2} E_o \cos \theta 
          \left( 1 - e^{-2(1+\delta) t} \right) $ \\[8pt]
 
 $q$    & $ -\frac{2}{1+\delta} E_o \cos \theta 
            \left( 1-e^{-(1+\delta) t} \right) $
        & $ -\frac{3}{2 (1+\delta)} E_o \cos \theta 
            \left( 1-e^{-2(1+\delta) t} \right) $ \\[6pt]
\end{tabular}
\end{ruledtabular}
\end{table*}

\subsubsection{Numerical Model} 
For larger applied fields, the nonlinear double layer capacitance forces us
to use numerical solutions to gain physical insight.  
Since the bulk electric potential is a time-varying harmonic function, 
it is most natural to numerically model the evolution equation for 
the potential using a multipole expansion with harmonic terms.  
Truncating (\ref{eq:phi_bulk_RC_time}) after a finite number of terms 
yields a discrete solution of the form:
\bea
  \phih(r,\theta,\tRC) = -E_o r \cos \theta 
    + \sum_{l=0}^{N} \frac{A_l(\tRC)}{r^{l+1}} P_l(\cos \theta)
  \label{eq:numerical_model_phi_weakly_nonlinear}
\eea
where the unknowns are the time-dependent coefficients in the expansion.
By using (\ref{eq:numerical_model_phi_weakly_nonlinear}) and enforcing 
that (\ref{eq:effective_charging_eqn_RC_time}) is satisfied at the 
collocation points $\theta_i = i \pi/N$, we derive a system of ordinary 
differential equations for the coefficients $A_l(\tRC)$:
\beq
  \mat{C} \mat{P} \frac{d \vec{A}}{d\tRC}  = 
    E_o \mat{P}(:,2) + \mat{Q} \vec{A}
  \label{eq:numerical_model_weakly_nonlinear_RC_time}
\eeq
where $\vec{A}$ is the vector of expansion coefficients, 
$\mat{P}$ and $\mat{Q}$ are collocation matrices that relate
the expansion coefficients to $\phi$ and $\partial \phi / \partial n$, 
respectively, and $\mat{C}$ is a diagonal matrix that
represents the double layer capacitance at the collocation points.
Note that $\mat{P}(:,2)$ (which represents the second column of 
$\mat{P}$ in MATLAB notation) is the discrete representation
of $P_1(\cos \theta) = \cos \theta$, so the term $E_o \mat{P}(:,2)$ 
accounts for the applied background potential.
The explicit forms for the collocation matrices and the discrete
double layer capacitance are given by 
\begin{widetext}
\bea
  \mat{P} = \left[ \begin{matrix}
     P_0(\cos \theta_0) & P_1(\cos \theta_0) & \ldots & P_N(\cos \theta_0) \\
     P_0(\cos \theta_1) & P_1(\cos \theta_1) & \ldots & P_N(\cos \theta_1) \\
     \vdots & \vdots & \ddots & \vdots \\
     P_0(\cos \theta_N) & P_1(\cos \theta_N) & \ldots & P_N(\cos \theta_N) 
  \end{matrix} \right] \ \ &,& \ \
  \mat{Q} = \left[ \begin{matrix}
     P_0(\cos \theta_0) & 2 P_1(\cos \theta_0) & \ldots 
       & (N+1) P_N(\cos \theta_0) \\
     P_0(\cos \theta_1) & 2 P_1(\cos \theta_1) & \ldots 
       & (N+1) P_N(\cos \theta_1) \\
     \vdots & \vdots & \ddots & \vdots \\
     P_0(\cos \theta_N) & 2 P_1(\cos \theta_N) & \ldots 
       & (N+1) P_N(\cos \theta_N) 
  \end{matrix} \right] \nonumber
  \\[6pt]
  \mat{C} &=& \diag \left( C(\Psi(\theta_0), C(\Psi(\theta_1), 
                       \ldots, C(\Psi(\theta_N) \right).
\eea
\end{widetext}
The system of ODEs for the expansion coefficients 
is easily solved using MATLAB's built-in ODE solvers by 
multiplying (\ref{eq:numerical_model_weakly_nonlinear_RC_time})
through by $(\mat{C}\mat{P})^{-1}$ and writing a simple function 
to evaluate the resulting right-hand side function. 

\subsubsection{Dipole-Dominated Charging}
\begin{figure}
\bc
\includegraphics[width=2.5in]{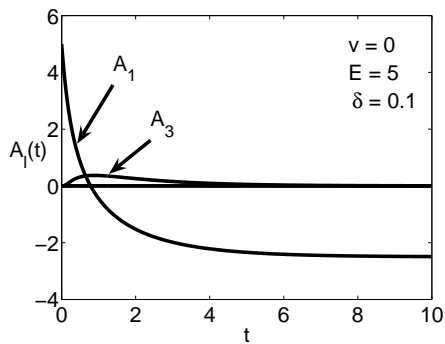}
\begin{minipage}[h]{3in}
\caption[Dipolar double layer charging in the weakly nonlinear regime]{
\label{figure:weakly_nonlinear_dipolar_charging}
Time evolution of the dominant coefficients in the Legendre polynomial 
expansion of the bulk electric potential in the weakly nonlinear regime 
at the RC time.  In this figure $v=0$, $E = 5$ and $\delta = 0.1$.
Notice that the dipolar term dominates the solution.
}
\end{minipage}
\ec
\end{figure}

From the numerical solution of the charging equation 
(\ref{eq:effective_charging_eqn_RC_time}), we see that when 
the sphere is electrically grounded, charging is dominated by the dipolar 
contribution to the response 
(see Figure~\ref{figure:weakly_nonlinear_dipolar_charging}).  
\begin{figure}
\bc
\includegraphics[width=2.5in]{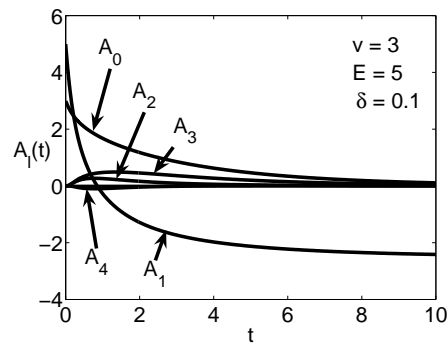}
\begin{minipage}[h]{3in}
\caption[Time evolution of dominant expansion coefficients for $\phi$ in 
the weakly nonlinear regime at the RC time when sphere has a nonzero 
applied voltage]{
\label{figure:weakly_nonlinear_dipolar_charging_v3}
Time evolution of the dominant coefficients in the Legendre polynomial 
expansion of the bulk electric potential in the weakly nonlinear regime 
at the RC time when the sphere has a nonzero applied voltage.  
In this figure $v=3$, $E = 5$ and $\delta = 0.1$.
Note that both symmetric and antisymmetric terms make non-negligible
contributions.
}
\end{minipage}
\ec
\end{figure}
While the nonlinear capacitance does in fact allow higher harmonics to 
contribute to the response, the higher harmonics only play a small role 
even at larger applied fields.  As expected, when the sphere is kept at zero 
voltage, the antisymmetry between the upper and lower hemisphere is not 
broken and only odd terms contribute to the series solution 
(\ref{eq:phi_bulk_RC_time}).  However, as shown in 
Figure~\ref{figure:weakly_nonlinear_dipolar_charging_v3},
if a nonzero potential is applied to the sphere, all harmonics contribute 
to the solution.  In this case, the dominant contributions come from the 
monopolar and dipolar terms.

The dipolar nature of double layer charging forms the foundation of
much of the work on the charging of colloid particles over the past half
century.  For instance, the non-equilibrium double layer is often 
characterized in terms of the induced dipole moment~\cite{dukhin1993}.  
As shown in (\ref{eq:phi_bulk_low_field_RC_time}) -- 
(\ref{eq:q_low_field_RC_time}), the monopole and dipole contributions are
the \emph{only} contributions in a linearized theory.  
Our numerical investigations demonstrate that, even for the nonlinear 
theory, the monopole and dipole response dominates the total response.
Our results provides additional theoretical support for the focus 
on the dipole response when studying colloid particles in applied fields.

\subsubsection{Extended Double Layer Charging}
\begin{figure}[htb]
\bc
\includegraphics[width=1.6in,height=1.4in]{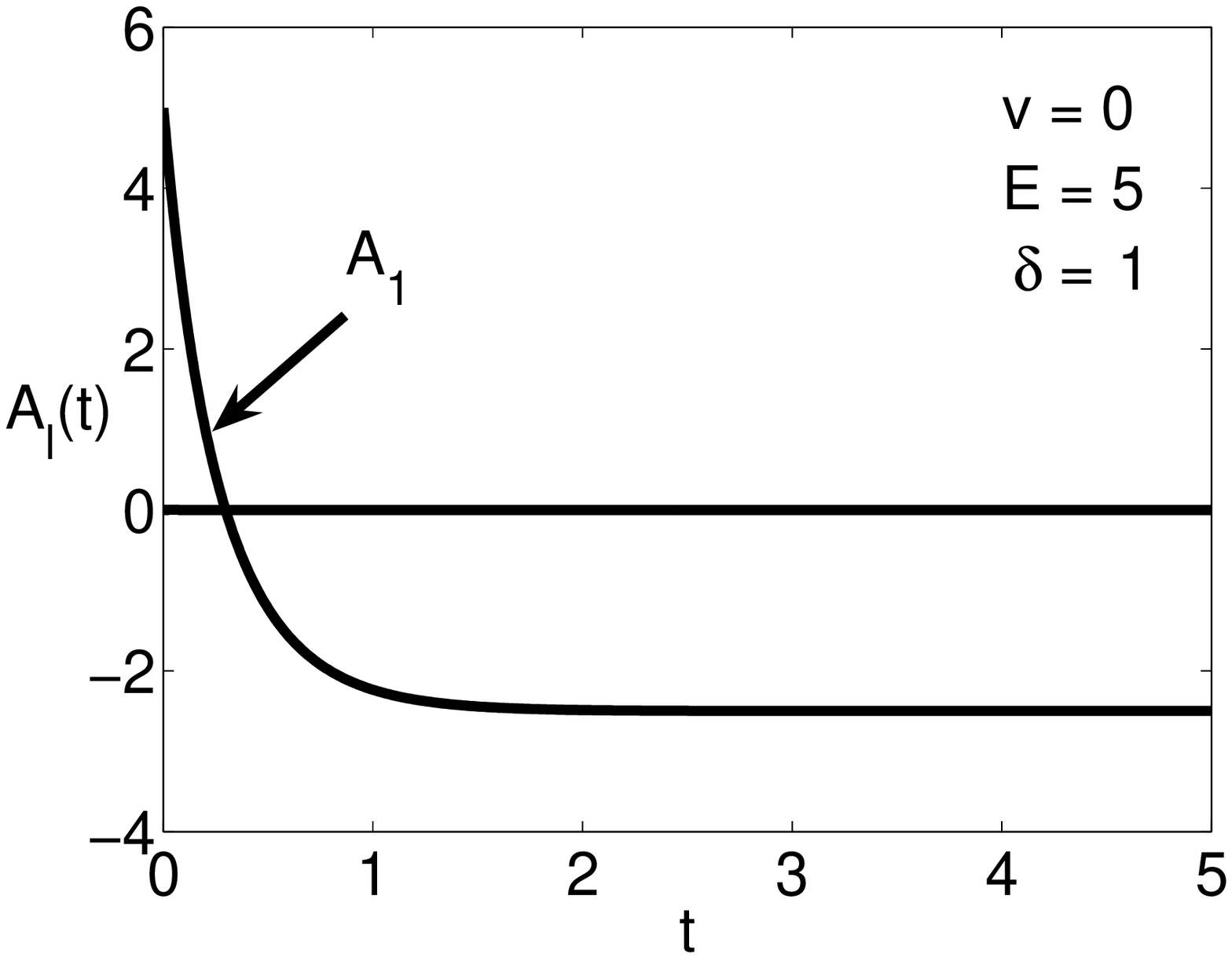}
\includegraphics[width=1.6in,height=1.4in]{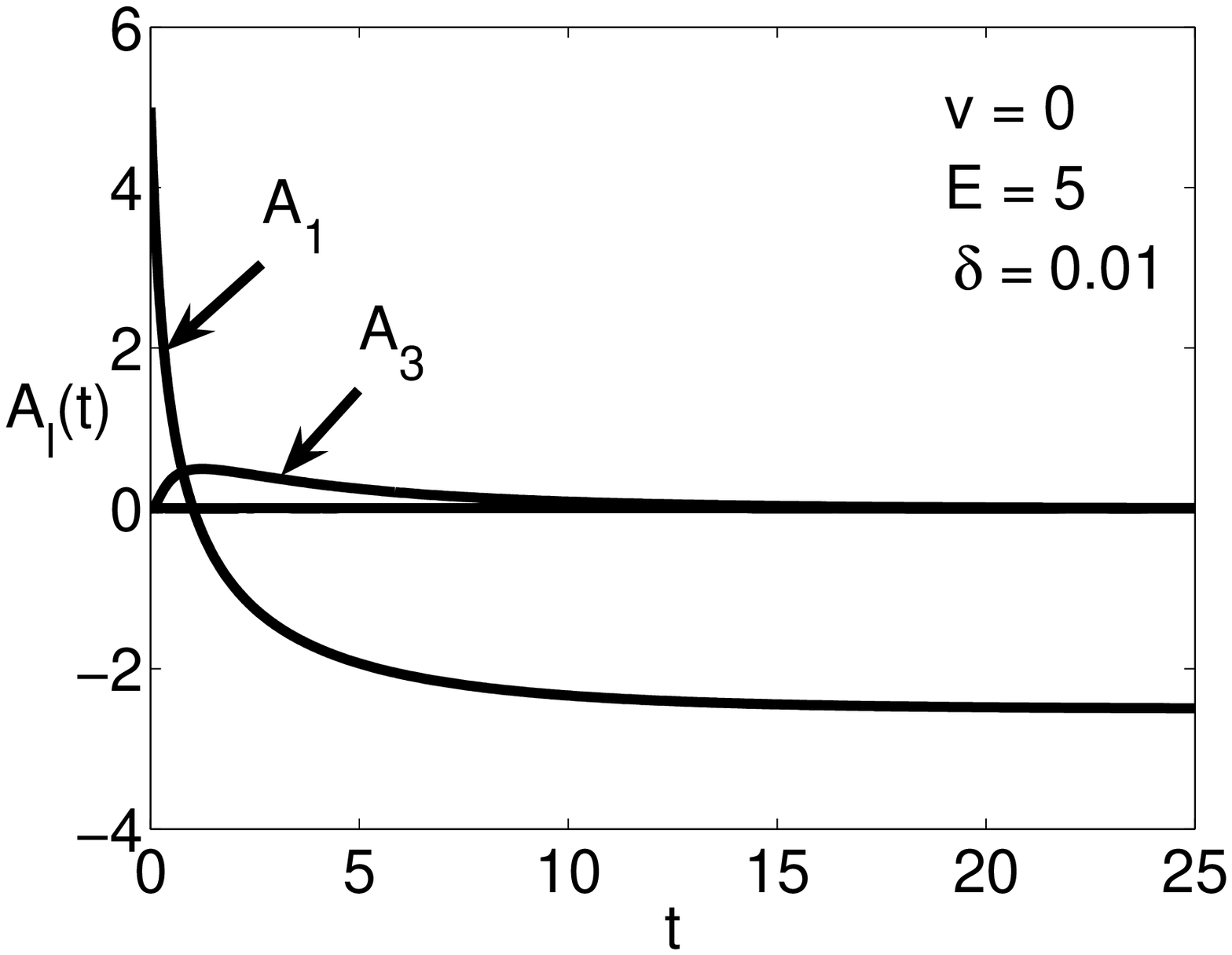}
\begin{minipage}[h]{3in}
\caption[Double layer charging in the weakly nonlinear regime at the 
RC time for varying $\delta$ values]{
\label{figure:weakly_nonlinear_retarded_double_layer_charging}
Time evolution of the dominant coefficients in the Legendre polynomial 
expansion of the bulk electric potential in the weakly nonlinear regime 
at the RC time for low (left) and high (right) Stern capacitance
values.  In these figures $v=0$ and $E = 5$.  
Note that double layer charging is retarded when $\delta$ is 
small but that the slowdown in double layer charging is suppressed
for larger $\delta$ values.
}
\end{minipage}
\ec
\end{figure}

The slowing of double layer charging is one important consequence of 
nonlinearity in the double layer capacitance (see 
Figure~\ref{figure:weakly_nonlinear_retarded_double_layer_charging}).
However, as shown in 
Figure~\ref{figure:weakly_nonlinear_retarded_double_layer_charging}
extended charging only occurs for $\delta \ll 1$.  
Mathematically, we only see slowed charging at
small values of $\delta$ because the double layer capacitance 
(\ref{eq:double_layer_capacitance}) can only become as small as 
$1/\delta$, which occurs at large zeta-potentials.  
For sufficiently small $\delta$, charging
is slowed at higher applied fields because the $\sech (\zeta / 2)$
term in the denominator of the double layer capacitance 
(\ref{eq:double_layer_capacitance}) becomes negligible when 
$\zeta \gg -2 \ln (\delta / 2)$.

\subsection{Dynamics at the Diffusion Timescale}
In this section, we examine the response of the system at the diffusion 
time scale.  We find that the only dynamic process is diffusion of neutral 
salt within the bulk in response to surface adsorption that occurred at the 
RC time scale.
Since the total amount of neutral salt absorbed by the diffuse charge layers 
during the charging phase is an $O(\eps)$ quantity, the bulk concentration 
only needs to decrease by $O(\eps)$ to compensate.   
Thus, we find that dynamics are not present at leading order;  
rather, they appear only in higher-order corrections. 
Also, at the diffusion time, surface transport, which is
negligible at the RC time scale, becomes important.

\subsubsection{Leading Order Bulk and Double Layer Solutions}

Substituting an asymptotic series into 
(\ref{eq:c_eqn_bulk_diffusion_time}) --
(\ref{eq:c_rho_poisson_eqn_bulk_diffusion_time}),
we obtain the leading order bulk equations:
\bea
  \frac{\partial c_0}{\partial t} &=& \lapl c_0 
  \label{eq:c0_eqn_bulk_diffusion_time} \\
  \lapl \phi_0 &=& 0.
  \label{eq:phi0_eqn_bulk_diffusion_time}
\eea
Applying the initial conditions (obtained by matching to
the solution at the RC time) and the leading order boundary conditions 
(derived from the effective flux boundary conditions 
(\ref{eq:q_evolution_eqn_GCS}) and (\ref{eq:w_evolution_eqn_GCS}))
yields the simple leading order solutions for a sphere
\bea
  c_0(\vec{x},t) &\equiv& 1 \\
  \phi_0(\vec{x},t) &=& -E_o r \cos \theta 
   \left(1 + \frac{1}{2 r^3} \right)
\eea
and a cylinder
\bea
  c_0(\vec{x},t) &\equiv& 1 \\
  \phi_0(\vec{x},t) &=& -E_o r \cos \theta 
   \left(1 + \frac{1}{r^2} \right).
\eea
Notice that there is no time dependence for either the concentration
or the electric potential.  It is worth mentioning that for a general 
geometry, the initial condition is a uniform concentration profile 
with the electric potential of an insulator in an applied field and the 
boundary conditions (which are consistent with the initial conditions) are
\beq
  c_0 \frac{\partial \phi_0}{\partial n} = 0 \ \ \mathrm{and} \ \ 
  \frac{\partial c_0}{\partial n} = 0.
\eeq

At the diffusion time scale, the double layer continues to remain in 
quasi-equilibrium, so its leading order structure is given by 
(\ref{eq:double_layer_structure}) with the bulk concentration set equal 
to $1$.  
However, unlike the double layer at the RC time scale, the leading
order zeta-potential is \emph{not} evolving, so the leading
order double layer structure is static in time.

\subsubsection{Higher-Order Bulk Diffusion} 
In order to see dynamics, we need to consider the first-order correction to 
(\ref{eq:c0_eqn_bulk_diffusion_time}) and
(\ref{eq:phi0_eqn_bulk_diffusion_time}):
\bea
  \frac{\partial c_1}{\partial t} &=& \lapl c_1  
  \label{eq:c1_eqn_bulk_diffusion_time} \\
  \lapl \phi_1 &=& - \nabla c_1 \cdot \nabla \phi_0
  \label{eq:phi1_eqn_bulk_diffusion_time}.
\eea
The boundary conditions for these equations are a bit more complicated.
Using (\ref{eq:q_evolution_eqn_GCS}) -- (\ref{eq:w_evolution_eqn_GCS})
and taking into account the leading order solutions, we find that the 
boundary conditions for the $O(\eps)$ equations are
\bea
  q_0~\delta^+(t) &=& \nabla_s \cdot \left( w_0 \nabla_s \phi_0 \right)
      - \frac{\partial \phi_1}{\partial n} 
    \label{eq:Oeps_q_evolution_eqn_GCS} \\
  w_0~\delta^+(t) &=& \nabla_s \cdot 
        \left( q_0 \nabla_s \phi_0 \right)
      - \frac{\partial c_1}{\partial n}
    \label{eq:Oeps_w_evolution_eqn_GCS}
\eea
where $q_0$ and $w_0$ are the leading order equilibrium surface
charge density and surface excess neutral salt concentration.
As mentioned earlier, at the diffusion time scale, these quantities
are static in time.
Also, note the presence of the delta-functions in time, which account for 
the ``instantaneous'' adsorption of charge and neutral salt from the bulk 
during the at the RC-time~\cite{bazant2004}. 

Mathematically, the appearance of the delta-functions is a consequence 
of the connection between the time derivative of double layer quantities at 
the two time scales in the asymptotic limit $\eps \rightarrow 0$.  
To illustrate this connection, consider the time derivative of the 
surface charge density, $q$.
Let $t$ and $\tRC$ be scaled to the diffusion and RC times, respectively, so
that $t = \eps \tRC$.
At these two time scales, the surface charge density can be written as
$q(t)$ and $\tilde{q} \left( \tRC \right)$ which are simply related by 
$q(t) = \tilde{q} \left( \tRC \right)$.  The time derivatives, however, 
are related by 
\beq
  \frac{\partial q}{\partial t}(t) = 
  \frac{1}{\eps}~\frac{\partial \tilde{q}}{\partial \tRC} 
    \left( t/\eps \right). 
  \label{eq:deriv_q_at_two_time_scales}
\eeq
Therefore, for nonzero $t$, $\frac{\partial q}{\partial t} = 0$ in the 
asymptotic limit because 
$\frac{\partial \tilde{q}}{\partial \tRC} \left( t/\eps \right)$ 
approaches zero faster than linearly as $\eps \rightarrow 0$ 
(as an example, see (\ref{eq:q_low_field_RC_time}) ).
In contrast, for $t=0$, $\frac{\partial q}{\partial t}$ is infinite
because $\frac{\partial \tilde{q}}{\partial \tRC}(0)$ has a fixed 
nonzero value.  
Next, consider the following integral with $t_2 > 0$:
\beq
  \int_{t_1}^{t_2} \frac{\partial q}{\partial t} dt 
  = \int_{t_1/\eps}^{t_2/\eps} \frac{\partial \tilde{q}}{\partial \tRC} d\tRC
  = \tilde{q} \left( t_2/\eps \right) - \tilde{q} \left(t_1/\eps \right).
\eeq
In the asymptotic limit, the integral approaches zero for
nonzero $t_1$ but approaches $\tilde{q}(\infty)$ when $t_1$ equals zero.
Putting the above properties together, we see that 
$\frac{\partial q}{\partial t}(t)$ is indeed a one-sided delta-function 
of strength $\tilde{q}(\infty) = q_0$.

In contrast to one-dimensional systems, the boundary layers play
a more active role in the evolution of the bulk concentrations because
surface conduction continues to play a role well beyond the initial
injection of ions at $t = 0$.  Note, however, that the surface 
conduction terms in (\ref{eq:Oeps_q_evolution_eqn_GCS}) and
(\ref{eq:Oeps_w_evolution_eqn_GCS}) are static, so they essentially 
impose fixed normal flux boundary conditions on the $O(\eps)$
bulk equations.

\subsubsection{Comparison with the Analysis of Dukhin and Shilov}
It is interesting to compare and contrast our weakly nonlinear analysis with 
the work of Dukhin and Shilov \cite{dukhin1969, shilov1970} on the 
polarization of the double layer for highly charged, spherical particles
in weak applied fields.
In both cases, bulk concentration variations and diffusion processes appear 
as a higher-order correction to a uniform background concentration
and are driven by surface conduction.  However, the significance of 
the surface conduction term arises for different reasons.
As mentioned earlier, the small parameters that controls the 
size of the correction is different in the two analyses.  
In Dukhin and Shilov's analysis, the small parameters are $\eps$ \emph{and} 
$E_o$.   Because they essentially use asymptotic series in $E_o$ as the 
basis for their analysis, they require that the double layer be highly 
charged in order for surface conduction to be of the same order of magnitude 
as the $O(E_o)$ normal flux of ions from the bulk.
In other words, because the size of the surface conduction is proportional 
to $E_o$, in order for surface conduction to be of the same order of
magnitude as the normal flux, the surface charge density \emph{must} be 
an $O(1)$ quantity: $\eps q = 2 \eps \sinh(\zeta_0/2) = O(1)$.
In contrast, we use asymptotic series in $\eps$ in our analysis and do not 
restrict $E_o$ to be small, so the surface conduction and normal flux of 
ions from the bulk have the same order of magnitude for small $O(\eps)$ 
surface charge densities regardless of the strength of the applied electric 
field (as long as it is not so large that the asymptotic analysis breaks 
down).  Thus, our weakly nonlinear analysis complements the work of 
Dukhin and Shilov by extending their analysis to stronger applied electric 
fields and the case where the surface charge density is induced by the
applied field rather than fixed by surface chemistry of the colloid 
particle.

\section{Strongly Nonlinear Relaxation
\label{sec:strongly_nonlinear_dynamics}}

\subsection{ Definition of ``Strongly Nonlinear'' } 

The weakly nonlinear analysis of the thin double-layer limit in the
previous section assumes that the dimensionless surface charge density
$\alpha$ and surface salt density $\beta$ remain uniformly small. For
our PNP model, we can estimate when this assumption becomes
significantly violated using Eq.~(\ref{eq:alpha}) with $\zeta =
E_0a/(1+\delta)$ and $C=C_0$, which occurs at fields above a critical
value at least a few times the thermal voltage,
\beq
E_0 > \frac{2kT}{z_+ea}\left(1 + \frac{\lambda_S}{\lambda_D}\right)
\log\left(\frac{a}{\lambda}\right).  \label{eq:strong1}
\eeq
As discussed in section~\ref{sec:surfproc}, this condition also
implies that surface 
adsorption of ions from the bulk is larger enough to trigger
significant bulk diffusion and surface transport through the double
layer, in steady steady state. However, as noted by Bazant
\etal~\cite{bazant2004},  weakly nonlinear asymptotics 
breaks down during relaxation {\it dynamics} at somewhat smaller
voltages, $\alpha/\sqrt{\pi\eps} < 1$, or (with units restored)
\beq
E_0 > \frac{kT}{z_+ea}\left(1 +
  \frac{\lambda_S}{\lambda_D}\right)\log\left(\frac{\pi
    a}{\lambda}\right),
\label{eq:strong2}
\eeq
since large surface adsorption can occur only {\it temporarily} at
certain positions. (The factor $\pi$ in this formula comes from a
one-dimensional analysis of bulk diffusive relaxation, which does not
strictly apply here, but the rough scale should be correct.

Beyond the weakly nonlinear regime, there are two main effects that
occur: (1) transient, local depletion of the leading order bulk
concentration and (2) surface conduction at the leading order.  As in
the case of a steady applied field, perhaps the greatest impact of
$\alpha = O(1)$ is that we must pay attention to factors of the form
$\eps e^{\zeta}$ or $\eps \sinh(\zeta)$, in addition to factors of
$\eps$, when ordering terms in asymptotic expansions.

In the thin double-layer limit, the boundary layers are still in
quasi-equilibrium, as long as the bulk concentration does not approach
zero, which would typically require Faradaic reactions consuming ions
at the conducting surface~\cite{bazant2005,chu2005}. Since we only
consider ideally polarizable conductors, bulk depletion is driven
solely by adsorption of ions in the diffuse layer, which is unlikely
to exceed diffusion limitation and produce non-equilibrium space
charge, although this possibility has been noted~\cite{bazant2004}.
Therefore, we would like to proceed as in the previous sections and treat the
bulk as locally electroneutral with effective boundary conditions.

\subsection{Leading-Order Equations} 

Unfortunately, the analysis of the leading order equations derived in
this manner does not appear to be as straightforward as the analysis
in the weakly nonlinear limit.  The main problem is that it seems
difficult to derive a uniformly valid leading order effective boundary
conditions along the entire surface of the sphere.  For this reason,
we merely present the \emph{apparent} leading order equations for the
strongly nonlinear regime and leave a thorough analysis for future
work.

At the leading order in the bulk, we find the usual equations for a neutral 
binary electrolyte:
\bea 
  \frac{\partial c_0}{\partial t} &=& \lapl c_0
  \label{eq:c_eqn_bulk_strongly_nonlinear}  \\
  0 &=& \nabla \cdot \left( c_0 \nabla \phi_0 \right) 
  \label{eq:rho_eqn_bulk_strongly_nonlinear} 
\eea 
with $\rho = O(\eps^2)$.  
The structure of the boundary layers is described by GCS theory with
the concentration and charge density profiles given by
(\ref{eq:double_layer_structure}). 
Effective boundary conditions for (\ref{eq:c_eqn_bulk_strongly_nonlinear}) --
(\ref{eq:rho_eqn_bulk_strongly_nonlinear}) are derived in the same manner
as for a steady applied field except that unsteady terms are retained.
Recalling that $q$ and $w$ grow exponentially with the zeta-potential,
we find that both the surface conduction terms and the time derivatives 
of the total diffuse charge and excess concentration appear in the leading
order boundary conditions:
\bea
  \eps \frac{\partial \qt_0}{\partial t} &=& 
    \eps \nabla_s \cdot \left ( \wt_0 \nabla_s \phi_0 \right ) 
    - c_0 \frac{\partial \phi_0}{\partial n} 
  \label{eq:O1_q_effective_bc_strongly_nonlinear} \\
  \eps \frac{\partial \wt_0}{\partial t} &=& 
    \eps \nabla_s \cdot \left ( \qt_0 \nabla_s \phi_0 \right ) 
    - \frac{\partial c_0}{\partial n} 
  \label{eq:O1_w_effective_bc_strongly_nonlinear}.
\eea

\subsection{Challenges with Strongly Nonlinear Analysis}
It is important to realize that these equations are mathematically much 
more complicated than the analogous equations in the weakly nonlinear regime.
First, the nonlinearity due to the electromigration term explicitly appears 
in the bulk equations at leading order; the nonlinearity is \emph{not} 
removed by the asymptotic analysis.  Furthermore, the diffuse layer 
concentrations depend on time explicitly through the bulk concentration 
at the surface in addition to the zeta potential:
\beq 
  \ct_\pm(t) = c_\pm(t) e^{\mp \psit(t)}.
\eeq
Already, these features of the equations greatly increases the challenge in
analyzing the response of the system.

However, the greatest complication to the mathematical model in the
strongly nonlinear regime is that effective boundary conditions
(\ref{eq:O1_q_effective_bc_strongly_nonlinear}) and
(\ref{eq:O1_w_effective_bc_strongly_nonlinear}) are not uniformly
valid over the surface of the sphere (or cylinder).  Near the poles,
the double layer charges quickly, so the time-dependent and surface
transport terms in the effective boundary conditions become $O(1)$ at
very short times.  In contrast, the amount of surface charge in the
double layer near the equator is \emph{always} a small quantity.
Thus, it would seem that near the equator, the only significant terms
in the effective boundary conditions are the normal flux terms.
Together, these observations suggest that the appropriate set of
boundary conditions to impose on the governing equations
(\ref{eq:c_eqn_bulk_strongly_nonlinear}) and
(\ref{eq:rho_eqn_bulk_strongly_nonlinear}) depends on the position on
the surface of the sphere.  Moreover, the position where the boundary
conditions switch from one set to the other depends on time as
double-layer charging progresses from the pole towards the equator.

\section{Conclusion \label{sec:conclusions}}

\subsection{ Predictions of the PNP Model }

In this paper, we have analyzed electrochemical relaxation around
ideally polarizable conducting spheres and cylinders in response to
suddenly applied background electric fields, using the standard
mathematical model of the Poisson-Nernst-Planck (PNP)
equations. Unlike most (if not all) prior theoretical studies, we have
focused on the nonlinear response to ``large'' electric fields, which
transfer more than a thermal voltage to the double layer after
charging. We have effectively extended the recent nonlinear analysis
of Bazant, Thornton and Ajdari~\cite{bazant2004} for the
one-dimensional charging of parallel-plate blocking electrodes, to
some new situations in higher dimensions, where the potential of the
conductor is not controlled. Instead, electrochemical relaxation in
our model problems is driven by a time-dependent background electric
field applied around the conductor, whose charges are completely free
to relax.  We are not aware of any prior analysis of nonlinear
response in such problems (whether or not using the PNP equations),
and yet it is important in many applications, in microfluidics,
colloids, and electrochemical systems, where applied fields or
voltages are often well beyond the linear regime. 

We have focused on the structure and dynamics of the double layer and
the development of bulk concentration gradients. A major goal has been
to move beyond the traditional circuit models commonly used to study
the linear response of electrochemical systems.  Through a combination
of analytical and numerical results, we have shown that significantly
enhanced ion concentration within the double layer is a generic
feature of nonlinear electrochemical relaxation around a conductor.
By interacting with the bulk electric field, the enhanced
concentration within the double layer leads to large surface current
densities.  In addition, because the double layer does not charge
uniformly over the surface, tangential concentration gradients within
the double layer itself lead to surface diffusion.  Due to their
coupling to bulk transport processes via normal fluxes of ions into
the double layer, these surface transport process drive the formation
of bulk concentration gradients.

We have also found that bulk concentration gradients are \emph{always}
present to some degree and play an important role in allowing the
system to relax to the steady-state.  For weak applied fields, they
are often neglected because they only appear as a first-order
correction to a uniform background concentration profile.  However,
for strong applied fields, the variations in the bulk concentration
becomes as large as the background concentration, so they cannot be
ignored.  These bulk concentration gradients lead to bulk diffusion
currents which result in net circulation of neutral salt in the region
near the metal colloid particle.

Another key contribution of this work is a careful mathematical
derivation of general effective boundary conditions for the bulk
transport equations in the thin double layer limit by applying matched
asymptotic expansions to the PNP equations.  We derive a set of
boundary equations that relate the time evolution of excess surface
(double-layer) concentrations to surface transport processes and
normal flux from the bulk.  An interesting feature of the effective
boundary conditions is that they explicitly involve the small
parameter $\eps$ and may have different leading order forms depending
on the characteristic time scale and the magnitude of double layer ion
concentrations.

It is beyond the scope of this article, but worth pursuing, to study
``strongly nonlinear'' dynamics in detail, where (according to the PNP
equations) the double layer adsorbs enough ions to significantly
perturb the bulk concentration and cause surface transport to bulk
transport of ions.  The challenging issues discussed at the end of
Section~\ref{sec:strongly_nonlinear_dynamics} need to be addressed in
order to gain a deeper understanding of the rich behavior of metal
colloid systems in this practically relevant regime.  Also, it would be
beneficial to validate solutions of the effective bulk and boundary
conditions in the thin double layer limit against solutions for the
full PNP equations.  The utility of the thin double layer
approximation cannot be fully appreciated until this comparison is
completed.

\subsection{ The Need for Better Continuum Models }

Unfortunately (or fortunately, depending on one's perspective), the
theoretical challenge of strongly nonlinear electrochemical relaxation
is much deeper than just solving the PNP equations in a difficult
regime: It is clear that the dynamical equations themselves must be
modified to better describe the condensation of ions in highly charged
double layers. As emphasized by Newman~\cite{newman_book}, the PNP
equations are only justified for dilute solutions, since each ion
moves in response to an independent stochastic force with constant
diffusivity and mobility, interacting with other ions only through the
mean long-range Coulomb force. Important effects in concentrated
solutions, such as short-range forces, many-body correlations, steric
constraints, solvent chemistry, and nonlinear mobility, diffusivity,
or permittivity, are all neglected.

It is tempting to trust the PNP equations in the case of a dilute {\it
  bulk} electrolyte around a initially uncharged surface. However, as
we have shown, the model predicts its own demise when a large electric
field is applied, due to enormous increases in {\it surface}
concentration in the diffuse part of the double layer, even if the
bulk concentration is small. Note that the condition for strongly
nonlinear relaxation (\ref{eq:strong1}) or (\ref{eq:strong2}) is
similar to the condition for the breakdown of the PNP equations --
both require applied voltages across the double layer only several
times the thermal voltage. In particular, the Gouy-Chapman solution to
the PNP equations predicts an absurd concentration of counterions of
one per $\AA^3$ at the surface for a surprisingly small zeta potential
around $5kT/z_+e \approx 0.2$ V even in a fairly dilute 0.01 M
electrolyte. This critical voltage is routinely applied to the
double layer in electrochemical systems.

Of course, electrochemists are well aware of this problem; in fact,
Stern originally proposed the compact layer of adsorbed ions, outside
the continuum region, as a way to cutoff the unbounded double-layer
capacitance~\cite{stern1924}. Since then, various empirical models of
the compact layer have appeared~\cite{delahay_book,bockris_book}, but
steric effects (or other nonlinearities) in concentrated solutions
have received much less attention. Borukhov, Andelman and Orland
recently postulated a continuum free energy for ions, taking into
account steric repulsion with the usual mean-field electrostatics and
minimized it to derive a modified Poisson-Boltzmann equation for
potential in the double layer~\cite{borukhov1997}. Their model
predicts a steady, equilibrium profile of ions with a condensed layer
at the steric limit for large zeta potentials. By extending this
approach to obtain the chemical potential, Kilic, Bazant and
Ajdari~\cite{kilic2006} have derived modified-PNP equations and
studied steric effects on double layer charging, but further model
development is still needed, not only for steric effects, but also for
field-dependent permittivity and/or diffusivity. The validity of a
continuum model at the scale of several atoms in the most condensed
part of the double layer must also be viewed with some skepticism.

Nevertheless, in spite of these concerns, it is a natural first step to
study nonlinear electrochemical relaxation using the standard PNP
equations as we have done. The details of our results will surely
change with modified transport equations and/or surface boundary
conditions, but we expect that some key features to be robust. For
example, steric effects could decrease the capacitance of the double
layer at large zeta potentials, but surface conduction and adsorption,
coupled to bulk diffusion, should still occur, albeit perhaps with
smaller magnitude for a given applied field. Also the mathematical
aspects of our boundary-layer analysis, such as the derivation of the
surface conservation law (\ref{eq:mubc}), could be applied to any
continuum transport equations.

\acknowledgments
This work was supported in part by the Department of Energy Computational 
Science Graduate Fellowship Program of the Office of Science and 
National Nuclear Security Administration in the Department of Energy
under contract DE-FG02-97ER25308 (KTC) and by the MRSEC Program of the
National Science Foundation under award number DMR 02-13282 (MZB and KTC).
The authors thank A.~Ajdari, J.~Choi and L.~H.~Oleson for many 
helpful discussions.
We are particularly grateful to L.~H.~Olesen for pointing out an important 
term that we had missed in deriving Eq.~(\ref{eq:effective_flux_bc_GCS}).

\bibliography{nonlinear_electrochemical_relaxation_conductors}

\end{document}